# AFFECTIVE TEACHABLE AGENT

# IN VIRTUAL LEARNING ENVIRONMENT

## (Book Draft)

**AILIYA BORJIGIN**

**2014**

# Preface

Teachable Agent (TA) is a special type of pedagogical agent which instantiates the educational theory of "Learning-by-Teaching". Soon after its emergence, research of TA becomes an active field, as it can solve the over-scaffolded problem in traditional pedagogical systems, and encourage students to take the responsibility of learning. Apart from the benefits, existing TA design also has limitations. One is the lack of enough proactive interactions with students during the learning process, and the other is the lack of believability to arouse student's empathy so as to offer students an immersive learning experience.

To solve these two problems, we propose a new type of TA – Affective Teachable Agent, and use a goal-oriented approach to design and implement the agent system allowing agents to proactively interact with students with affective expressions. The ATA model begins with the analysis of pedagogical requirements and teaching goals, using Learning-by-Teaching theory to design interventions which can authentically promote the learning behaviors of students. Two crucial capabilities of ATA are highlighted –Teachability, to learn new knowledge and apply the knowledge to certain tasks, and Affectivability, to establish good relationship with students and encourage them to teach well. Through executing a hierarchy of goals, the proposed TA can interact with students by pursuing its own agenda.

When a student teaches the agent, the agent is performed as a "naive" learning companion, and when an educator teaches the agent during the design and maintenance time, the agent can perform as an authoring tool. To facilitate the involvement of educators into the game design, we develop an authoring tool for proposed ATA system, which can encapsulate the technical details and provide educational experts a natural way to convey domain knowledge to agent's knowledge base.

An agent-augmented educational project, Virtual Singapura, is developed as a show case to illustrate how the ATA model is implemented in a real educational system. Formative and summative assessment results are provided to show that ATA can assist students to reflect on their learning process, and consequently improved student's learning experience and encourage them to take the learning responsibility.

# Table of Contents

# List of Figures

# List of Tables

# Chapter

# 1

# Introduction

With the development of personal computer devices, internet connection, and 3D virtual world, computer-aided education is at its best time with an alluring and challenging prospect. Researchers from education, computer engineering, as well as psychology are working together and engaged in offering students a flexible, personalized, and immersive learning experience. As a consequence, various pedagogical agents have been developed for agent-augmented educational software. Among those efforts, a special type of agent called Teachable Agent (TA) emerged around the 1990s. Soon after its emergence, research on TA has become an active field. A Teachable Agent is designed as a naive learner which needs students to teach and in this way, help students improve their learning capability. Rather than traditional pedagogical agents, which perform as a tutor to deliver knowledge to students, this type of agents may stimulate students to be more proactive and willing to take the learning responsibility for TAs.

## 1.1 Why Teachable Agents

The idea of a Teachable Agent comes from a famous educational approach, Learning-by-Teaching [1]. Learning-by-Teaching is an educational method for students to learn by teaching their peers. This theory has got many supports from the research in peer-assisted tutoring [2], reciprocal teaching [1], self-explanation [3], and many others [4, 5]. A famous experiment from Bargh and Schul [6] observed that people who prepared to teach others on paragraph comprehension had better quiz results about the paragraph understanding than those prepared to do the comprehension for themselves. A similar situation was also reported by Gaustard [7], who found that student tutors often benefited as much or more than their tutees. These interesting phenomena bring educators a novel viewpoint for motivating students to learn. Particularly for the E-learning domain, researchers can exploit intelligent agent technology to build a virtual student for students to teach. Students can act as instructors to teach the "naive" virtual agent. As a result, the students can directly benefit from the Learning-by-Teaching process.

Currently, most of the pedagogical agents play the role of an experienced tutor [8-10] or a knowledgeable learning companion [11]. These efforts in some degree may be considered as the replication of lecture-based teaching or an automation of the conventional teaching process through a computer-based interface [12]. Among those approaches, one of the problems is how to utilize the full scope and potential of pedagogical agents to authentically elicit student's learning behavior and highly motivate students [13]. The instructions given directly to the students may reduce student's self-exploration and self-discovery [14]. The experience of taking the proactive in learning is important for students to master the learning skill and can precisely encode the information to their memory [15]. The TA is purposely designed for authentically arousing a student's initiative and avoiding a passive learning process.

According to [15], the benefits of a TA are mainly from three perspectives:

- *Structuring the student's knowledge organization.* Teaching processes, such as providing explanations or receiving feedbacks, bring teachers an opportunity to reorganize his/her knowledge structure and gain deeper understanding of the domain knowledge. The preparation of lessons and the interactive communication during lessons pushes the teacher to think more deeply and thoroughly in order to express his/her ideas in a concise and correct way. Thus, the teaching *per se* will highly improve the teacher's learning. According to Fantuzzo's work [16], TAs can help students structure and reorganize their knowledge.

- *Taking responsibility.* When playing the role of a teacher, the students need to take the responsibility of tutoring the learners; they should decide how to learn a particular piece of knowledge and judge whether the content is relevant or not. In [17], researchers mentioned that the challenge of teaching others highly motived people of all ages because of the sense of responsibility. Thus being a teacher is a strong motivation for students to engage in learning and willingly take the responsibility of their own learning.

- *Enhancing reflection and metacognition.* Reflection is a powerful tool for teachers to achieve effective tutoring. Some educational researchers [5] reported that teachers often reflected on their communications with students during and after their lessons. For TAs, Roscoe & Chi [18] verified that TAs can increase student's knowledge reflection and self-explanation. When students interact with TAs, the interaction will provide a spur to reflection related to the learning material. This may help students deeply understand the knowledge and improve their metacognitive capability.

The Teachable Agent, as a new carrier of the learning-by-teaching theory, attempts to realize these educational benefits in a virtual learning environment, and help students to learn in a novel and interesting way.

## 1.2 Research Scope and Educational Requirements

In order to achieve the benefits mentioned above, what educational requirements should a TA fulfill? As a pedagogical system, the most important responsibility of TAs is to facilitate students' learning. In detail, there are three essential facets to achieve this educational purpose, including students' interest towards learning, students' learning performance and students' self-efficacy after using a TA.

- *Students' interest towards learning.* Motivation is crucial for learning. Once a student becomes interested in learning, he/she is already halfway to mastering the learning essence. Students' interest towards learning in our research refers to learners' disposition towards the task of applying knowledge they have learned and towards whether they would like to work with TAs. If a TA can attract a student to engage in learning voluntarily and to willingly spend more time, rather than being passively involved in the learning process, it will be a good tool to motivate students.

- *Students' learning performance.* Learning performance can be measured in various ways. In this research, we specifically consider students' ability to apply learned knowledge to complete tasks/tests, and the ability of recalling learnt material and inherent relationships. The ability of using what has been taught to solve problems is the ultimate goal of education. We will measure their performance in tests, and consider whether they apply enhanced reasoning to think deeply, and whether they build up knowledge association between current learning content and their prior knowledge. The next perspective, knowledge retention is also a fundamental factor of learning. In order to help students obtain a deep understand, the TA should allow students to become immersed in the learning scenario, and induce students to encode information in a multi-dimensional space so as to obtain a more comprehensive and integrated understanding. Neither recall nor the capability of applying knowledge should be ignored when assessing students' learning performance. With the two dimensions together, we can use TAs to reflect students' learning proficiency.

- *Students' self-efficacy after using a TA.* Self-efficacy represents the belief of one's own competency towards the completion of a particular task and the reaching of a goal. It is the perspective of students' self-evaluation of his/her learning capacity. Different people have different judgment for themselves. If a TA system promotes students to have the feeling of achievement and make them more sensitive to their improvements during the learning process, it will provide a positive influence and bring students encouragement and confidence.

All above are the three key perspectives that we need to consider when designing TAs. As a promising research direction, researchers working on TAs have won many achievements. At the same time, existing TAs also have large room to improve in many aspects. Our research aims to improve the TA design through disclosing the limitations of previous studies and further enhance students' learning interest, learning performance and their self-efficacy.

## 1.3 Research Problems & Objectives

As a flourishing research field, teachable agent has been studied for more than twenty years. At the early stage, researchers were trying to build TAs on specific and well-defined problems such as mathematical problems that ask students to define computation rules and sequences by choosing the right solutions from multiple answers. The example is MCLS (Math Concept Learning System)[19]. Later on, a system called DENISE (Development ENvironment for an Intelligent Student in Economics) [20] was developed to use causal qualitative models in learning economics. Unlike the rules and sequences in MCLS, DENISE asked students to construct sequences of causal relations via Socratic strategy – the teacher probed a student by asking questions. Based on above early attempts (other related literatures can be found in Chapter 2), researchers began to realize the importance of shared representations of knowledge. Shared knowledge representation means the knowledge is represented in a viewable, explicit, and clear way through all the interactions between students and agents. To achieve this, a new system was developed by the corporation of AAA Lab in Stanford University and the Teachable Agent Group in Vanderbilt University, that developed "Betty's Brain" [21]. The system adopted the Concept Map as a niche graphical tool to represent knowledge and applied exhaustive search of concept maps to reason and thus answer students' questions. Similar to Betty's Brain knowledge representation and interface design, the SimStudent Group [22] in Carnegie Mellon University introduced two new extensions to explore TAs, consisting of developing TAs for educational researchers to explore learning theories, and using TAs as a domain knowledge authoring tool.

Apart from achievements discussed, current TA development was also reported with limitations. The most obvious two drawbacks are the lack of enough initiative during the interactions with students [15], and the lack of believability to arouse students' empathy so as to immerse students in the learning experience [23].

## 1.3.1 Enhance TAs with Proactive Interactions

For the first perspective, the proactive interaction provided by TAs is important, because the interaction between TAs and students is the core that determines whether a TA can facilitate students' learning. The interaction contains a bidirectional exchange. Even though TAs perform as a learner, they need to avoid responding to students passively[15, 24]. The next question is how to define the initiative of a TA? In our research, we focus on three aspects.

- The ability of TAs to learn new knowledge from students in order to encourage students to reflect on the learning materials.
- The ability to apply the learnt knowledge, and provide feedback to students in order to give them an opportunity to validate and rethink their teaching.
- The ability to establish good relationships with students and encourage them to teach well in order to promote students to take the responsibility of learning.

In this work, we aim to use a goal-oriented approach to overcome the lack of proactive interactions, by enabling TAs with agent's goal settings. Goal orientation is one of the key features in agent systems. Agents are goal oriented [25], since goal selection and concrete behavior selection form the foundations of an agent's initiative. Only an agent has the capability to choose its own goals and act to achieve its goals, the agent is called "active". Therefore, a goal-oriented teachable agent has more flexibility to provide students with highly proactive interactions. To achieve this, we will bring to TAs an autonomous mechanism to organize various activities and selectively carry out the right activity at the right time.

## 1.3.2 Enhance TAs with Affective Capability

The second area of limitation is the existing TAs lack believability to immerse students in the learning scenario. Does a TA's emotion influence students' interest toward the agent as well as the related learning tasks? Many works gave a confirmed answer. Baylor et al. [26] demonstrated in their work that when a pedagogical agent showed its own emotions, students felt that the agent was more inviting and engaging. Mori et al. [27] also mentioned that students viewed the tasks more enjoyable if the agent had emotions. Kim et al. [28]conducted experiments and found that student's interest in the learning task and their feelings towards an pedagogical agent were positively influenced by the agent's positive affect. She also mentioned that when agents paid attention to the emotional states of students, the students were more willing to interact with the agent during learning. It is a fact that the more a TA behaves like a real person, the more a learner will be engaged. That is because when people treat the agent as a social mate, they tend to care and are more likely to help [29]. A TA with emotional reactions can give a student an impression that he/she and the agent are "in things together" [30]. It can engage the student to care about his/her teaching performance, and spend more effort on improving the agent's learning progress. Therefore, emotions play an important role.

Considering a student's learning performance (we define the learning performance as the capability of retention and problem solving using previous knowledge), many researchers argued that the efficiency of information processing and information retrieval were influenced by the combination of cognition and affect [31-33]. For information processing, people tend to remember emotional events better, regardless of whether the emotion is good or bad [34]. Therefore, when a TA with emotional capabilities involves a student into an emotional learning scenario, the student is likely to have a better remembrance than that in a plain scenario. For information retrieval, people tend to have different processing styles when one's affective states are different [35]. For instance, positive emotions likely activate a "heuristic, creative and top-down" style of information processing. Negative emotions, on the contrary, tend to foster a "detail-oriented, systematic and bottom-up" style of information processing. Similar findings were also mentioned in [36] that people who received positive feedback were happier, and consequently more committed and productive in their learning processes.

In real life, the teacher's expression of emotions affects the way a student reasons over the causes of a success or a failure [37]. For example, when a student did not provide a right answer for the question, other people's opinion may cause the student to interpret the same situation in different ways, such as attributing the failure to his lack of intelligence or to the lack of efforts. At this stage, if somebody encourages him to reflect deeply on why the answer is wrong and helps the student to find out the reason for giving a wrong answer, the student will realize that as long as he carefully analyzes the problem he will get the right answer. This kind of self-efficacy is important for students' learning. A well-designed TA has the potential to influence learners' self-efficacy by responding to student positively. Based on the analysis above, we argue that a TA needs the capability of generating emotions when confronting various situations during the Learning-by-Teaching process.

## 1.4 Summary of Contributions

In light of the above discussion, our research objective is to propose an affective teachable agent (ATA) which can proactively interact with students with affective expressions to increase students' learning motivation, improve student's learning performance, and enhance their self-efficacy. The main contributions of this research include five aspects.

(i) *Goal-oriented ATA Model.* The formalization of an agent system and the definition of corresponding concepts can help us to analyze the problem domain, and clearly specify the tasks to be solved. The goal-oriented modeling approach models the TA deliberation process from the TA's goal setting to goal selection, which can enable a TA with proactive behaviors and proactive feedbacks to students. The goal hierarchy of agent design provides flexibility in changing the system design or enhancing the capability by extending current system implementations.

(ii) *Teachability Reasoning Model.* Regarding the educational aspect, one of the most important characteristic of TA is to make its internal states transparent to students. The proposed TA system lets students teach TAs through an interface with explicit knowledge representation, and allows students easily to develop the level of structured knowledge embodied in the agent. Meanwhile, our TAs can practice what they have learnt from students in the 3D virtual world to give students a direct impression on how good their teaching is and how workable their approaches are. The design can help students reflect on their teaching/learning and restructure their knowledge network.

(iii) *Affectivability Reasoning Model.* The proposed affective teachable agent is specifically designed to extend traditional TA systems with the capability to elicit emotions and the capability to clearly represent the emotional process, making the internal state of the teachable agent transparent to students. OCC (Ortony Clore Collins) [38], as the most recognized emotional model, is used as the foundation of emotion elicitation. To bring OCC theory with a concrete computational representation, we used Fuzzy Cognitive Map (FCM) to do the quantitative modeling, because it is a convenient modeling tool that can transform an OCC-based scenario to a computable causal graph, and it has the potential to merge other newly coming elements as sub-graphs to easily depict any emerging complex situations.

(iv) *ATA Game Authoring Approach.* Owing to the Teachability characteristic, ATAs have been used in two ways by different "teachers": when a student teaches the agent, the agent performs as a "naive" learning companion, but when a teacher teaches the agent during the designing process, the agent performs as an authoring tool which can provide teachers with a natural way to convey domain knowledge to the agent's knowledge base. In this way, our teachers can be easily involved in the design process. The authoring process can hide the game programming details from educators, and the setting of learning goals and educational contents constitute a well-defined game structure since it works as a map that tells developers what contents should be involved.

(v) *Deploying ATA in Virtual Singapora Project.* The proposed ATA has been implemented in the project Chronicles of Singarpura. Our objective is to realize the educational benefits based on the teachability and affectivability of ATA, and also bring teachers an authoring tool to design educational game scenarios.

## 1.5 Organization of the Book

The rest of chapters are organized as follows:

- Chapter 2 provides the literature review of TAs and highlights the limitations of existing TA design. As we intend to model a TA's emotions, related work on emotional modeling is also introduced.

- Chapter 3 focuses on the proposed ATA model, which is designed to achieve two main capabilities of the TA, to learn from students and to be affective during the interaction process.

- Chapter 4 introduces the system design of the ATA model. By considering the involvement of educators in game design, an authoring tool is developed for encapsulating the technical details.

- Chapter 5 exemplifies the ATA model via the development of a real educational project with formative and summative assessments of the ATA system.

- Chapter 6 concludes the research work and discusses potential directions for future work.

**Chapter**

# 2

## Literature Review

## 2.1 TA – An Interdisciplinary Research Field

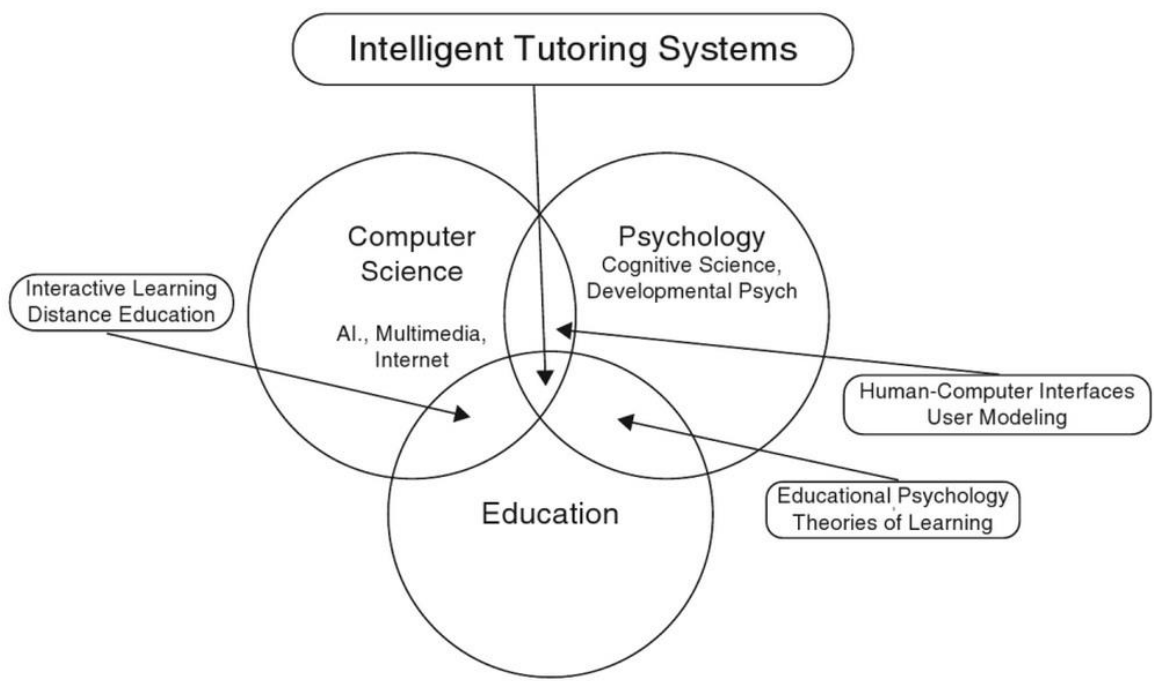

Figure 2.1: Intelligent Tutoring Systems as an Interdisciplinary Research Field [13]

In the early 1970s, a few researchers adopted the human tutor as a metaphor for their computer-aided educational models to enhance traditional lecture-based instruction through designing "intelligent" interaction between virtual tutors and students [39]. The related research involves theories and approaches from multiple disciplines. According to [13], research on Intelligent Tutoring System is the intersection of three main disciplines which are shown in Figure 2.1. *Education* emphasizes the design of teaching strategy, and aims to figure out what is the best pedagogical intervention for each individual; *Artificial Intelligence*, a subfield of *Computer Science*, addresses how to build up an "intelligent" system to teach students; and *Cognitive Science*, the subfield of Psychology, provides insights into how we learn and feel.

The design of a teachable agent system also needs cooperation from these three areas. From the perspective of *Education*, Learning-by-Teaching (LbT) theory provides the educational fundamental for TA design. Findings from LbT may guide the intervention design of TA systems, such as the educational requirements of the system, the key problem of domain knowledge representation, the student's preference of TA features, the advice on the communication style between TA and students, etc.

All those educational design issues finally should be realized through a computer-based system, and *Artificial Intelligence (AI)* becomes necessary at this stage. AI brought TA the potential to understand students. By monitoring the learning environment in real time, TA can reason about how to provide proper intervention to which student and at what time.

Finally, *Psychological Theories* may help researchers to understand the learning process of students and help simulate TAs with human-like behaviors. In this book, we use affective theories in psychology to simulate affective behaviors of TAs to make the system more attractive and make it easier to immerse the students in the learning environment.

With this interdisciplinary structure for a TA, many researchers have contributed from different perspectives. The related research centers and projects are introduced in the following Section 2.2.

## 2.2 Teachable Agents

Two thousand years ago, Roman philosopher Lucius Annaeus Seneca mentioned "docendo discimus" in his letter to a friend in Latin, which in English means "we learn by teaching" [40]. Nowadays, many educational researchers are continuously working on it. German Professor Jean-Pol Martin systematically built an organized theory called "Lernen durch Lehren" (LdL) in German (learning by teaching in English) which has been widely used in language teaching and other courses. Based on this pedagogical progress, researchers on Artificial Intelligence also attempted to build E-learning systems with the capability of being taught by students.

### 2.2.1 "Virtual Students" – the Early Attempts of TA

Several early projects on TA systems focused on simulating "virtual students". The main target was not to build a virtual avatar that looks like a student, but to explore the learning capability of human beings and simulate the basic learning functions. One example was the project "Math Concept Learning System (MCLS)" in 1989 [41]. This project built a TA for well-defined mathematical problems. The system let students define computational rules and sequence for some linear equations. The system generated a set of general rules via an inductive machine-learning algorithm "Iterative ID3" based on the inputs from students. Both the student and the system would solve a linear equation separately, and if the solutions were not the same, the system would let the student choose the right from wrong. For this TA system, the way of generating computational rules by computer is not visible to students. In other words, the student could not directly know the internal reasoning mechanism of the TA system, so they would face a problem when they wanted to find the logic flaw of the TA.

In the early 1990s, another system DENISE (Development ENvironment for an Intelligent Student in Economics) [20], was designed to teach economics through qualitatively modeling the causal relationship between economic concepts. Unlike the rules and sequence in MCLS, DENISE asked students to construct sequences of causal relations via a Socratic strategy – the teacher probes the student by asking questions (as Figure 2.2). The system used a "reversed Socratic strategy" to let the "virtual student" ask the student tutor multiple questions and in this way, the virtual student learns the student's knowledge and promote the student to reflect on their knowledge learning process. The participants reported difficulty in recalling concepts and relationships they had taught DENISE earlier [42]. This may be a consequence of the invisible representation of the internal reasoning process, because without a visual representation, it could be difficult for students to understand why DENISE answered a question in a particular way.

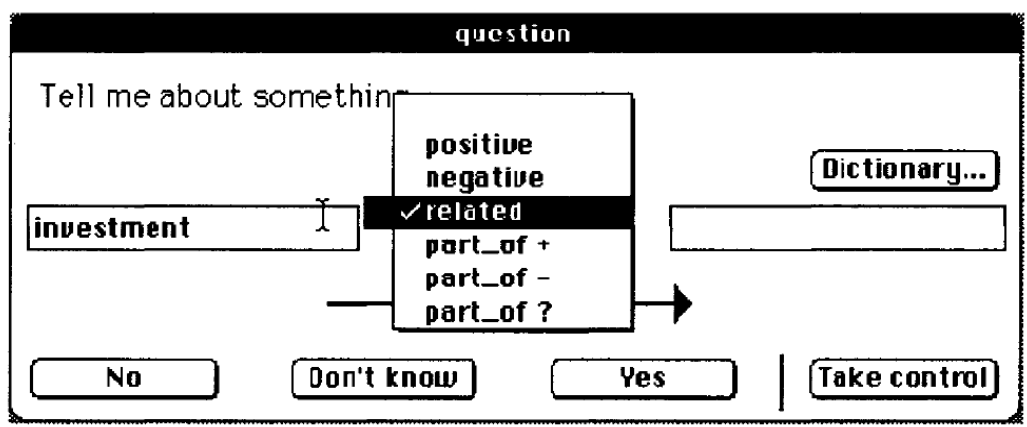

Figure 2.2: The Syntax of DENISE [20]

Apart from mathematics and causal relationship inquiries, a TA system was also designed for learning computer program. Project "Diagnosis-Hint Tree" (DHT) [43] was a diagnosis tool for a certain programming language. It worked as a debugging-tool, providing students several possible program solutions with a tree structure. It asked students to guide the computer on how to program through diagnosing the trees. The project also used comparative experiments to demonstrate the pedagogical benefits of Learning-by-Teaching theory on learning computer programming. The interface is shown as Figure 2.3.

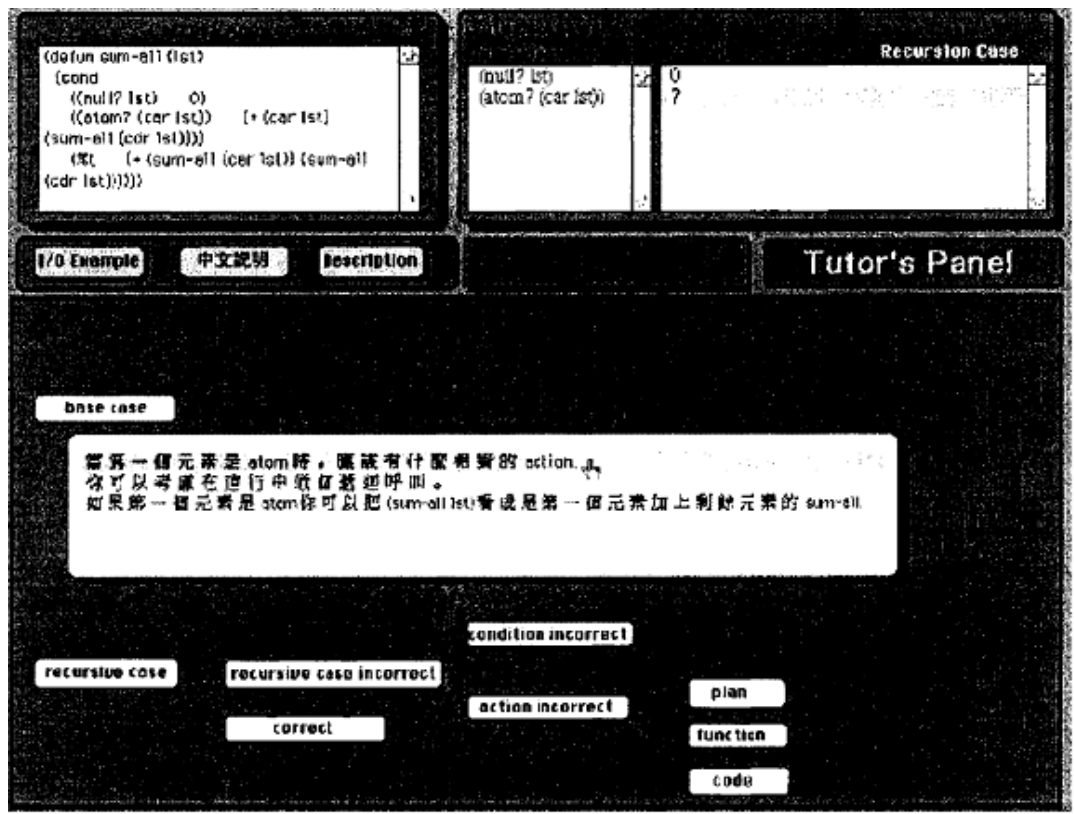

Figure 2.3: The Interface of DHT [43]

This system gave the appearance of being taught by students, but actually the agent had full domain knowledge of the program. Therefore, the system can compare student's teaching with the correct answers. Similarly, another Pseudo TA was designed in – "A Virtual Classroom" project [44]. Students were asked to teach agents and after the teaching they could observe the agents' performance in a virtual classroom. The observation of TA's performance could give students the feedback on their teaching, but the problem is that it is still an implicit representation of the teaching performance. Based on those early attempts, researchers began to realize the importance of shared representations.

Shared knowledge representation means the knowledge is represented in a viewable, explicit, and clear way for all the interactions between students and agents such as teaching, reasoning, and answering questions. Knowledge representation directly affects interactions between students and computer agents and it is the basis of students' knowledge organization and reflection. Clear knowledge representation can immediately help students understand an agent's reasoning process and find their own errors. Otherwise, it is difficult for students to establish a smooth communication with computers. To offset this limitation, a new teachable agent system called "Betty's Brain" was built by two research groups, the "AAA lab" at Stanford University and the "Teachable Agents Group" at Vanderbilt University.

## 2.2.2 Betty's Brain

To build a TA system with shared representation, two major research centers emerged at the end of 1990s. One is the AAA lab at Stanford, and the other is the Teachable Agent Group at Vanderbilt University. These two groups have made several prototypes and tests. One of the most successful prototype is Betty's Brain [15] which was designed to teach middle school students the interdependence and balance among entities in a river ecosystem. As the system should be easy to understand for students who have very limited knowledge of teaching or the learning materials, it utilized a widely acceptable technique known as a Concept Map (CM) to do the knowledge representation. The interface of Betty's Brain is shown in Figure 2.4. The big square frame on the right upper side is the display of the river ecosystem CM. The CM uses arcs with arrows to connect concepts in small rectangular frames to represent the interdependence relationships between each entity.

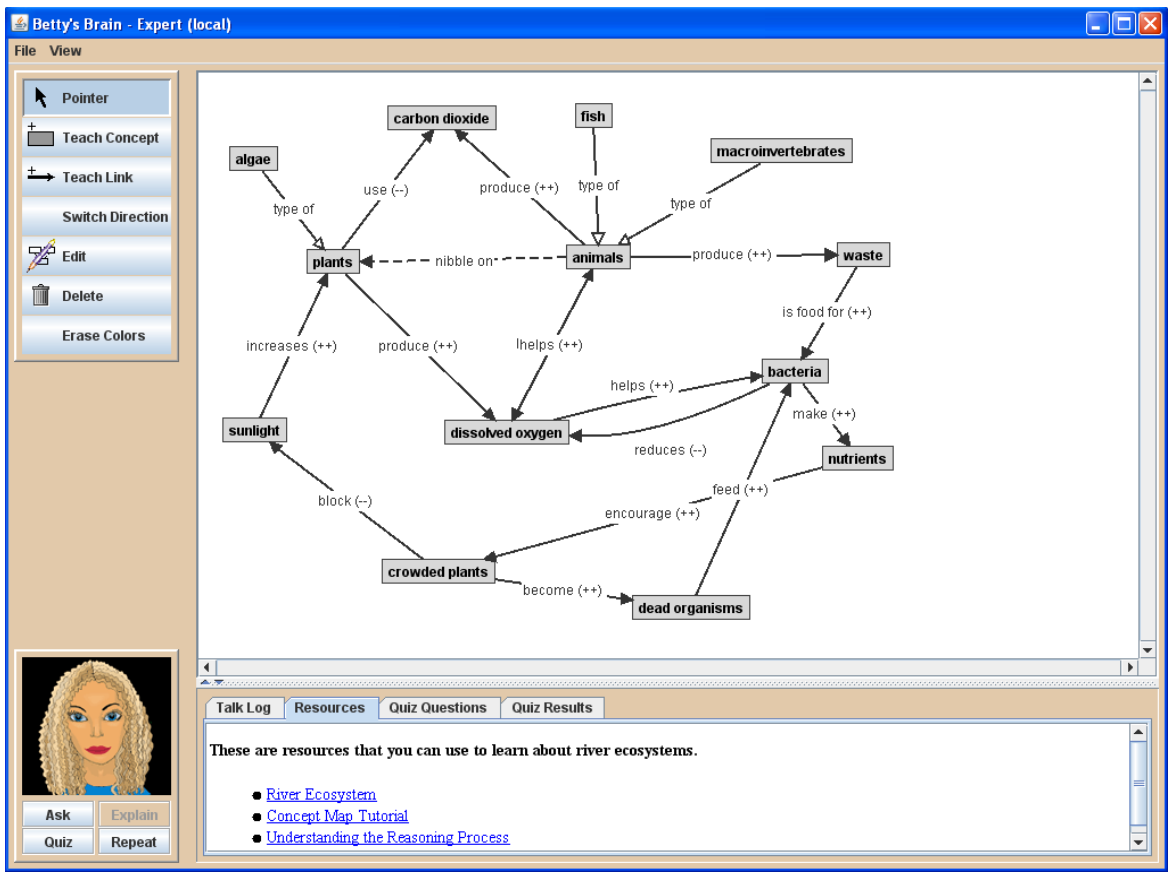

Figure 2.4: Software Interface of Betty's Brain [15]

The agent Betty is presented at the left lower side of the interface as a female avatar. The interaction with Betty contains three phases.

- The first phase is *Teach Betty*. In this phase students teach Betty by creating a Concept Map. They can drag the rectangle of "Teach Concept" to add concepts, and drag the linkage of "Teach Link" to add linkages (relationships) between concepts.

- The second phase is *Query Betty*. The software provides two types of question templates for students to put their questions to Betty. One is "If a [CONCEPT] [CHANGES], what happens to another [CONCEPT]", and the other is "Tell me about a [CONCEPT]". Betty will find all paths that lead from the source concept to the destination concept according to the CM provided by her pupil teacher in order to answer the corresponding questions.

- The third phase is *Quiz Betty* which provides the pupil teacher a choice to let Betty take a quiz and to observe how Betty performs. The system grades the quiz and provides hints to help the student to debug the CM if Betty's answer is not that expected by the student.

The aim of this system is to create an effective teaching environment with shared knowledge representation and reasoning mechanism. To aid and motivate learning, formative assessment is also provided in the teaching and quiz modes, whereas overall evaluation or summative assessment is covered in the test mode. The backstage of the system used exhaustive search for the concept map to do reasoning and answer corresponding questions.

With this platform, how can Betty's Brain realize the benefits of learning-by-teaching theory? We examine this from three aspects corresponding to the three advantages of learning-by-teaching. Firstly, Betty helps students develop structured networks of knowledge. A CM helps students avoid doing complex programming by dragging graphical nodes and arcs to construct domain knowledge. It helps students to meaningfully organize knowledge, get better memory, and easily apply knowledge to new situations. Secondly, it provides opportunities for students to take responsibility for teaching. When sending Betty to do the quiz, students can observe learning results and make changes to the CM on their own. They need to dominate all the teaching process. The third aspect is that teachable agent helps students to develop meta-cognitive skills. When they monitor Betty's learning status, such as asking her questions or sending her a quiz, it also provides the opportunity for students to monitor their own learning status. The reflection of knowledge can help them to double check what they have done, reorganize their knowledge structure and practice their rethinking habits.

### 2.2.3 Multi-dimensional Research Based on Betty's Brain

Based on the Betty's Brain system many studies were carried out, such as the design of learning environments [45-47], meta-cognition [48, 49], agent feedbacks[50, 51], student's behavior modeling[52, 53], self-regulated learning[54], etc.

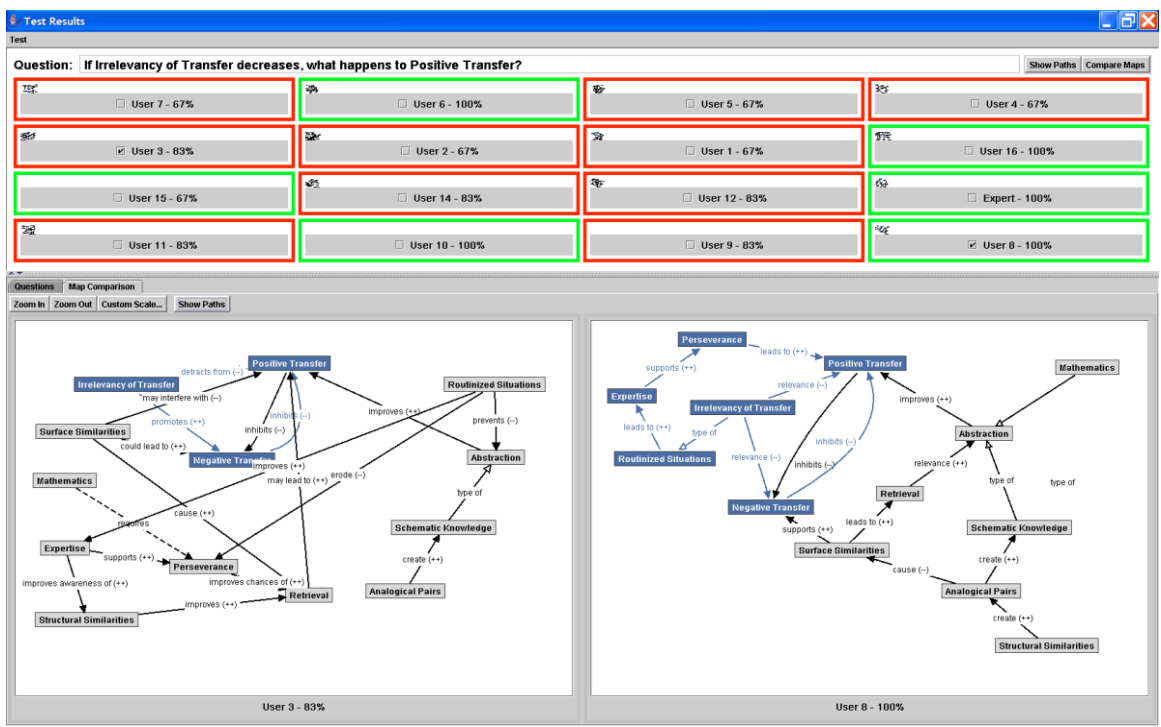

Figure 2.5: Front of Class Quiz System [55]

The *learning environment* Front of class quiz system [55] (as in Figure 2.5) is one of the extensions of Betty's Brain. The classroom teacher sends one question across the system to assess all the agents in the classroom at the same time. Results for each agent are displayed in the top panel of the tool. Red means the answer is wrong, otherwise the answer is highlighted in green. The classroom teacher can select two agents to view their difference, and set up a discussion in the classroom.

The teachable agent also can be embedded into pedagogical videogames [56]. Figure 2.6 shows a screenshot of the videogames. The teachable agent appears in the virtual world as a virtual avatar. Students are helping their agents on how to grow pumpkins, so that the virtual agent can win the pumpkin contests. The teachable agent can not only answer questions, but also take actions in the virtual world. For example, the avatar can reason about what types of fertilizer is needed so that their pumpkins can grow. Students can observe the agent's actions and can directly perceive the consequences of the agent's actions as the pumpkin grows or wilts.

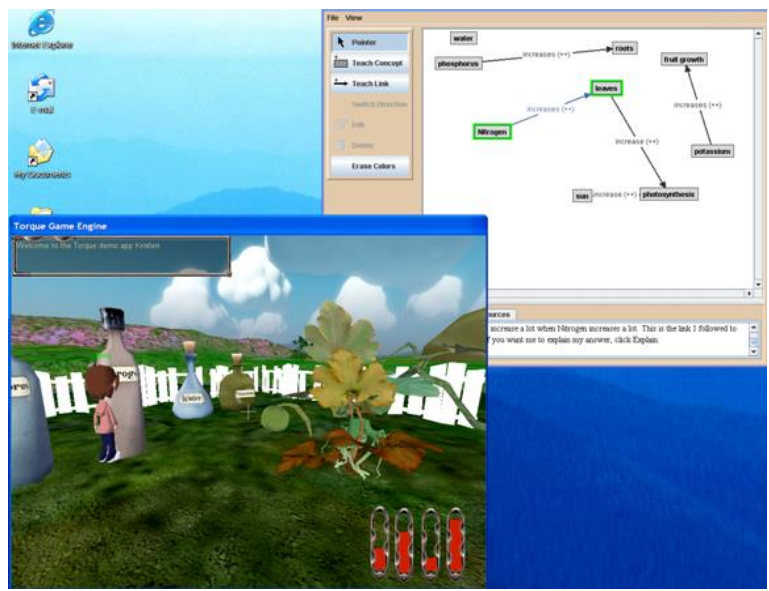

Figure 2.6: Betty in a Guided-Discovery Video Game [56]

There are many types of interactions and learning resources in the game environment. For instance, students can self-experiment to know about the function of nitrogen and phosphorous; they can observe other agents or can be informed by other agents about knowledge related to "energy". By incorporating TAs into games, students can be easily motivated. Moreover, a game-like learning approach allows multi-dimensional features and resources to be integrated coherently into one learning environment for problem solving.

*Feedback techniques* involved in teachable agent are related to maximize the computer's potential for generating interactive feedback for learners [57]. Some of the applications which use feedback technologies are illustrated together in Figure 2.7a in the All-Possible-Questions matrix, which uses automated scoring to indicate a teachable agent's accuracy for all questions, where green indicates correct, red indicates wrong, and yellow indicates correct results with wrong reasoning path. The system in Figure 2.7b is called Front-of-Class display, in which the classroom teacher can set up a classroom discussion by simultaneously quizzing multiple teachable agents in a virtual classroom environment. Figure 2.7c is called "Game Show", in which students can make their agents participate and compete in an online game show. Figure 2.7d is called "Lobby", where students can customize their agents, chat, draw concept maps, and play games.

a
b

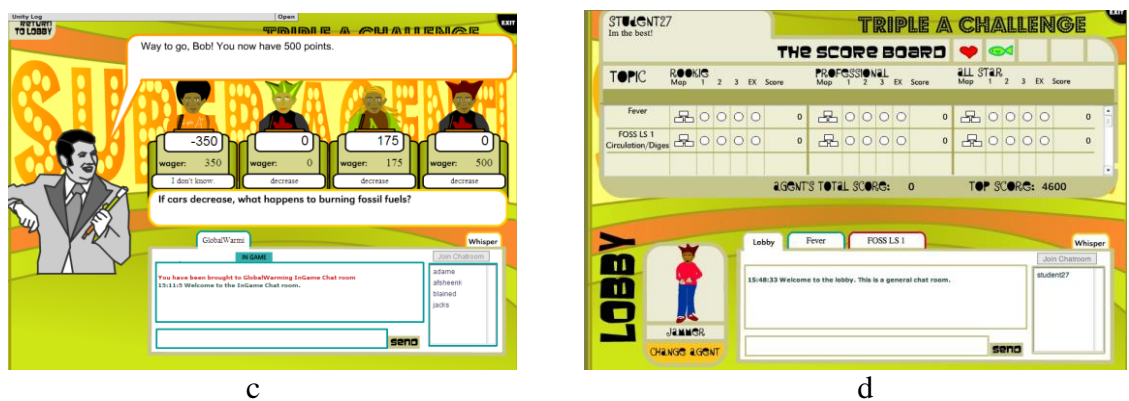

c　　　　　　　　　　　　　　d

Figure 2.7: Providing Affiliated Feedbacks with Multiple TAs [58]

As teachable agent system is highly relevant to the students' behaviors, *student behavior models* cannot be ignored in this system. Some papers have discussed on how to determine students' patterns of behavior. Jeong [59] has utilized hidden Markov models (HMMs) to analyze the students' pattern when using computer-based learning-by-teaching environments. This paper discusses how analysis techniques, and presents evidence that HMMs can be used to effectively determine students' pattern of activities.

To aid the communications between users and teachable agents, several papers [55, 60] focus on the *speech and animation* of teachable agents. Bodenheimer's work [60] aims to design agents expressing their emotions through spoken dialog and facial expressions. Their method is to record a sequence of video with a subject speaking facial images, corresponding to phonemes. The agent with this mechanism has the lip movements synchronized with speech and some basic facial expressions linked to learning tasks. However, the emphasis of the research is the speech synthesis and animation generation, the system does not provide a specialized emotion elicitation model to guide the selection of agent's emotional behaviors. Instead, the way to control and select animations is hard-coded.

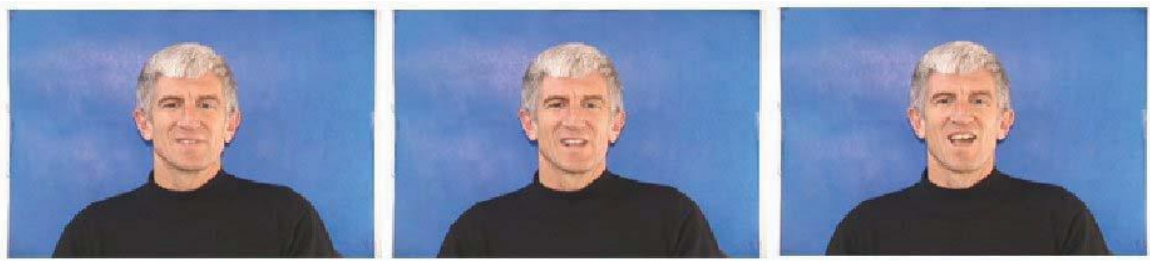

Figure 2.8: Animation Generating Strategy: Interpolating The Left and Right Images
From the Middle Image [60]

Another work related to animation focuses on the animation of thoughts [55]. Instead of displaying a situation, the teachable agents show an animation of the thinking process that a person may use to infer what is happening in a situation. Taking an example of an ecosystem, the teachable agents can animate the reasoning along the ecosystem's causal chains.

## 2.2.4 SimStudent

Apart from these traditional research directions on teachable agents, a new teachable agent project "SimStudent" [22, 61] at Carnegie Mellon University investigated two new ways to utilize teachable agent. One is providing educational researchers simulated students (teachable agents) to explore theories of learning; the other is building teachable agents as cooperated tutees for a cognitive tutoring system. In this research, attention has been paid to the design of SimStudent in terms of student-agent interaction and the method of knowledge representation.

SimStudent is an artificial intelligent agent that is able to learn procedural skills by observing student's input examples. The SimStudent has been integrated into an online learning environment with gaming features. An illustration of an online environment, called APLUS (Artificial Peer Learning environment Using SimStudent), is shown in

Figure 2.9. APLUS provides an interface for students to interact with SimStudent, as shown in the left corner of Figure 2.9. SimStudent is designed to facilitate students in mastering the skills of solving linear algebraic equations. Students are supposed to tutor the agent by collaboratively solving a problem with the agent step by step. It starts with the student to ask the SimStudent a mathematical equation. Before each move, the agent will ask the student whether a certain operation is correct. The student can choose "Yes" or "No" to lecture the agent. Since students are also new to this mathematical problem, the student can refer to example lists in the interface.

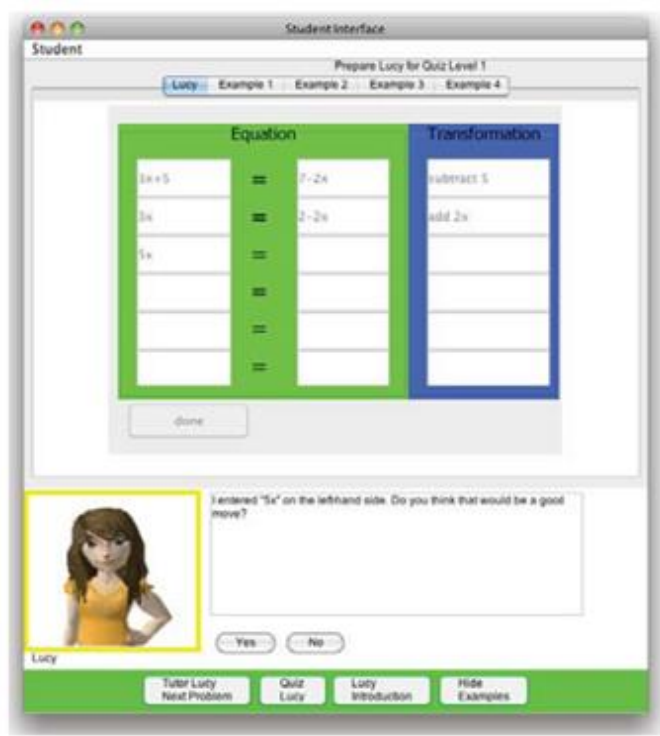

Figure 2.9: A Screenshot of APLUS with SimStudent [61]

The examples are essentially self-learning materials for students to master relevant procedural skills. If the student thinks the agent's move is incorrect, the SimStudent will attempt another action. If the student agrees with the agent's move, the SimStudent will continue to solve the problem. If the agent cannot provide any more steps, it will ask the student for a hint. A hint is a move input by the student. By such an iterative process, the agent will store the student's procedure pattern in its knowledge base. The SimStudent is implemented with an inductive mechanism such that it can produce a set of rules that simulate the student behaviors.

After tutoring, students in the classroom have a competition with each other to evaluate their agents' performance. An examination will be used to access SimStudent learning performance. The agent will use the similar procedure pattern to solve the quiz problems, and the typical quiz results are shown in Figure 2.10, where the correct procedure will help the agent to complete solving a problem, whereas a "shallow feature" learnt from the student will produce an incorrect and incomplete solution.

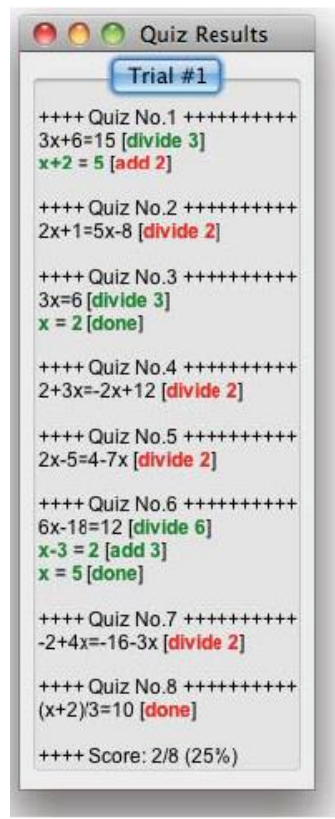

Figure 2.10: An Example of Quiz Results of SimStudent by Replicating a Student's Procedure Pattern [61]

## 2.2.5 DynaLearn's Teachable Agent

The educational software project, DynaLearn [62], aims to offer an interactive learning environment, allowing students to construct and assess their conceptual knowledge. Conceptual knowledge is different from the procedural knowledge stated in SimStudent in the fact that such knowledge involves understanding the conceptual interpretations of system behavior entailed in subjects like Biology, Physics or Environmental Science. Students are required to identity system entities with structural information, understand system processes and capture quality causal relationships between entities. The knowledge enables students to answer what-if questions about the underlying system.

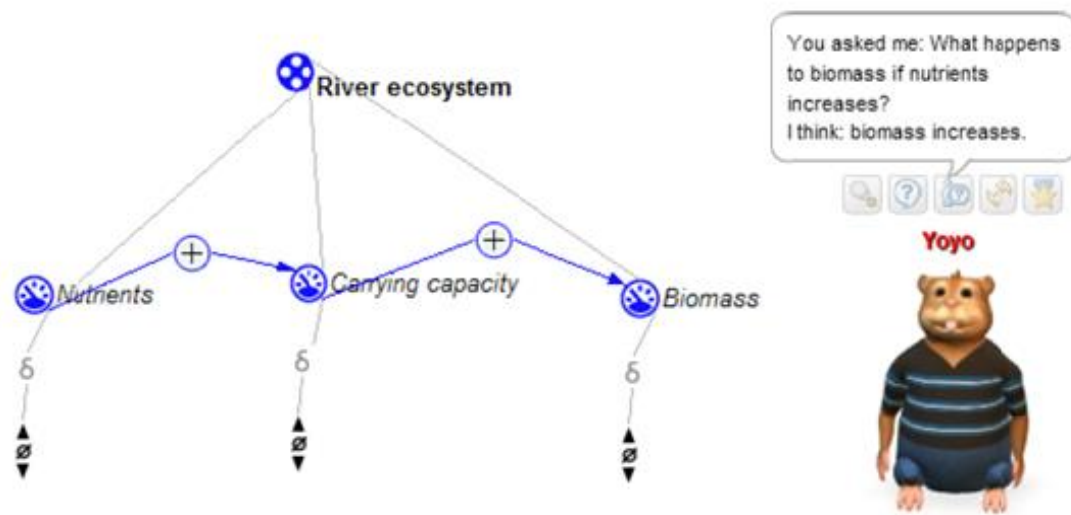

Figure 2.11: A Screenshot of DynaLearn with Teachable Agent [62]

To facilitate student learning of conceptual knowledge, DynaLearn consists of three main components in their system design, namely Conceptual Modeling, Qualitative Reasoning and Virtual Characters. The system (shown in Figure 2.11) first requires students to input their knowledge with Conceptual Modeling, which is a graphical editor for students to model their knowledge in the form of diagrammatic representations (as shown in the left of Figure 2.11).

Based on the models entered by the students, the system will assign a virtual character (as shown in the right of Figure 2.11) to the students. A student can choose the appearance and name his/her virtual character as a pet. Motivated by research finding in education, the system particularly designs the virtual character with low-competency, and slow-responding nature. It is purposely design to encourage students to build personal relationship with their pets, and emotionally motivate students to keep modifying their knowledge model for their pets. The student can ask a pet to answer one's questions in the form of dialogue with drop-down menus. The reasoning of the virtual agent is running on Qualitative Reasoning. The reasoning component contains an engine to extract behavior graphs from scenarios, initial values and system assumptions. The engine has been implemented for multiple domains, including Physics, Ecology and Economics. The agent answers the student's question based on the student's quality reasoning model.

If the answer of the pet is correct, the student will be notified directly with a positive feedback. Otherwise, the student has to re-examine the models he created previously. To clarify why the pet cannot answer his question, the student can ask the agent to explain its answer in detail. The agent will explain its reasoning process, which usually involves many steps. In order to make reasoning path transparent to the student, a history of the pet's explanations is also displayed to the student. The student can quickly check the agent's reasoning process and identify the flaws with ease.

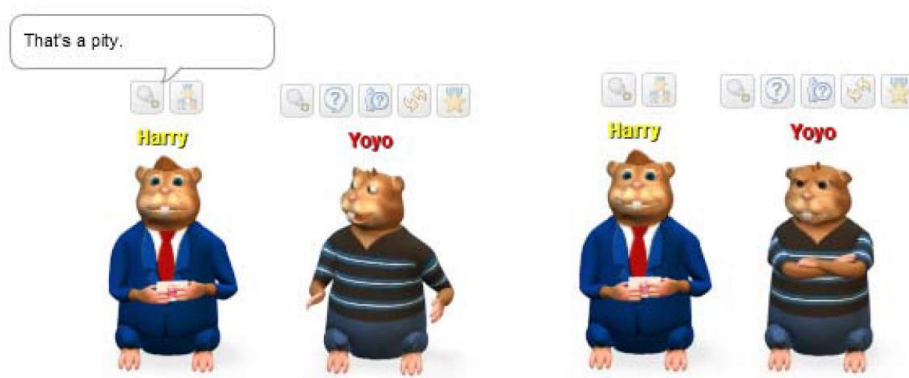

Figure 2.12: Emotion Expression in DynaLearn Teachable Agents [62]

At the end, a quizmaster (a virtual character) appears and starts to assess the pet. The ask-and-explain process between agents is similarly to that between students and agents. According to the correctness of its answers, the pet will express different emotions (as shown in Figure 2.12). If the answer is correct, the emotion will be joy. Otherwise, the agent will behave like fighting hard to obtain the right answer. This will stimulate the student's joy or empathy towards the agent, so that a tight relationship binds the student and his pet.

## 2.2.6 Summary of Existing Teachable Agents' Limitations

Through the history of teachable agent development, this type of pedagogical agent system has gradually improved with the deeper understanding of learning-by-teaching theory and the development of artificial intelligence. The emerging of 3D virtual learning environments and the constant evolution of agent technology bring teachable agents new growing space nowadays. However, current TAs also have limitations, the most obvious two drawbacks being the inability to take initiative in interactions with students [15], and the lack of believability to arouse student's sympathy and further immerse students in the learning experience [23]. The interaction between TAs and students is important for facilitating a better learning, so a TA should be more active and should avoid responding to students passively [24, 63].

On the other hand, some existing research on TAs involved emotional factors, such as bringing facial expressions to TA [60] and making the reasoning process in TA's "brain" visible[55]. However, these applications are not real affective teachable agent systems because they do not provide specific emotional models to examine the background generation process of TA's expressions. They only focused on how to vividly present different emotions through facial expression and speech, whereas the process of emotion elicitations is simple and predefined by several if-then rules. Taking the latest research work on animated TAs [60] as an example, the emotion elicitation process was hard-coded based on several predefined rules, such as "happy when performing well " or "sad when performing poorly". These rules are simplistic and based on unsystematic observation of human experiences and thus do not have psychological theory support. As a consequence, the agent without a specifically designed emotion elicitation mechanism cannot adapt to dynamic situations nor can it provide flexible and rich responses. These problems will hamper the effective interactions between users and agents, and counteract student's motivation for using the pedagogical software. Therefore, there is a need to design teachable agents which have the capability of generating emotions when confronting all kinds of situations encountered during a learning-by-teaching process.

## 2.3 Theoretical Background of Affective Modeling

With the increasing popularity of research on affective agent, numerous emotional models have been proposed. A summary made by Ortony, Revelle and Zinbarg in 2007 [64], divided the related state of art into four parts. The first part deals with cognitive and perceptual aspects of emotion, which are the "input" of emotions [38, 65, 66]; the second part concentrates on action tendencies, which are the "output" of emotions [67]; the third part discusses the facial expressions of emotions [68, 69]; and the final one looks into the affective neuroscience, with emphasis on the brain structures and mechanisms [70-72]. These theories reflect different aspects of emotion. The emotion generation needs the all the four parts to work together, as emotion is essentially a complex and multi-component system. The cognitive aspect is most important for studying affective agents. Therefore, We provide a review on cognitive theories of emotion bellow.

## 2.3.1 Various Appraisal Theories

Ortony et al. proposed that appraisal theory is the most crucial element of emotion. Appraisal refers to the evaluation of antecedent events that result in a particular emotion [64]. Many scholars have worked in this area [38, 65, 66, 73-75].

Roseman [65] proposed a cognitive emotional model, which is based on the appraisal of 6 variables such as unexpectedness, control potential, or agency. Although this model does not produce as many emotions as the OCC model does, it expands the appraisal to more variables and introduces the idea of focus. Altering the focus may lead to different emotions. For example, if one focuses on the performance in an exam of his/ her own, the generated emotion could be pride or shame, but if the focus is on the teacher who sets the questions, the emotion could be appreciation or anger. Therefore, this model is probably more comprehensive. A computational framework based on this model can be found in [76]. However, the limitation of this model is that it cannot give a complete view of the motional process. It lacks a method by which perceived events can be categorized. In addition, the reasoning process is deeply interwoven with the emotional process, and thus both external events and internal states can trigger emotions.

Sloman and Croucher [77] proposed a very interesting emotional model which includes three layers of emotion formation. At the very bottom, a reactive layer refers to the limbic system, which is in charge of primary emotions. Above it, there is a deliberate layer, capable of planning, evaluating options, making decisions and allocating resources. As such, emotions induced by goals are processed by this layer. On the highest level, there is a self-monitoring layer, which deals with emotions involving a self-identity, such as shame or grief. From low to high, each layer actually represents a stage in the evolution of our species.

The most famous cognitive model of emotion is OCC, and it has been largely used for recognition of users' emotions in computational systems and for implementation of emotions in machines. It was created by Ortony, Clore, and Collins [38], and is known as OCC due to the initial letters of the authors' name. It explains "the origins of emotions by describing the cognitive processes that elicit each of them" [78]. It provides a classification scheme for common labels of emotion based on a valence reaction to events and objects in the light of agent goals, standards, and attitudes [68].

In order to describe emotion, OCC, on one hand, defines a hierarchical organization of emotion types, which classifies emotions into distinct groups; on the other hand, OCC works out the particular factors that influence the emotions' intensity. In other words, they try to answer two main questions: what emotions is it, and how strong or what degree of intensity is it. Besides, for analyzing the generation of emotions more accurately, OCC discussed goals in a more refined manner. OCC grouped goals into three categories: precondition goals to goals of higher levels, implicit goals (well-being, life preservation), and explicit goals of short life span (attaining water, food, sleep). The OCC hierarchical structure of emotions is shown in Figure 2.13.

Many researchers used the OCC model as the basis for generating emotion elicitation rules. Typical OCC-implementations include the Tok Project [76] and the Emile Project [79].

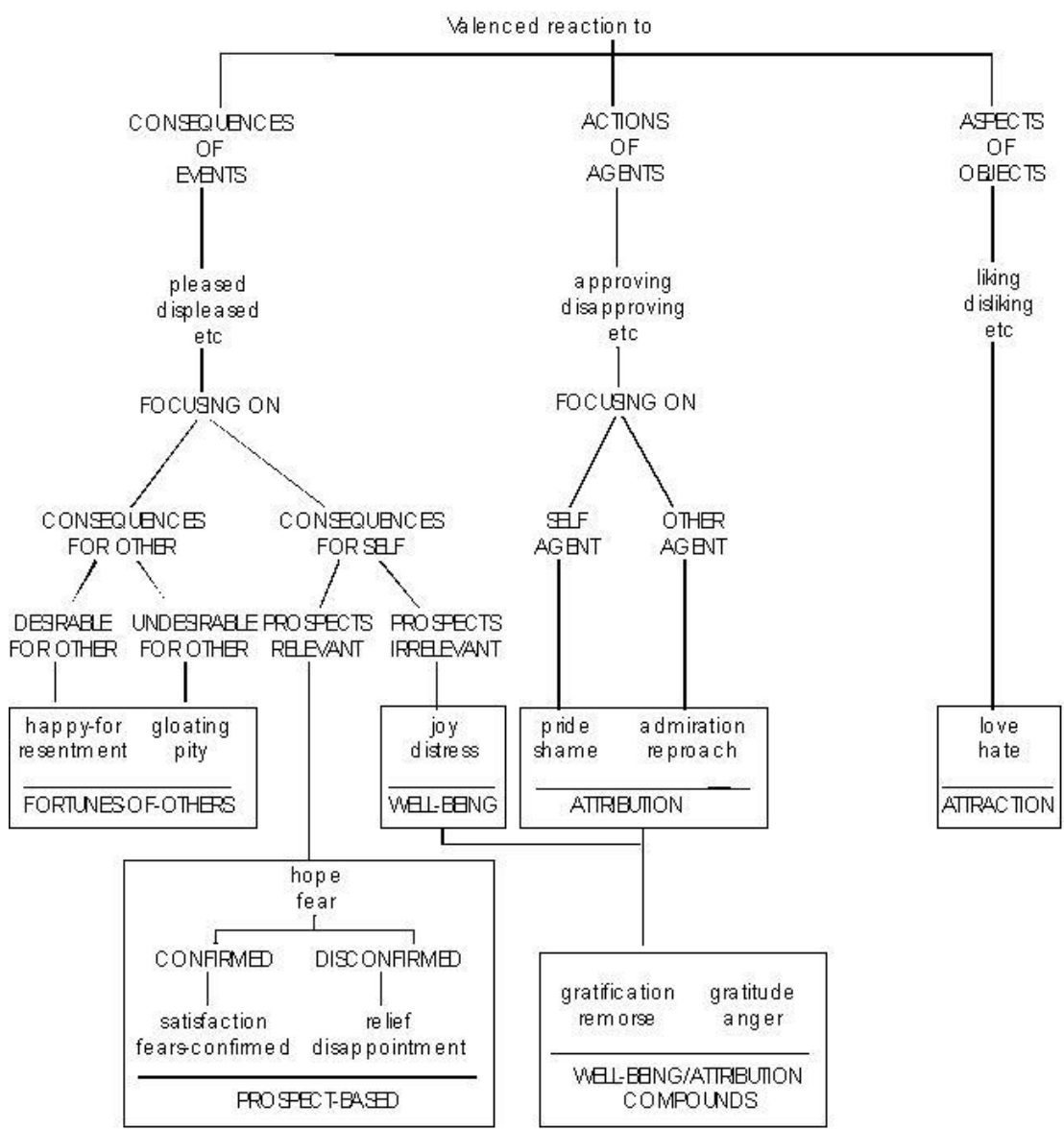

Figure 2.13: Hierarchical Structure of Emotions in OCC Model [38].

Tok Project was developed at Carnegie Mellon University by Joseph Bates, Bryan Loyall and W. Scott Reilly. Tok combines a reactivity element called HAP [77] to handle agent's goals and a module called Em [80] to generate emotion (based on the OCC model) and memory. The project has been specifically designed for use in non-real time worlds[81]. Tok handles the behavioral aspect of the Oz world inhabitants, and its relationship to Em and HAP is shown in Figure 2.14.

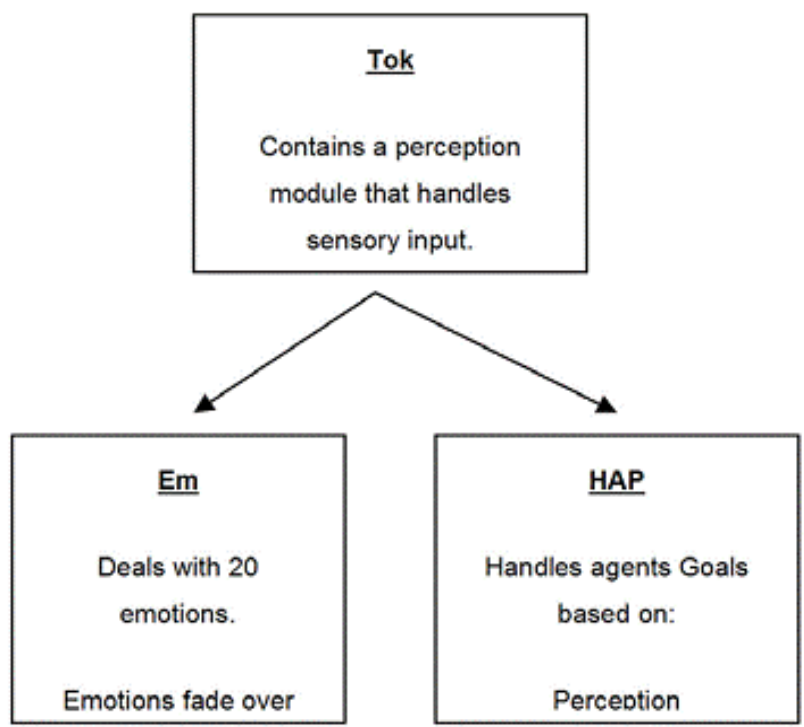

Figure 2.14: Architecture of the Three Modules in Tok [82].

Émile is built upon both Em architecture and the Affective Reasoner project (Elliot 1992). It combines two foundamental aspects of emotion modeling: appraisal and coping [82]. The appraisal process calculates the emotional states that result from given stimuli for the agent; the coping process then models the emotional influence on the action selection [83]. Em allows agents to observe the emotional states of other agents and alter behaviors accordingly[84].

The most important principle of emotional modeling, which can be used in AI, is computability, because the main job of embedding an emotional model into intelligent agents is to combine a psychological model with a computer environment so that computational requirement can be adapted within this environment. OCC theory is one of the best known emotional models to be implemented computationally [85]. Thus, my research focus is on enhancing the computability of the OCC model.

## 2.3.2 OCC Model

The OCC model assumes that emotions can arise from the evaluation of three aspects of the world: events, agents, and objects. The perceptions of goodness and badness of these aspects can be respectively judged by goals, standards, and attitudes. In this way, we can get three rules to begin the emotion elicitation:

1. By evaluating an event with goals, an event can be judged as desirable or undesirable, which is the foundation for eliciting event-based emotions.

2. By evaluating an agent's behavior with a hierarchy of standards, an agent can be judged as praiseworthy or blameworthy, which is the foundation for eliciting agent-based emotions.

3. By evaluating an object with a person's attitude, an object can be judged as appealing or unappealing, which is the foundation for eliciting object-based emotions.

In light of these elicitation rules, 22 types of emotions can be induced. The following paragraphs explain the main categories of emotion and issues related to computation in more details.

▪ **Event-based emotions**

We start by discussing the reactions to events for achieving one's goals. These affective reactions arise when a person construes the consequences of an event as being desirable or undesirable, so that judged desirability is the most important (central) variable that affects the intensity of all these Event-based emotions.

Table 2.1: Event-Based Emotions

| Event-based emotions | Examples | Eliciting Conditions | Variables Affecting Emotion Intensity |
|---|---|---|---|
| about the prospect of an event | Hope | (pleased about) the prospect of a desirable event | (1) the desirability of the event |
| | Fear | (displeased about) the prospect of an undesirable event | (2) the likelihood of the event |
| about the confirmation of a prospect | Satisfaction | (pleased about) the confirmation of the prospect of a desirable event | (1) the intensity of the attendant hope or fear |
| | Fears-confirmed | (displeased about) the conformation of the prospect of an undesirable event | (2) the effort expended for attaining or preventing the event |
| about the disconfirmation of a prospect | Relief | (pleased about) the disconfirmation of the prospect of an undesirable event | (3) the degree to which the event is realized |
| | Disappointment | (displeased about) the disconfirmation of the prospect of a desirable event | |

Event-based emotions include 12 types of emotions, these are: (1) Two Well-being emotions: joy and distress; (2) Four Fortunes-of–others emotions: happy-for, pity, gloating, and resentment; (3) Six Prospect-based emotions: two are centered, around the prospect of an event (hope and fear); two focus on the confirmation of a prospect (satisfaction and fears-confirmed); and two are about the disconfirmation of a prospect (relief and disappointment). The details of these emotions—the examples, the eliciting conditions, and the intensity variables are listed in Table 2.1 and Table 2.2.

Table 2.2: Event-Based Emotions of Well-Being and Fortunes-of-Others Emotions

| Event-based emotions | Examples | Eliciting Conditions | Variables Affecting Emotion Intensity |
|---|---|---|---|
| two well-being emotions | Joy | (pleased about) a desirable event | (1) the desirability of the event |
| | Distress | (displeased about) an undesirable event | |
| fortunes-of-others emotions<br><br>(Good-will emotions) | Happy-for | pleased about an event presumed to be desirable for someone else | (1) the desirability of the event for oneself<br>(2) the desirability of the event for the other person<br>(3) the deservingness of the other person<br>(4) one's liking for the other person |
| | Pity | displeased about an event presumed to be undesirable for someone else | |
| fortunes-of-others emotions<br><br>(Ill-will emotions) | Gloating | pleased about an event presumed to be undesirable for someone else | |
| | Resentment | displeased about an event presumed to be desirable for someone else | |

▪ **Agent-based emotions**

The second group of emotions is caused by the actions of agents, when they are viewed as being either praiseworthy or blameworthy. For instance, if the actions of an agent are approved by the emotion holder, then the agent is judged as praiseworthy and the emotion holder will have the potential to pride oneself when the agent is her/himself, or have the potential emotion of admiration when the agent is another party. The details can be found in Table 2.3.

Table 2.3: Agent-Based Emotions

| Agent-based Emotions | Examples | Eliciting Conditions | Variables Affecting Emotion Intensity |
|---|---|---|---|
| two attribution emotions on the (extended) self as agent | Pride | approving of one's own praiseworthy action | (1) the degree of praiseworthiness or blameworthiness of the agent (which include variables like effort, responsibility, and intention) |
| | Shame | disapproving of one's own blameworthy action | (2) the strength of the cognitive unit, where the agent is not oneself but a person or institution with whom one identifies |
| | | | (3) deviations from the person-based or role-based expectations of the agent |
| two attribution emotions on the agency of others | Admiration | Approving of someone else's praiseworthy action | (1) the degree of praiseworthiness or blameworthiness of the agent |
| | | | (2) deviations from the person-based or role-based expectations of the agent |
| | Reproach | Disapproving of someone else's | |

| | | blameworthy action | |
|---|---|---|---|
| | | | |

- **Compound emotions**

The compound emotions (as Table 2.4) focus on both the agent of the event and the desirability of the outcome. In other words, they are the combination between the wellbeing emotions (within the event-based group) and the agent-based group. The compound emotions are shown below,

Admiration + Joy ⟶ Gratitude

Reproach + Distress ⟶ Anger

Pride + Joy ⟶ Gratification

Shame + Distress ⟶ Remorse

Table 2.4: Compound Emotions

| Compound Emotions | Examples | Eliciting Conditions | Variables Affecting Emotion Intensity |
|---|---|---|---|
| Compound emotions on both the agent of the event and the desirability of the outcome | Gratitude | combining the approval of an agent's action with pleasure at the desirable outcome | (1) the variables that affect the related well-being emotions |
| | Anger | combining disapproval of an agent's action with displeasure at the undesirable | (2) the variables that affect the related attribution |
| | Gratification | combining approval of one's own action with pleasure at the desirable outcome | |

| | Remorse | combining disapproval of one's own action with disapproval of the outcome | emotions |
|---|---|---|---|

- **Object-based emotions**

This group of emotions is called attraction emotions, which are caused by reactions to objects, or aspects of objects, in terms of their appealingness. If an object fits the taste or attitude of the emotion holder, it will be judged as appealing, and the emotion holder will have the potential of liking experience. The details are in Table 2.5.

Table 2.5: Object-Based Emotions

| Object-based Emotions | Examples | Eliciting Conditions | Variables Affecting Emotion Intensity |
|---|---|---|---|
| two attraction emotions | Love | liking an appealing object | (1) the degree to which the object is appealing or unappealing |
| | Hate | disliking an unappealing object | (2) the familiarity of the object |

- **Variables affecting the intensity of emotions**

The variables can be generally divided according to their extensions into two categories. The variables affecting the intensity of all the emotions are called *global variables*, which includes four types. *The sense of reality* depends on how much one believes that the emotion-inducing situation is real. *Proximity* depends on how close in psychological space one feels to the situation. *Unexpectedness* depends on how surprised one is by the situation. *Arousal* depends on how much one is aroused to the situation.

The other type of variable is called *local variable*, which is tied to and affects only a particular group of emotions. These variables are listed based in the relevant emotional groups.

The event-based emotions are all affected by the *desirability* variable. For the fortune-of-others emotions, there are 3 more variables, including *Desirability-for-other* which reflects how one evaluates desirability for the other person's goal, *Liking* which reflects how much one is attracted to the other person , and *Deservingness* which depends on how much one thinks the other person deserved what happened.

The variables for the prospect-based emotions consist of *Likelihood* which reflects the degree of belief that an anticipated event will occur, *Effort* which reflects the degree to which resources were expended in obtaining or avoiding an anticipated event and *Realization* which depends on the degree to which an anticipated event actually occurs. The attribution emotions which are affected by the central praiseworthiness variable including *Strength of cognitive unit,* reflects how much one identifies with the person or institution who is the agent of the emotion-inducing event and *Expectation-deviation* showing how much the agent's action deviates from expected norms. Furthermore, the attraction emotions are relevant to the *familiarity of the object*, and the *appeallingness*.

Although the OCC model is computation oriented, its computational strategies are still on the level of abstraction. To interpret OCC model as a computational model, the essential requirement is to rebuild OCC in a computational representation, which is more suitable to use.  In this book, we propose to use a fuzzy tool to realize the causal logic of OCC.

## 2.3.3 Emotional Modeling based on OCC

Since OCC provides a common computational approach for implementing emotions, it can be directly applied to various computational models, which can be classified into layered mapping models, rule-based system, fuzzy logic, and Bayesian networks. In the following section, key contributions in each category are highlighted.

- **Pleasure-Arousal-Dominance Mapping of Emotions**

To enhance the believability of a conversational agent towards a human-like character, researchers [86] focused on the model of dynamics of emotions over time. They interconnected two psychological concepts by linking short-term emotion state with long-lasting mood state, and allowed the derivation of categorical emotional terms as the output. The agent then could generate emotional expressions based on the symbolic output of the system. There were two components in the emotion system, one for the course of emotions over time as well as the interactions between emotions and mood, and the other for categorizing symbolic emotional terms by mapping in the PAD space.

The PAD (Pleasure-Arousal-Dominance) emotional state is a three-dimensional approach to measure emotions, which assigns a variety of emotions to coordinate in the emotional space. The study in [86] mapped 22 OCC emotions into PAD emotions, which then have 3-tuple values to represent an OCC emotion. Then the computations of vectors can be directly applied.

The mapping was based on ontology matching, which first categorized OCC emotions into 8 varieties of PAD emotions, and then from the semantics of emotion determined the final position of each emotion in the three-dimensional emotion space.

The translation from abstract emotions to numerical emotions which are computable can be achieved using third party ontology software, and the application is then straightforward by manipulating vectors. However, it is hard to be accurate since different individuals have different interpretations of emotions, so it is very risky to constrain ambiguous emotions into a limited dimension of space.

A mass-spring model was adopted to simulate the time course of emotions as shown in Figure 2.15, which was expected to better than linear or exponential alternatives. Emotion and mood are simulated as two independent spiral springs. The forces are calculated as if the two sprints were anchored in the origin and attached to the reference point. By adjusting the spring constants and the inertial mass of the point reference, various characters can be simulated. To express the feeling of boredom (i.e. the inactivity of stimuli), the researchers defined an epsilon neighbourhood around the origin. Outside of the area, the value of boredom fell to zero, while -1 if the agent was most bored.

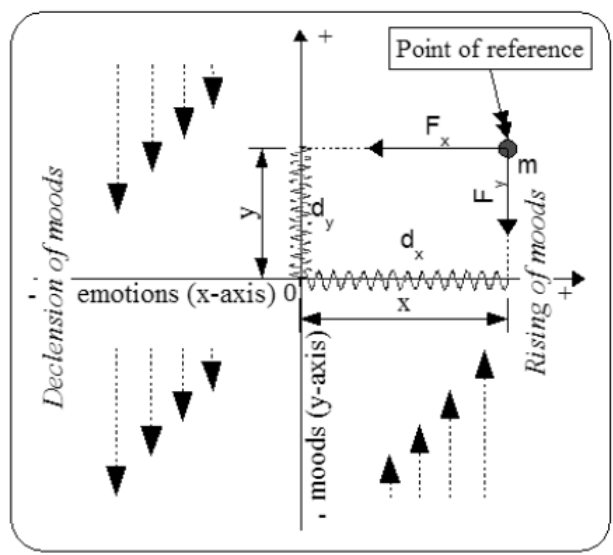

Figure 2.15: The Mass-Spring Modeling of Emotion and Mood over Time [87]

After the dynamic emotion states were calculated, the system categorized the emotion into a symbolic representation. The mapping was done by transforming the readings in a PAD space.

$$D(t) = (x_t, y_t, z_t), \text{where } x_t = [-1,1], y_t = [-1,1], z_t = [-1,0]$$

The reading D(t) at time t contained the emotional state $x_t$, the mood valence $y_t$, and the degree of boredom $z_t$. The triple was converted to an emotional category by a defined mapping function as follows [87],

$$K(x_t, y_t, z_t, t) = \left(p(x_t, y_t), a(x_t, z_t), d(t)\right), \text{where } p(x_t, y_t) = \frac{1}{2}(x_t, y_t) \text{ and } a(x_t, z_t)$$
$$= |x_t| + z_t$$

The mapping function produced a triple in the PAD space. Pleasure function $p(x_t, y_t)$ was assumed to be the overall valence and thus computed as the mean of valence of both emotion and mood. Arousal function $a(x_t, z_t)$ was assumed to be any kind of emotion that was with high valence, and therefore the absolute value was considered with negative offset of boredom. Dominance function d(t) was used to describe agent's feelings of control and influence over situation. For example, a high dominance value distinguished angriness from fear. The dominance value was modelled as a function changing over time and independent from emotions and moods. The computed triple in PAD space was used to identify the closest category of emotions. An emotion category was activated if the reference point fell within a predefined range of the center of a category.

- **Rule-Based Emotion Inference**

The work by Jaques and Viccari [88] applied OCC to provide emotional support to students through an animated pedagogical agent, called "PAT". The purpose of this work was to motivate and encourage students, and to make them believe in their self-ability, by encouraging a positive mood for the students, in the hope of increasing their learning efficiency.

In order to achieve that, PAT needs to recognize the student's emotions from his/her observable behavior, which comprises predefined user actions in the system's interface. These inputs allow the system to analyze user's emotions based on the inference from the OCC model. The scheme representing appraisal for joy and distress is shown in Figure 2.16. It is clear that the emotional analysis process is built upon the OCC model and is actually a tree-like rule base structure following which an emotion, joy or distress, is inferred from the user inputs. Hence, the important steps to appraise an emotion include defining the events that can occur in the system, defining user's goals for understanding the desirability of an event, and classifying the events to identify the emotions.

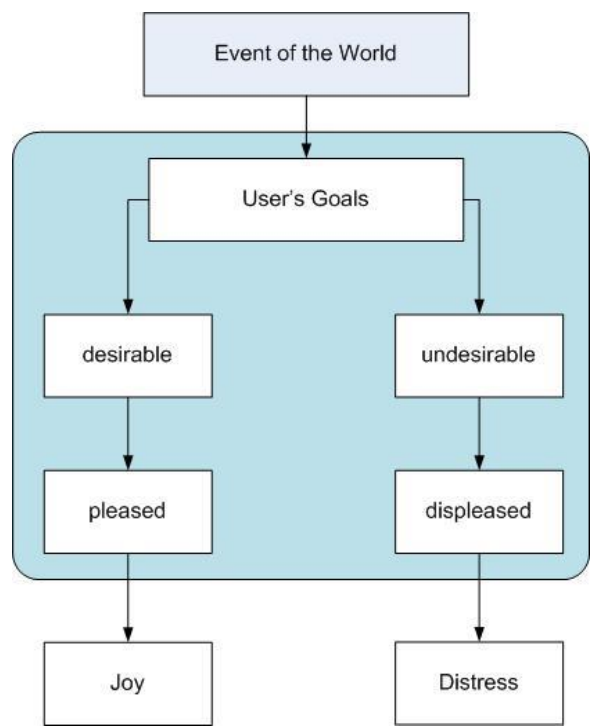

Figure 2.16: Appraisal Scheme for Joy and Distress of "PAT" [89].

The events in the educational system are caused either by a student or by a PAT. In their work, the system was quite predicable and limited to the number of events which are listed in Figure 2.17. Similarly, student goals were classified as mastery goals which were oriented towards developing new skills and abilities, and improving their level of competence as well as learning new things and performance goals which were demonstrated by students who would feel better when they pleased the teacher or did better than others. The individual goals were pre-assessed based on a questionnaire and entered as user inputs before running the system. A set of rules was defined to classify different events into known emotions by comparing the event contents and user's goals. For example, if a student is performance-oriented, then the event "did not accomplish the task correctly" is undesirable. Therefore, from the event type, it is known that whether a task is accomplished correctly or not would elicit joy or distress, so this event would cause the student to have distress emotion.

|   | Events |   |
|---|---|---|
| 1 | Student provided an incorrect task answer | Event that can elicit joy/distress emotions |
| 2 | Student provided a correct task answer | |
| 3 | Student did accomplish the task | |
| 4 | Student gave up the chapter | |
| 5 | Student finished the chapter | |
| 6 | Student asked for help | Elicit shame emotion |
| 7 | After agent's help | Events that can elicit anger/ gratitude emotions |
| 8 | Student denied agent's help | |
| 9 | Student accepted agent's help | |
| 10 | Student disabled agent | |
| 11 | Student enabled agent | |

Figure 2.17: Sample Events of "PAT" [89].

Interestingly, this work did not only appraise the student's emotion, but also produced corresponding tactics to help the student. The tactics were also predefined and stored in the system as a strategy rule base. The affective tactics were delivered as messages displayed to students or directly as a help by an animated PAT.

This work formed a set of rules based on the OCC model to infer user emotions as well as provide tactics to motivate and encourage students. However, the inference of user emotions based on the OCC model is difficult and sometimes inaccurate, because it heavily relies on how sound and complete the model is. Unfortunately, human emotion is very complex and hard to generalize as one computational model. Moreover, since the implemented rule based system lacks dynamic support in the uncertain and unpredictable environment, a more sophisticate computational model which can handle uncertainties is preferred to a set of static rules.

Apart from the above project, another pedagogical project called FearNot! [90] is also using OCC-based rules. It was a program developed to tackle and reduce bullying problems in schools. Bullying happened when a child was hit, kicked or indirectly excluded socially or hurt by maliciously rumour spreading. The objective of the system was to build an anti-bullying demonstrator to students who were 8 to 12 years old. Students acted as an invisible observer of a victim and through interaction with the virtual victim, the students discussed the underlying problems and proposing coping strategies. It was critical for such system to work by gaining empathy from the students, and therefore emotions and the dynamics of emotions were carefully modelled to achieve believability. Building an environment that engages new ways for children to learn requires the embedding of autonomous characters that are believable and can gain empathy from children.

The emotion modelling was referred to OCC theory, and the attributes considered in the emotion modelling were shown as follows [90].

| Attributes | Explanation |
| --- | --- |
| Type | The type of the emotion being experienced |
| Valence | Denotes the basic types of emotional response. Positive or negative <br><br> value of reaction |
| Target | The name of the agent/object targeted by the emotion |
| Cause | The event/action that caused the emotion |
| Intensity | The intensity of the emotion |
| Time-stamp | The moment in time when the emotion was created or updated |

The intensity of emotion was assigned with different values, which was computed based on different situations. The intensity of emotion was not constant over the entire course, and the changes of emotion intensity reflected the dynamics of the emotion itself. The researchers achieved the dynamic modelling of emotion by incorporating a decay function for each emotion to capture the intensity as a function of time. The formula was defined as follows [90],

$$Intensity(em, t) = Intensity(em, t_0) \times e^{-bt}$$

The intensity of an emotion $em$ at time $t$ was calculated referring to the value of the intensity of the emotion when it was generated. After some time $t$, the value of intensity decayed and reached a defined threshold value close to zero. When this happened, such emotion was removed from the character, implying that the specific emotion would no longer exhibit as agent's emotional state. The constant $b$ chartered the rate of decay over time, and different emotion might have different decay rates.

- **Fuzzy Logic Adaptive Model of Emotion**

In order to generate agent emotions dynamically rather than previous static rules or pre-determined domain knowledge, researchers [91] proposed a Fuzzy Logic Adaptive Model of Emotion (FLAME) as a new computational model of emotions to incorporate memory and experience into the emotional process. This model used a fuzzy logic approach to map events to emotions, which is based on the OCC model to build fuzzy rule bases. In addition, this mapping was learned based on users' emotional status and the events encountered.

The importance of a goal and the impact of an event on a goal were considered to be fuzzy concepts, and thus a set of membership functions was designed as in Figure 2.18. Then the mapping was given as a fuzzy rule like:

IF       Impact ($G_1$, E) is $A_1$ AND Impact ($G_2$, E) is $A_2$ ….. AND Impact ($G_k$, E) is $A_k$

AND    Importance ($G_1$) is $B_1$ AND Importance ($G_2$) is $B_2$ … AND Importance ($G_k$) is $B_k$

THEN   Desirability (E) is C

A Mamdani defuzzification approach computed the final measure of a particular event's desirability. Consequently, with a set of defined quantitative measures of emotion, an emotion's intensity can be directly calculated. For example, according to some equations formulated by psychologists Price et al. [92], "Hope" is calculated as $Hope = (1.7 \times \sqrt{expectation}) + (-0.7 \times desirability)$, where *expectation* takes probability measure of the corresponding event.

The novelty of this work is that fuzzy inference method and adaptive learning were applied to extend computation of OCC for a more realistic and dynamic environment. However, this approach is strictly based on OCC, so the limitation of the OCC model such as lack of temporal description of emotions still applies.

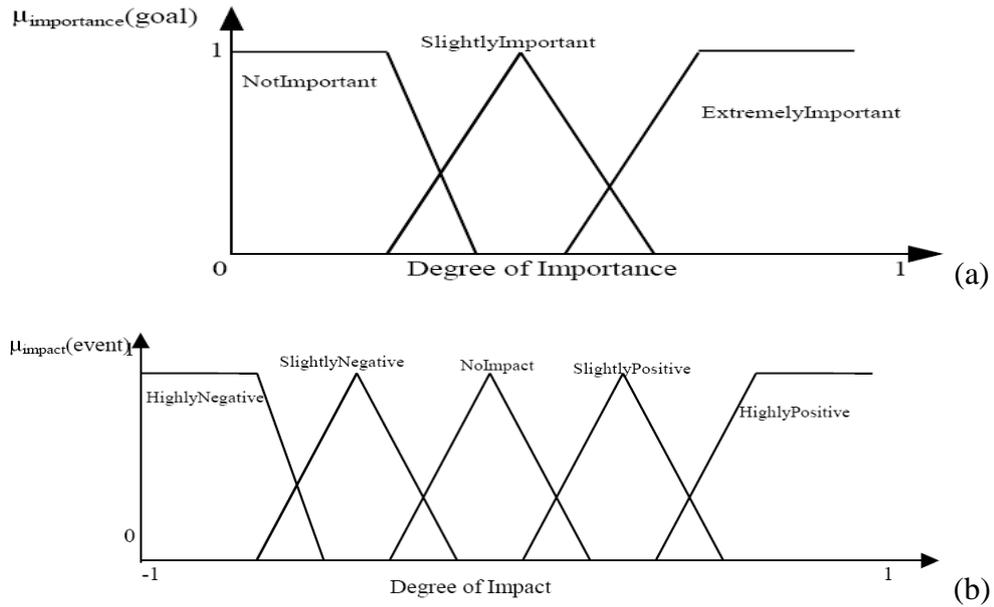

Figure 2.18: Memberships for Importance of Goal (a) and Impact of Event (b) in FLAME [91].

## 2.3.4 Other Emotional Modeling Approaches

Supervised learning method as a common solution to classification problem has been well known. Researchers found the same method could be used for emotion classifications. The study [93] adopted two fuzzy approaches to classify the motions expressed in music, and the emotion model used for the study was adopted from Thayer's model. As shown in Figure 2.19, the emotion space was presented in 2D space and divided into 4 quadrants by the values of valence and arousal of an emotion. In this classification problem, totally there were 4 classes of emotions, each of which corresponded to one quadrant.

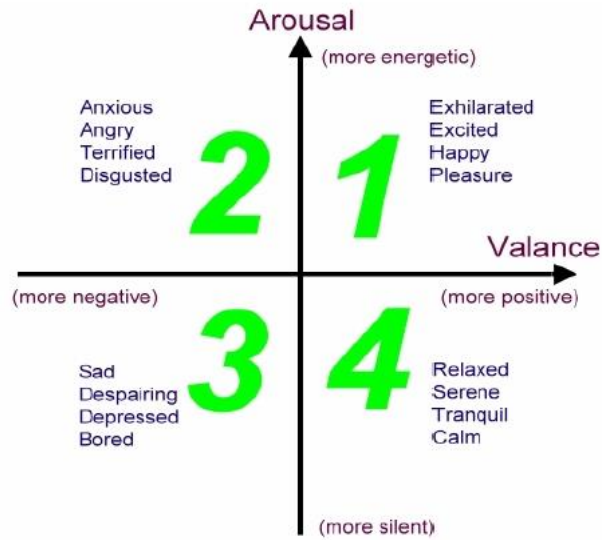

Figure 2.19: Thayer's Model of Emotion [93]

The training samples were from 243 collected songs. From each 25-second segments were used to be labelled by subjects into the four emotion classes. Fuzzy vectors were adopted as the labelling outputs. Each element denoted the membership of this element belonging to that class. The testing samples were then classified by two types of fuzzy classifiers, with references to the training samples. One classifier was Fuzzy k-NN classifier, which derived from the original k-nearest neighbourhood algorithm. The classification process can be summarized as follows,

$$\mu_{uc} = \frac{\sum_{i=1}^{k} w_i \mu_{ic}}{\sum_{i=1}^{k} w_i} \text{ and } w_i = d_{iu}^{-2}$$

A fuzzy membership $\mu_{uc}$ for an input $x_u$ to each class c was a linear combination of the weighted k-nearest training samples. The number of neighbours k was selected empirically. The weight was computed as inversely proportional to the distance $d_{iu}$ between input $x_u$ and $x_i$. $\mu_{ic}$ was the membership of training sample $x_i$ to the class c, and generated based on following equation.

$$\mu_{ic} = \begin{cases} \beta + \left(\frac{n_c}{K}\right) * (1 - \beta), & if\ c = v \\ \left(\frac{n_c}{K}\right) * (1 - \beta), & otherwise \end{cases} \text{ and } \beta \in [0,1]$$

The membership of training sample was determined after the subjects labelled the sample. The majority voted class v was the class which had the most votes, and $n_c$ was the number of samples belonging to that class, while K was the empirically selected number of neighbours and could be different with previous k.

The researchers also proposed a more accurate classifier, i.e. Fuzzy Nearest-Mean classifier, to classify the music emotion label. They first computed the mean of each feature extracted from the music segment by,

$$\mu(c, f) = \frac{1}{N_c} \sum_{n=1}^{N_c} F_{c,f,n}$$

$\mu(c, f)$ was the mean of the feature f in class c, and $F_{c,f,n}$ was the value of the feature f of the segment n in class c. $N_c$ was the total number of segments in class c. The distance of an input x to each class c was calculated as the sum of squared errors between the features of x and the mean of each class, where

$$d_{xc} = \sum_{f=1}^{N_f} \left( x_f - \mu(c, f) \right)^2$$

$N_f$ was the number of extracted features. The final fuzzy vector of the input sample to each class was obtained by computing the inverse of the distance,

$$\mu_{uc} = \frac{d_{xc}^{-n}}{\sum_{c=1}^{C} d_{xc}^{-n}}$$

For classification purpose, the maximum element in the fuzzy vector was chosen as the final emotion class.

Another research project works on representing the meaning of words in the context of emotions. Researchers [94] developed a systematic approach to extract word models from crowd sourcing. They adopted Interval Approach to encode emotional worlds by type-2 fuzzy sets. The idea was to reduce a large vocabulary space of emotions to a much smaller set of possible emotion vocabulary. The emotion words were recorded in an emotion codebook, which was a set of words and a function that mapped emotion words to their corresponding region in the emotion space.

Rather than mapping to a single point in the emotion space, uncertainty was modelled using type-2 fuzzy sets. Type-1 fuzzy sets extend crisp sets by adding membership grade to be a point in [0,1]. Type-2 fuzzy sets further extend type-1 by defining the membership grade as a function ranging from [0,1] at a given point. Type-1 assumes the membership grade as a fix point, whereas type-2 allows for uncertainty within a range. An example type-2 fuzzy membership is shown in Figure 2.20, where a normalized membership can be defined by a tuple of 9. The shaded area represents the possible fuzzy membership grades of a point.

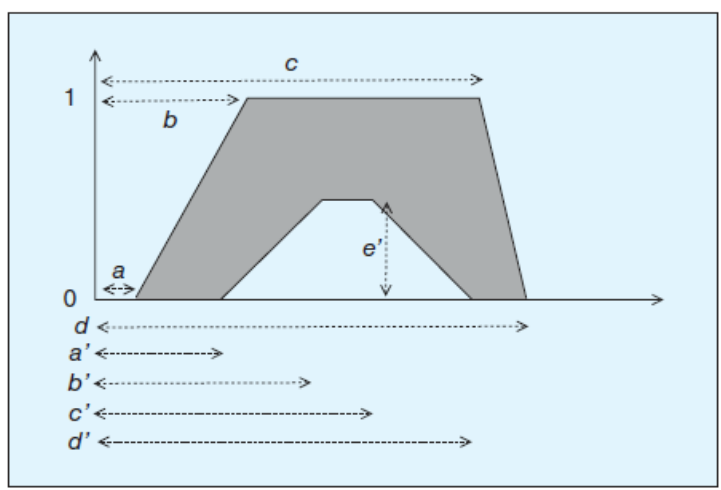

Figure 2.20: An Example of a Trapezoidal Type-2 Fuzzy Membership [94].

In order to determine each membership function of each emotion word, the 9 algebraic properties of fuzzy sets were computed from the data collected in interval approach surveys. In an interval approach survey, subjects gave their ratings on words by abstract scales. Instead of selecting a single value on the scales, subjects chose an interval range on the scales. Later, the chosen intervals were used for calculated the properties of the membership functions.

The mapping from one vocabulary to the emotion vocabulary was proposed as choosing the word from the output language such that the similarity was maximized for a given input word, that is,

$$w_{output} = arg\ max_{w_2 \in W_2} sm\left(eval_{C_2}(w_2), eval_{C_1}\left(w_{input}\right)\right)$$

The similarity function was defined based on the Jaccard Index for type-2 fuzzy sets, and we have,

$$sm_J(A, B) = \frac{|A \cap B|}{|A \cup B|} = \frac{\sum_{i=1}^{N} min\left(\overline{\mu}_A(x_i), \overline{\mu}_B(x_i)\right) + \sum_{i=1}^{N} min\left(\underline{\mu}_A(x_i), \underline{\mu}_B(x_i)\right)}{\sum_{i=1}^{N} max\left(\overline{\mu}_A(x_i), \overline{\mu}_B(x_i)\right) + \sum_{i=1}^{N} max\left(\underline{\mu}_A(x_i), \underline{\mu}_B(x_i)\right)}$$

where $\overline{\mu}$ and $\underline{\mu}$ were the upper and lower membership functions. The experiment obtained the similarity between an emotion word and a vocabulary and illustrated in Figure 2.21. The emotion vocabulary was in bold.

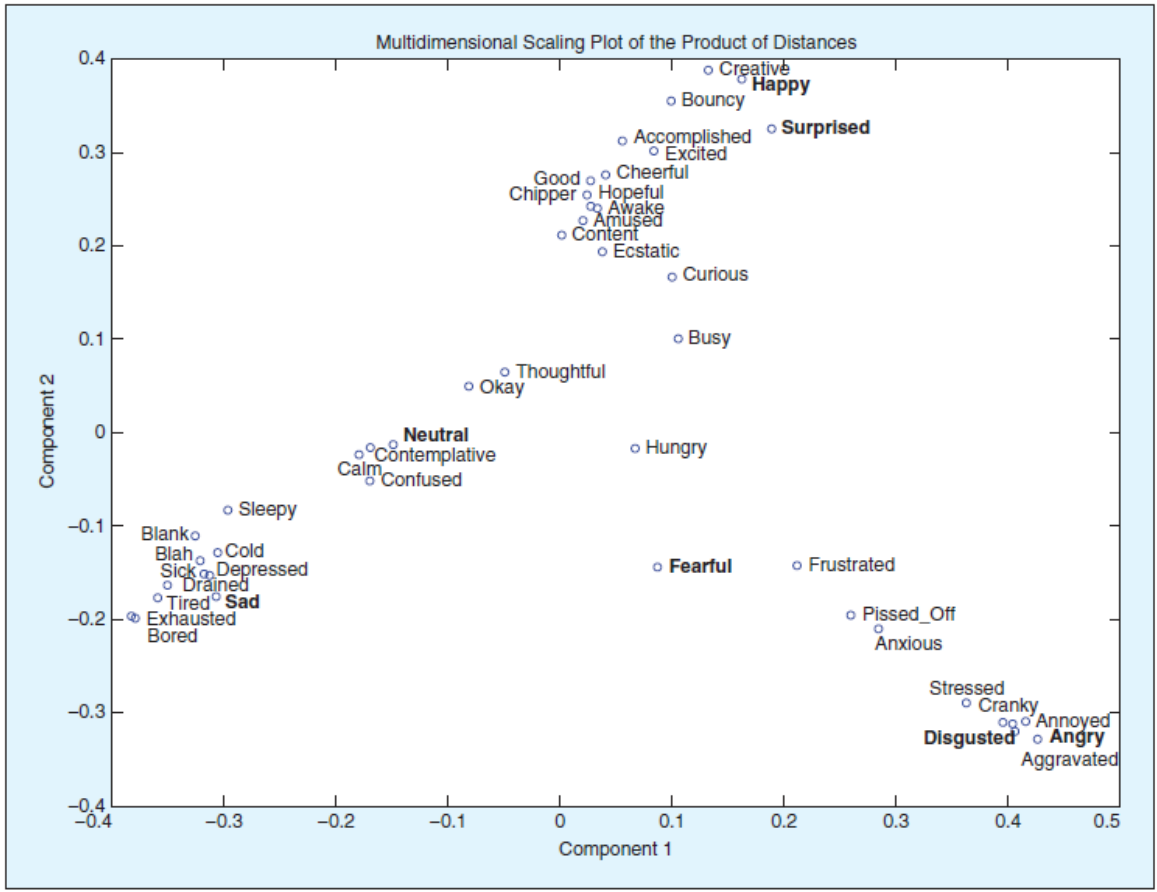

Figure 2.21: Similarity of Emotion Words Mapped by Type-2 Fuzzy Sets [94]

Another application using emotional model focuses on drivers. Driver's emotion was closely studied to assert its impacts on driving behaviors and road safety. It was clearly shown that some strong emotions would be related to road accidents and endangering driving safety. Researchers [95] modelled the relationships between traffic conditions at intersections, including waiting time and different road alignments, and driver emotions using a set of fuzzy rules. The fuzzy inference relied on the definition of a driver's emotional space, which was a 2D space with 4 quadrants for four different types of emotions. The sketch of the emotional space is shown in Figure 2.22.

In the emotional space, an emotion had its coordinates in the system as [x, y] . For any coordinate that was outside the unit cycle, normalization was applied. The coordinates could be represented with polar coordinates as e = (ρ, θ) where ρ = $\sqrt{x^2 + y^2}$ and θ = arc tan $\frac{y}{x}$. The modelling of driver's emotion divided the space according the angle θ equally into 4 ranges, and emotions fell within the range were classified as one of the four emotions (happy, nervous, angry and relief). The length of the vector ρ was used to represent the intensity of emotions.

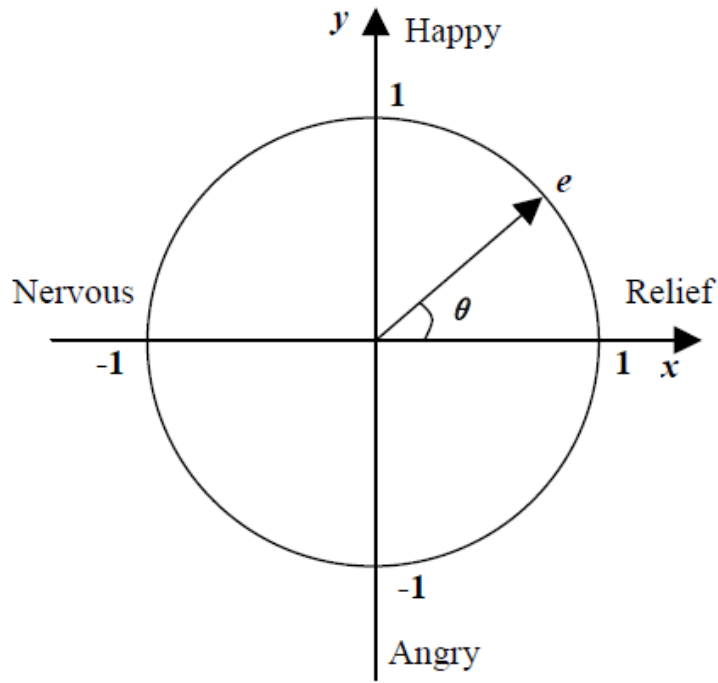

Figure 2.22: Emotional Space of a Driver [95]

Their approach was to do fuzzification for the input variables into 5 regions. For example, waiting time was divided into very long, long, medium, short and very short, while horizontal radius of road was divided into very small, small, medium, big and very big. Corresponding fuzzy membership function was created as the triangular functions shown in Figure 2.23.

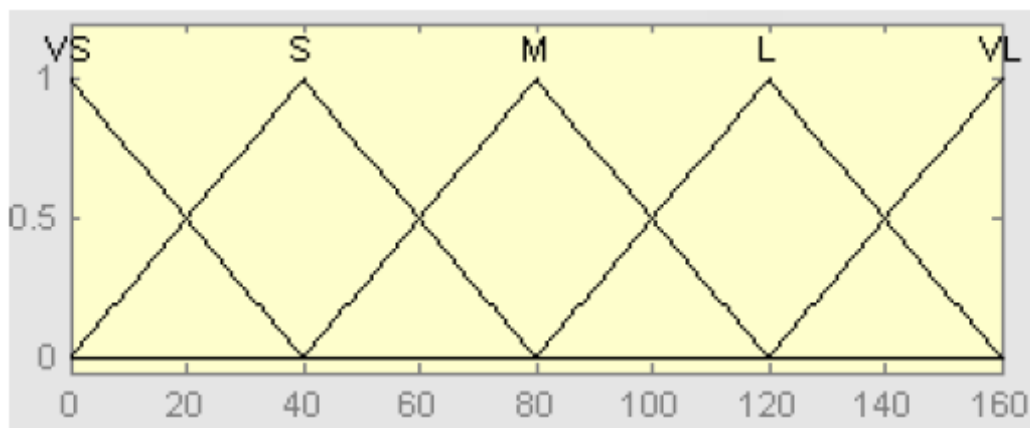

Figure 2.23: The Membership Function for Waiting Time [95]

The study assumed that the waiting time only caused the change of emotion in the vertical direction, and the road alignment only caused the change of emotion in the horizontal direction. Based on the assumption, they defined a set of fuzzy rules for the input and output mapping. For example, a very long waiting time at the intersection would cause a very big negative change in the vertical direction of emotion. The rules allowed for the calculation of emotional changes in both horizontal and vertical direction based on the input variables. The output of emotional changes was derived from the defuzzificaiton with centroid method which converted fuzzy variable into its crisp representation

## 2.4 Summary

In this chapter, we first gave a broad view of the TA research as an interdisciplinary research field, which depends on the efforts from computer engineering, education, as well as psychology. Based on this background, we introduced various existing TA systems, and summarized the limitations. Our research will focus on the related limitations of existing TA design and try to solve the problems. As we intend to involve affective modeling into TA design, theoretical background of affective modeling is provided.

**Chapter**

# 3

# Modeling Affective Teachable Agents

In this chapter, we will introduce the proposed Affective Teachable Agent which is designed based on a goal-oriented modeling approach. To realize the Learning-by-Teaching strategy and help students to actually obtain the learning benefits, the TA design is focused on two perspectives. First is the ability to learn new knowledge and apply the knowledge to certain tasks. Second is the ability to establish good relationships with students and encourage them to teach well. These two perspectives of capability of a TA are built into the Affective Teachable Agent system with a goal-oriented modeling approach. The goal-oriented TA can reason and decide by itself how to select the next goal and what actions to take. Through executing a hierarchy of goals, the proposed TA can proactively interact with students by pursuing its own agenda, and provide students with believable communication and a motivated learning experience. In the following sections, we first discuss what features of a TA should be designed in order to attain the teaching goal, and propose the Affective Teachable Agent model with a formalized representation. After that, we utilize a goal-oriented modeling approach to instantiate the theoretical agent modeling to a practical implementation. To support a TA with emotional capability, a computational approach for emotion elicitation is proposed.

# 3.1 From Teaching Goal to Intervention Design

For all the educational systems, the ultimate goal is to facilitate student's learning and to stimulate their learning interest. We need to understand the learning process clearly to design effective educational interventions. We can achieve this by answering two questions:

- What activities will produce learning?
- How an intelligent agent system can help us to achieve that?

According to Learning-by-Teaching theory, the teaching process may help students form a clear understanding of what they are teaching and further obtain a deep understanding and long lasting retention of the related knowledge. In other words, the teaching process *per se* will produce learning, because teaching may promote students to think deeply, repeatedly, and orally [13]. The learning happens when students know clearly what to teach; when they represent their understanding; when they think about the same problem repeatedly; and when they change their angle of viewing the problems.

The next question is how to use the TA system to enlighten students to achieve these learning processes. The fundamental requirements for TA systems are summarized as follows. First, the TA should at the very beginning give students enough hints to let them know what they need to teach what task they need to fulfill. This may help students immerse themselves into the teaching role effectively. Second, the TA should have an explicit representation of appropriate knowledge, which can help students clearly represent their knowledge to TA and reciprocally understand the TA's reasoning process. Third, the TA should follow students' direction and give students the feeling of direct control. According to [96], if students have direct control of their activities and can see the immediate effect of their activities, they can be more motivated. In TA designs, what the TA learnt is what the student taught. Hence, the student can review his own knowledge structure through the knowledge representation of the TA. Fourth, the TA should provide diverse feedback to students throughout the learning process in order to show its learning progress to students and indicate the teaching performance of students. The feedback may enhance student's reflection and influence their self-monitoring and self-evaluation. Lastly, the TA should establish a good relationship with students so that they may form a strong empathy with the TA and take the responsibility to help the TA. To summarize, we divide the capabilities a TA into two categories, ***Teachability*** and ***Affectivability***.

***Teachability***, *in our ATA model, refers to the ability that an agent can record the teaching content taught by students to its knowledge base, do reasoning according to the learnt knowledge, practice the learnt knowledge in virtual world and provide proper feedback throughout the learning process.*

***Affectivability,*** *refers to the ability that an agent can generate and express emotions as an additional type of interaction.*

Based on the above analysis, the proposed Affective Teachable Agent can be defined as a Teachable Agent with teachability and affectivability. Next, we will propose this new type of teachable agent, and use a goal-oriented approach to realize the agent's mental model according to educational requirements.

## 3.2 Formalization of an Affective Teachable Agent

To define the proposed TA, we begin from the basic definition of a generic agent system. According to [97], an agent is a system situated within an environment that senses that environment and acts on it, over time, in pursuit of its own agenda and so as to affect what it senses in the future (as Figure 3.1). Therefore, we have:

**Definition 1**: *An agent, Ã, is formally specified as a tuple Ã = (**E**, **P**, **K**, **R**, **A**), where*

**E** *is the set of environment states that can be perceived by the agent;*

**P** *is the set of perception states that the agent detects about the environment;*

**K** *is the set of knowledge points that the agent possesses;*

**R** *is the set of reasoning mechanisms that the agent follows;*

**A** *is the set of actions that the agent behaves.*

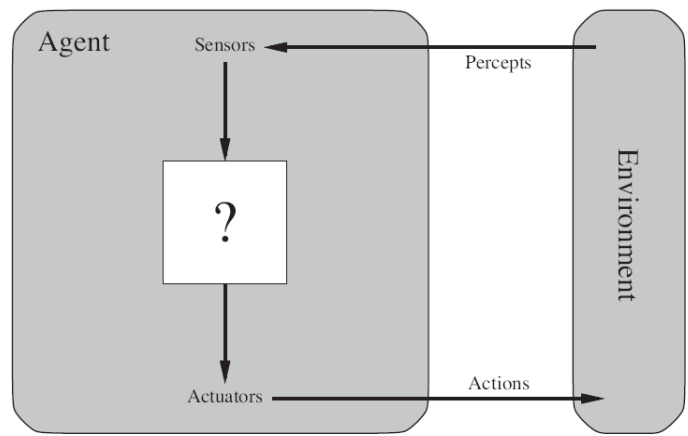

Figure 3.1: Agent Model Proposed by [98]

As a specific type of agent, the modeling difference between a TA and any other type of agents lies in the reasoning mechanism. TAs are designed for students to teach, and therefore its reasoning is mainly on how to learn from student's input.

**Definition 2**: *A teachable agent,* **TA***, is an agent that is defined as a tuple* **TA** = (**E**, **P**, **K**, **Rt**, **A**)*, where* **E***,* **P***,* **K** *and* **A** *follow the definitions in Definition 1, and* **Rt** *is the set of reasoning mechanisms which enable TAs to learn from students and improve students' learning in return.*

The generic TA model is illustrated as Figure 3.2.

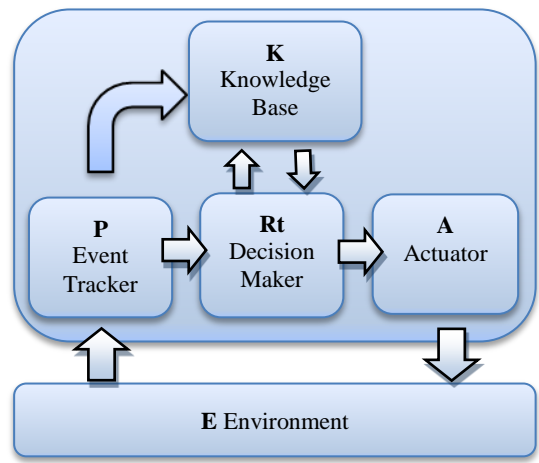

Figure 3.2: Generic Teachable Agent Model [15]

This generic teachable agent model represents a teachable agent executing "Perceive – Reason – Act" (PRA) cycle. Taking the system Betty's Brain as an example, the agent can perceive the input from students, record the input into its knowledge base, do causal reasoning based on the its learnt knowledge, and release various types of feedbacks for students. The model enables the agent to be taught but it lacks a proactive mechanism for the agent to do interaction with students.

Goal orientation is one of the key features in agent systems. To a certain degree, agents are goal oriented [25], since the goal selection and the further concrete behavior selection are the foundation of an agent's initiative. Only if an agent has the capability to choose its own goals and act to achieve its goals, is the agent called "active". Thus, a goal oriented teachable agent has more flexibility to interact with students proactively. Based on the existing TA model, we define the proposed TA with a goal selection component. An agent with goal selection capability should have a planning unit to find the appropriate sequence of actions to reach its goals. The question is that before doing the goal selection, what goals should a TA have? In our opinion, there are two main goals for a TA. The first and the most natural one is having the capability *to be taught*, as learning by teaching is the essence of the pedagogical benefits. The second, we propose, is the capability *to be affective*, because emotional factors are crucial for almost every aspect of students' learning [99-101] and emotions deeply influence the efficiency and effectiveness of interactions [102]. The TA needs to reason and decide what actions it should take so that its goals can be fulfilled. Therefore, in this research we propose a new type of TA which is called Affective Teachable Agent.

**Definition 3:** *An affective teachable agent,* **ATA***, is an agent that is defined as a tuple* **ATA** = (**E**, **P**, **K**, **G**, **Rt**, **A**)*, where* **E***,* **P***,* **K***, and* **A** *follow the definitions in Definition 2,* **Rt** *is a tuple* **Rt** = (Tr, Ar)*, and* **G** *is a set of goal selection mechanisms which can support the agent to pursue its goals.*

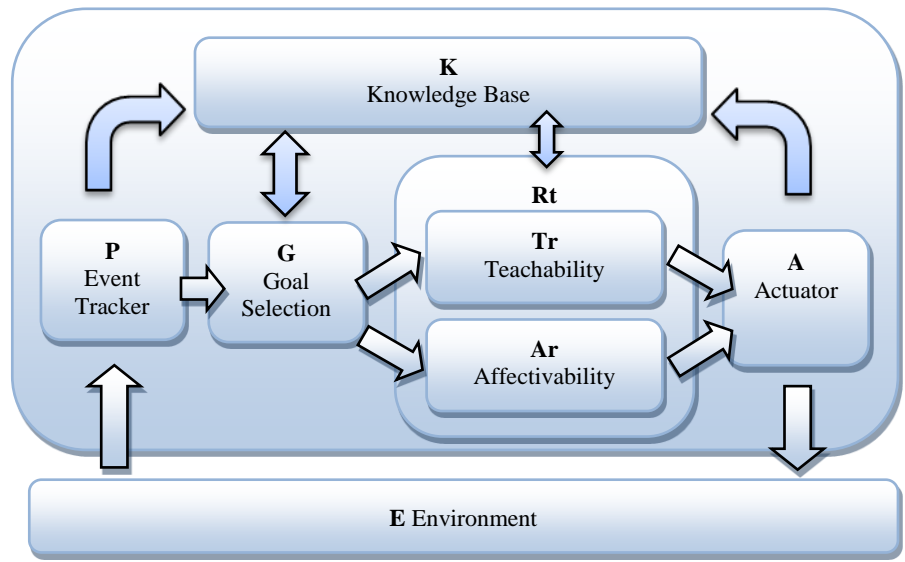

Figure 3.3: The Proposed Affective Teachable Agent Model [103]

**Definition 3.1: Rt** *is defined as a tuple* **Rt** = (Tr, Ar), where

> **Tr** *the abbreviation for* **Teachability Reasoning**, *is the set of reasoning mechanisms that can make the agent learn from students.*
>
> **Ar** *the abbreviation for* **Affectivability Reasoning**, *is the set of reasoning mechanisms that can make the* agent generate and express emotions.

For the *Teachability Reasoning*, the agent has two main types of tasks, which are operating repeatedly and form two running cycles,

Learning Cycle: *PK*

1. **Perceive**: The agent perceives the teaching knowledge taught by students.

2. **Knowledge storing**: Storing the knowledge that teachable agent perceives.

Acting Cycle: *PGTr(K)A*

1. **Perceive**: The agent perceives the environment and senses student's behaviors such as whether the student get close, whether he teach the knowledge, etc.

2. **Goal Selection**: The agent selects a learning goal according to its current situation and the information it perceived.

3. **Teachability Reasoning**: The agent does the reasoning on certain task selection and behavior selection according to student's learning progress such as whether to ask for help from students or whether to practice by themselves.

4. **Action**: The agent acts according to the reasoning outcome.

For the *Affectivability Reasoning*, the Affective Cycle is always operating with the Learning cycle and Acting Cycle in parallel because the emotional responses need to be available at any time.

Affective Cycle: *PArA*

1. **Perceive**: The agent perceives the environment and the student's behavior.

2. **Affectivability reasoning**: The agent reasons and elicits the emotional tendencies based on the perceived environmental stimuli.

3. **Action**: The emotional tendencies are expressed as a type of feedback to students.

The Teachability Reasoning is to simulate the learning capability of TA, whereas the Affectivability Reasoning focuses on generating TA's emotions. These two components are working concurrently with each other.

**Definition 3.2:** *Knowledge* **K** *of ATA includes three subsets, namely* **Domain Knowledge**, **Goal Structure**, *and* **Runtime Data**, *where*

**Domain Knowledge** *refers to the expert knowledge, or the expert's performance about certain subject in a domain. (Details will be introduced in Chapter 4.1.)*

**Goal Hierarchy** *contains a set of hierarchical structures that are designed for achieving various goals of the ATA. (Details will be introduced in Chapter 3.3.)*

**Runtime Data** *stores the temporary status and historical process of the data transfer of the ATA.*

The ATA may store or search the *domain knowledge* related to the learning content to interact with students about the learning tasks. The ATA also can take the initiative to pursue its own agenda through executing the *goal structure*, such as practicing some skills based on the learnt knowledge or starting an inquiry to make uncertain understandings clear. All the processing data of the agent system are stored as *Runtime Data* that can track the agent's behavior and assist system control.

## 3.3 Goal-Oriented Modeling of Affective Teachable Agents

An agent system is an active entity that continuously perceives the environment, makes decisions, and acts on the environment. This "perceive—reason—act" sequence helps an agent properly react to the environment. However, forming proper reaction is only one perspective of being an "active entity". The more important perspective is the self-determination, or the proactive pursuit of the agent's own agenda. A TA also needs to have an autonomous mechanism to organize various activities and selectively carry out the right activity at the right time.

Goal orientation is one of the approaches to achieve an autonomous mechanism in agent system. To some degree, agents are goal oriented [25], since the goal selection and the further concrete behavior selection are the foundation of agent's initiative. If an agent has the capability to choose its own goals and act to achieve its goals, then the agent is called "active". Thus, a goal oriented teachable agent has more flexibility to provide students highly active interactions.

From the perspective of a goal oriented teachable agent, what goals should a teachable agent possess? As we have mentioned, the ultimate goal of a TA is to facilitate students' learning and to stimulate their learning interest. Within this educational principle, there are two sub-goals, which are corresponding to the two capabilities that the proposed ATA needs to attain: the Teachability and Affectivability. Each of the sub-goals also has multiple requirements and functions to be fulfilled. Thus, we need to select a hierarchical goal-setting tool to model the agent system. Goal Net [25] is an agent modeling methodology to do this.

The practical implementation of our TA is based on a goal-driven agent modeling methodology, Goal Net [104], in which an agent's goal is achieved through the completion of a sequence of sub-goals.

**Definition 3.3**: **Goal Net**, *defined as a 4-tuple* (**S, Arc, T, Br**), *is a graphical representation of agent goals and actions (as Figure 3.4), where*

**S**        *a set of goal states that the agent wants to achieve, represented as the nodes $S_0$ to $S_5$. Special states include the start state* ***S*** *and the terminated state* ***E****.*

**Arc**    *a set of arcs, defines how a goal state connects with another. The arcs, shown as* ***arrows*** *in Fig 3.4, connect separated goal states to a unified goal structure.*

**T**     *a set of transitions, specifies what tasks the agent needs to fulfill before moving from one goal state to next goal state. Therefore, each arc has a transition to define the certain tasks. Transitions are represented as* **vertical bars**, *t₁ to t₇ in Fig. 3.4.* *t₂ is a specific Choice Transition to indicate the selectable path from one sate (S₀) to another (S₁ or S₂).*

*Specifically, transition T has four kinds of parameters* **Tt, Tf, S$_{pre}$** *and* **S$_{post}$**,

**T = $f$(Tt, Tf, S$_{pre}$, S$_{post}$)**, where

> **Tt** *describes the trigger of a function Tf;*
>
> **Tf** *defines the function of an agent behavior;*
>
> *The successful triggering of Tf transfers the goal state of GoalNet G from the prior state* **S$_{pre}$** *to the next state* **S$_{post}$**.

**Br**     *a set of branches, denotes the boundary of Goal Net (as the* **dashed lines**). *A goal state can be decomposed as a Sub Goal Net when it is too complicated to be represented as a single state. The Sub Goal Net is laid out through the branches from the higher goal state.*

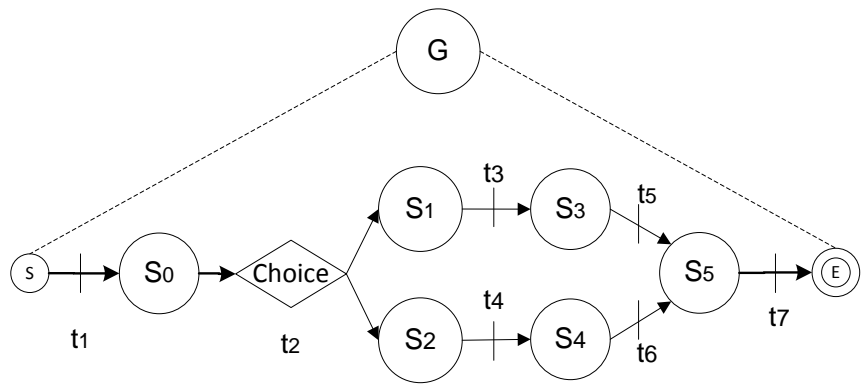

Figure 3.4: An Illustration of Goal Net

**Definition 3.4**: *A **Goal**, which is defined as* **G** $\in$ {**Root**} $\cup$ {**Atom**} $\cup$ {**Comp**}, *is a state that the agent wants to achieve, where*

**Root**    *refers to the set of root goals in Goal Net. The root goal is the start of the whole Goal Net structure. Each Goal Net has only one root goal.*

**Atom**    *refers to the set of atomic goals in Goal Net. The atomic goal is a single state, which cannot be decomposed.*

**Comp**    *refers to the set of composite goals in Goal Net. The Composite goal is a state which can be further decomposed to other goal states through the Branches Br in Goal Net and form a sub Goal Net with a set of sub goal states and transitions.*

All of these agent goals are represented as goal states S in Goal Net. The Goal Net modeling has the following properties,

- Hierarchy— A goal in Goal Net can be either atomic or composite. A composite goal comprises of a hierarchy of sub goals. A composite goal representing a higher level of agent behaviors is always decomposable into sub goals, which represent more concrete agent implementation at a lower level.

- Temporal— Goal Net transitions associate with each other in a temporal manner. Two transitions can be in one of the four basic relationships, including sequence, concurrency, choice and synchronization.

- Polymorphism— Each transition can be realized by specific actions. There are three types of action selection strategies in Goal Net, namely, direct, conditional and probabilistic.

An agent tries to achieve a set of goals in a complex, dynamic environment. It reasons and decides on its own how to select the next goal and what actions it should take so that its goal can be successfully achieved namely goal selection and action selection.

**Definition 3.5: Goal selection** *is to identify a sequence of goals, highlighting the selection of next arc to reach the next goal for agents to attempt*.

**Definition 3.6**: **Action Selection** *focuses on the selection of transitions on an arc. It is used to define what tasks to do in order to successfully go through the arc and let the agent reach the next goal.*

Based on the aggregated path reward, an agent optimizes the goal selection by maximizing the reward. For action selection, there are three types of strategies, including,

- Sequential execution: agents directly select an action according to the fixed order of actions, without considering other factors.
- Rule-based inference: agents apply a set of rules to decide which action to take. A rule fires when premises of the rule are satisfied. It is useful when the information of action selection is completed and unambiguous.

Probabilistic inference: agents adopt Bayesian networks that can infer the relations between agent states and the optimal actions. It relies on probabilities when information is uncertain and incomplete.

We adopted Goal Net, because it is natural to describe an agent's various tasks as sub goals, which align well with the agent's highest goal, "to facilitate students' learning ". The Goal decomposition allows this higher-level learning goal to be seamlessly transformed to a sequence of lower-level actionable goals. This Goal hierarchy bridges the gap between theoretical agent modeling and the concrete functional design.

Beginning with a root goal, and then subsequently connecting all the sub-goals and states identified for achieving the root goal, an instantiated "net" structure will be created.

These GoalNet structures will be stored in the agent's knowledge base as the Goal Structures.

## 3.4 ATA Routine Modeling with Goal Net

During the lifetime of a Teachable Agent, it repeatedly senses, reasons, and acts on the environment. The repeated cycle forms the main routine of the agent and it can be represented as a Goal Net as shown in Figure 3.5.The description of the Goal Net is as follows,

Table 3.1: Goal States in ATA Routine Model

| Goal State ID | Description | Type |
|:---:|:---|:---:|
| $S_0$ | To Execute the Main Routine of TA | Root |
| $S_1$ | User is far away | Atom |
| $S_2$ | User is coming | Atom |
| $S_3$ | Teachability Goal Pursuing | Composite |
| $S_4$ | Affectivability Goal Pursuing | Composite |
| $S_s$ | Start State | Atom |
| $S_e$ | End State | Atom |

Table 3.2: Transitions in ATA Routine Model

| Transition ID | Description | $(T_f, S_{pre}, S_{post})$ |
|:---:|:---|:---|
| $T_1$ | Detect user | $(f_{check\_user}, S_s/S_1, S_1)$ |
| $T_2$ | Initiate the learning goal pursuit | $(f_{init\_sub\_goal}, S_2, S_3)$ |
| $T_3$ | Initiate the affective goal pursuit | $(f_{init\_sub\_goal}, S_2, S_4)$ |
| $T_4$ | Finish | $(f_{finish}, S_3,/S_4, S_e)$ |

We have,

**Root** = {$S_0$}, **Atom** = {$S_1$, $S_2$, $S_s$, $S_e$}, **Comp** = {$S_3$, $S_4$}

At the highest abstraction, this Goal Net explains the root goal of the agent, i.e. "To execute the main routine of a TA". The agent repeatedly senses the environment at a pre-determined sampling frequency and decides whether a user has been detected. If a user is present, it starts both the Teachability Pursuit and the Affectiability Pursuit processes; if a user is absent, the agent will continue sensing the environment.

This main routine is designed for achieving the proactivity of TAs, and it is the top-level goal of an agent to achieve its goals. Under this main routine, two sub goal hierarchies are designed to fulfill the specific learning tasks and finally to achieve Teachability and Affectivability. To fire the composite goal *Teachability Goal Pursuing* and *Affectivability Goal Pursuing*, an iterative execution of the Sub Goal Net is required.

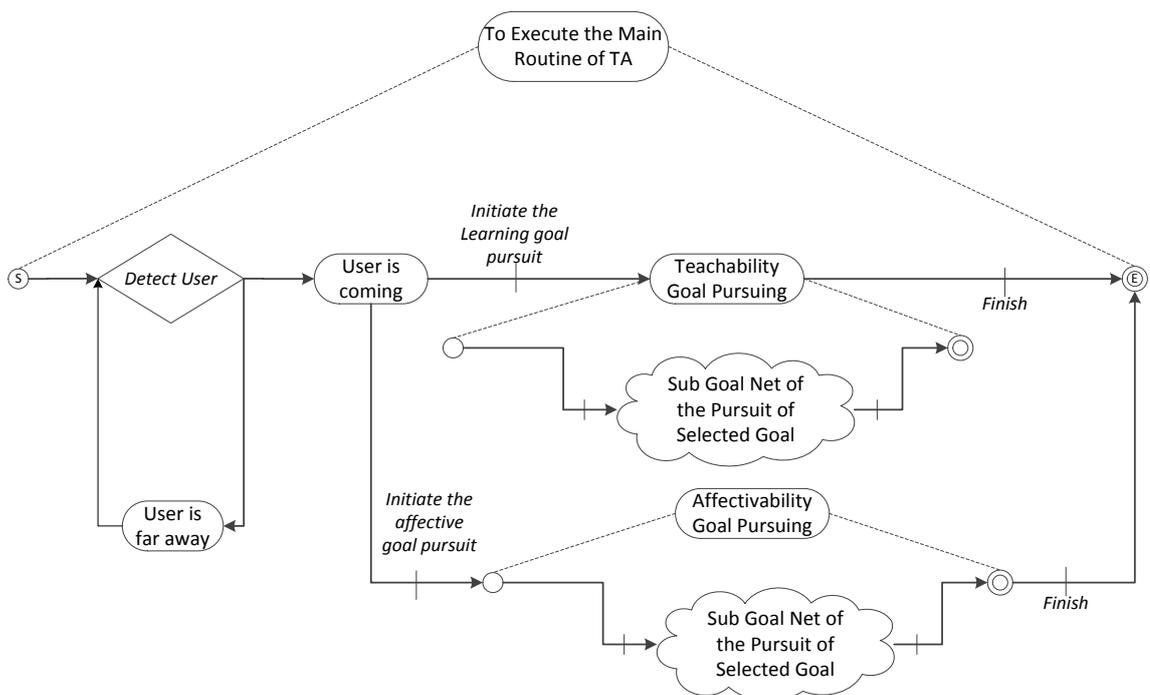

Figure 3.5: Model of ATA's Main Routine [105]

Viewed from a broad perspective, an agent repeats the routine cycle that starts with detecting the environment and activates the two concurrent sub goal nets to deal with the specific tasks. The Sub Goal Net will guide the TA to respond to the environment with concrete behaviors. In detail, the agent repeatedly senses the environment at a predetermined sampling frequency and detects users. If a user is present, it analyzes the user's current learning stage (whether the user has taught some knowledge before or what teaching stage he reached last time) and triggers the corresponding teaching/learning goal structures through the pursuit of the Sub Goal Net with the composite goal *Teachability Goal Pursuing*. At the same time, the agent also triggers the Sub Goal Net with composite goal *Affectivability Goal Pursuing* to generate TA's emotional expressions according to the event happened at that moment. On the other hand, if an event is absent, the agent turns the cycle back to continuously observing the environment. Note that the information of Goal Nets is stored and retrievable from agent's implementation repository. After that, the agent executes the Goal Net until the end of the main routine is reached.

There are three Sub Goal Nets which can be loaded in the agent's main routine, and their goals are "to learn from user", "to practice", and "to be affective", respectively. We will explain the details of the goal-oriented models for reasoning in teachability and affectivability in the following sections.

## 3.5 Teachability Modeling with Goal Net

A Teachable Agent achieves teachability through learning knowledge from users. The learning process is triggered when the agent detects that a user is approaching and the agent has not learnt valid knowledge from that user. The Teachability modeling involves two Sub Goal Nets, "to learn from user" and "to practice". The first one is designed for TA to learn from students and store the learnt knowledge into the domain knowledge base. The second one is designed to provide students with more types of feedback that can illustrate the learning performance of the TA.

### 3.5.1 Modeling the Learning Process of the ATA

In the Goal Net with the composite goal "to learn from user" (Figure 3.6), the first step of the TA is to ask for help from students in order to attract students to teach. During this process, the TA needs to ensure that the students can understand clearly about what they need to teach. Therefore, the agent initiates the conversation by requiring teaching from the user. When considering how to properly ask help from students, the agent's question selection is based on the user log of what the user has/has not taught before. There are three situations: 1) TA omits the material that has been taught, 2) highlights the knowledge that is wrong, and 3) connects the learnt knowledge point with current learning. The Goal Net of "to learn from user" (as ) is described as follows,

Table 3.3: Goal States in ATA's Learning Model

| Goal State ID | Description | Type |
|---|---|---|
| $S_0$ | To Learn from User | Root |
| $S_1$ | Response Received | Atom |
| $S_2$ | User Agree | Atom |
| $S_3$ | User Reject | Atom |
| $S_4$ | Teaching Panel Displayed | Atom |
| $S_5$ | Knowledge Received | Atom |
| $S_6$ | Error Found | Atom |
| $S_7$ | No Error | Atom |
| $S_s$ | Start State | Atom |
| $S_e$ | End State | Atom |

Table 3.4: Transitions in ATA's Learning Model

| Transition ID | Description | $(T_f, S_{pre}, S_{post})$ |
|---|---|---|
| $T_1$ | Require teaching | $(f_{message\_teaching}, S_s, S_1)$ |
| $T_2$ | Check response | $(f_{check\_response}, S_1, S_2/S_3)$ |
| $T_3$ | Choose a teaching approach | $(f_{show\_approach}, S_2, S_4)$ |
| $T_4$ | Perceive knowledge | $(f_{perceive\_input}, S_4, S_5)$ |
| $T_5$ | Check error | $(f_{check\_errors}, S_5, S_6/S_7$ |
| $T_6$ | Alert user | $(f_{message\_alert}, S_6, S_s)$ |
| $T_7$ | Save knowledge | $(f_{save\_knowledge}, S_7, S_e)$ |
| $T_8$ | Finish | $(f_{finish}, S_3, S_e)$ |

We have,

**Root** = $\{S_0\}$, **Atom** = $\{S_1, S_2, S_3, S_4, S_5, S_6, S_7, S_s, S_e\}$, **Comp** = $\phi$

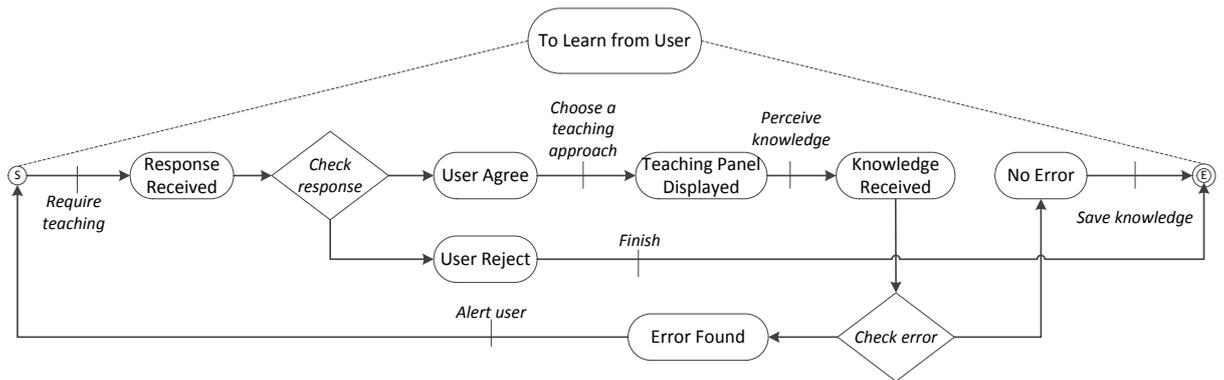

Figure 3.6: Model of ATA's Learning

When TA asking for help, the students may reject the agent's request, and this simply ends the Goal Net execution and triggers an affective expression through affectivability reasoning. Once the user agrees to teach the agent, the system will select a teaching approach from the teaching strategy database, which is stored as domain knowledge in TA's knowledge base. A clear, explicit knowledge representation is one of the most important aspects of the whole system, since it influences the efficiency of human computer interaction and the teaching experience of students [13]. Thus, the system should provide a shared knowledge presentation to facilitate student's teaching. In our design, two kinds of teaching approaches are designed. The first is using a concept map, and the other is using an experiment simulation. The concept map is utilized as an interface tool for the agent to grasp structured data from the student. The agent tracks the changes of the concepts and relations on the concept map drawn by the student. An error checking mechanism is used to alert the student, if any syntactic error is detected. Otherwise, the agent analyzes the received structured input and saves the input as its knowledge learnt from the student. Both concept maps and knowledge representation are application-dependent and are defined by software designers. A use case will be given in Chapter 4 for more details.

## 3.5.2 Modeling the Practices of ATA in Virtual World

Regarding the educational aspect, one of the most important characteristic of a teachable agent is to keep its internal states transparent to students, because the explicit reasoning process can largely help students learn from the activity of agents, and allow students to form structured knowledge embodied in the agent. As mentioned before, the TA should follow students' direction and give students the feeling of direct control. What the TA learnt is what the student taught. In this way, the TA's performance should be consistent with the learning content. Based on this kind of design, we attempt to use the agent's practice performance in the virtual learning environment to provide students with more feedback on their teaching quality. For example, the agent can answer questions correctly, have high score in the assessment, or be able to deduct appropriate actions upon a situation. The feedbacks provided by the activities of the TA in virtual world can indicate the gap between a TA's current performance and the desired level of performance. With the demand of making up this gap, more teaching effort from students can be motivated [13]. Additionally, watching a TA's performance also can reduce student's uncertainty about how good (or poor) their teaching is.

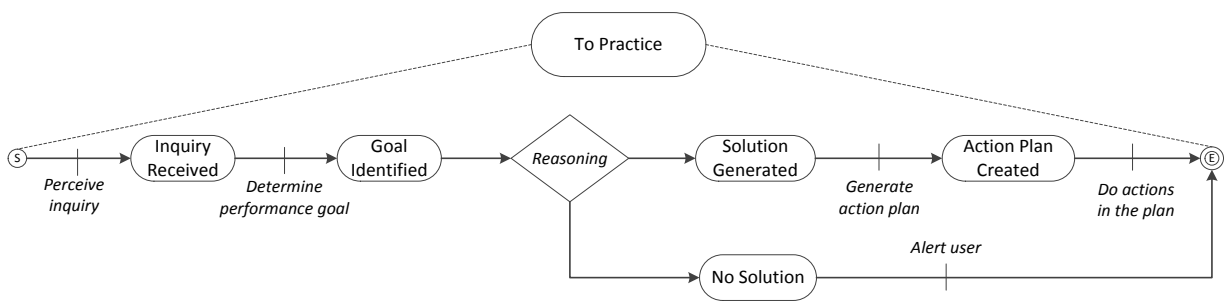

Figure 3.7: Model of ATA's Practice in Virtual World

The performance is content-dependent, and aligns with teachable agent's practice through applying the learnt knowledge for reasoning. This process is controlled by another Goal Net, with the root goal "to practice" (as

Figure 3.7). The description of the Goal Net is listed in Table 3.5 and Table 3.6.

Table 3.5: Goal States in ATA's Practice Model

| Goal State ID | Description | Type |
|---|---|---|
| $S_0$ | To Practice | Root |
| $S_1$ | Inquiry Received | Atom |
| $S_2$ | Goal Identified | Atom |
| $S_3$ | Solution Generated | Atom |
| $S_4$ | Action Plan Created | Atom |
| $S_5$ | No Solution | Atom |
| $S_s$ | Start State | Atom |
| $S_e$ | End State | Atom |

Table 3.6: Transitions in ATA's Practice Model

| Transition ID | Description | $(T_f, S_{pre}, S_{post})$ |
|---|---|---|
| $T_1$ | Perceive inquiry | $(f_{perceive\_input}, S_s, S_1)$ |
| $T_2$ | Determine performance goal | $(f_{identify\_target}, S_1, S_2)$ |
| $T_3$ | Reasoning | $(f_{reasoning}, S_2, S_3/S_5)$ |
| $T_4$ | Generate action plan | $(f_{generate\_plan}, S_3, S_4)$ |
| $T_5$ | Do actions in the plan | $(f_{execute\_plan}, S_4, S_e)$ |
| $T_6$ | Alert user | $(f_{message\_alert}, S_5, S_e)$ |

We have,

**Root** = $\{S_0\}$, **Atom** = $\{S_1, S_2, S_3, S_4, S_5, S_s, S_e\}$, **Comp** = $\phi$

The goal of this Goal Net is "to practice" by reasoning over the learnt knowledge. To achieve this goal, the agent starts by perceiving inquiries from the system. Questions are one type of inquiry, if the agent is evaluated in an ask-and-answer manner. In a more interactive way, an inquiry may require the agent to perform a certain task, such as doing an experiment. For both cases, the agent can get feedbacks from the environment, and the latter allows the students to gain a more insightful view by observing how the agent behaves. Students can initiate the inquiry, but more importantly, an inquiry also can be started by the system *per se*. Once the agent has enough knowledge to exploit in the environment, it will practice by itself. The agent reasons over the knowledge base for solutions. It is possible that due to inaccurate or insufficient knowledge learnt from the student, the agent cannot obtain any solution. In these cases, the agent issues a notification to the user and begs the user to teach more related knowledge. If a solution is found, the agent realizes the solution through actions that can be directly carried out by the agent. A plan can be as simple as a single action, such as "choose Answer A", or as complex as a sequence of actions. With the decomposed Goal Net, the agent will conduct the intended actions in the environment. Note that a generated solution may not necessarily guarantee a good feedback, and the student is expected to observe agent's performance feedback and to learn from agent's behavior accordingly.

## 3.6 Affectivability Modeling with Goal Net

In the pursuit of a good relationship with students, a TA needs to establish empathy with students. The scope of affectivability of a TA is achievable through the elicitation of the agent's emotions. It is natural to have emotional expression together with the learning and practicing behaviors of the TA. Therefore, besides the thread for teachability, it is necessary to have a thread for affectivability running in parallel. In the TA routine Goal Net, there are two parallel threads: one handles the events related to teachability; and the other listens to the events that may elicit emotions. In this part, we will explain how to build up a goal-oriented emotion elicitation model with Goal Net.

### 3.6.1 Qualitative Emotion Elicitation for ATA

According to a well-known emotional model, Ortony Clore Collins (OCC) [38], emotions can be classified into three groups, namely goal-oriented emotions, standard-oriented emotions, and attitude-oriented emotions. For each group of emotions, there is a type of trigger, or in other words, a causation of the group of emotions:

$$Emo\_Causation \in \{Event, Agent, Object\}$$

Where:

$Event$ is the trigger of goal-oriented emotions

$Agent$ is the trigger of standard-oriented emotions

$Object$ is the trigger of attitude-oriented emotions

Among the three groups, event-based emotions are the most complicated, yet they occur most frequently in a virtual learning environment. Since virtual worlds are commonly modeled as event-driven and implemented by event listeners, a virtual agent's behavior is usually considered as a certain reaction to virtual world events. Therefore in this book we only consider cases where $Emo\_Causation \in Event$. The OCC model uses the cognitive conditions to elicit emotions and defines the goal-oriented emotions as a type of emotional state that is generated based on the relations between current event and agent's goals. In this research, we will discuss how to generate the event triggered, goal-oriented emotions to enrich the interactions between a TA and students.

The general logic flow of the OCC includes four steps. The first stage is to identify the current event and agent's goal, since the elicitation of goal-oriented emotions is based on the relation between these two factors. An event at time t is defined as $event_t = \langle event\_content_t, event\_endurer_t \rangle$, where the consequence of $event\_content_t$ affects $event\_endurer_t$. The endurer can be an agent or the student. The goal is represented as $goalof(event\_endurer_t)$, which refers to the goal of the event endurer. For example, the goal of the student is to teach the agent, and the goal of an agent is to get help from students.

**Step 1**: Identify $event_t = \langle event\_content_t, event\_endurer_t \rangle$

$$goal_t of (event\_endurer_t)$$

The next step is to analyze the desirability of a TA for the current event. According to OCC, the desirability can be defined as $desirable(event_t, Goal) \in \{True, False\}$. For example, when a TA's goal is getting help from student, and the current event is getting acceptance from student, then, the desirability of TA on the event getting acceptance from student is true.

**Step 2**: Analyze $desirable(event_t, Goal) \in \{True, False\}$

Moreover, a virtual agent who is experiencing the emotion is known as $Emo\_Holder$. Note that $event\_endurer_t$ can be the $Emo\_Holder$ (consequences for the self) or other agents (consequences for other). If $event\_endurer_t$ is not the $Emo\_Holder$, subgroup fortunes-of-others is raised. Then we need to consider the will of $Emo\_Holder$ towards $event\_endurer_t$. If the emotion holder has a good will to the event endure, then the emotion holder will feel *happy for* the event endurer when good things happen, and feel *sad* when bad things happen. If the emotion holder has a bad will to the event endure, the emotion holder feels *annoying* when good things happen to the event endurer, and feels *gloating* when bad things happen to the event endurer. Thus, we have,

**Step 3**: Identify $Emo\_Holder$, and if $Emo\_Holder \neq event\_endurer_t$, identify

$$will(Emo\_Holder, event\_endurer_t) \in \{good\_will, \ ill\_will\}$$

For the situation, $event\_endurer_i = Emo\_Holder$, we need to consider the "prospect relevance". We call an event *prospect relevant* if the event has a delayed, rather than immediate, consequence. Whereas we call an event *prospect irrelevant* if the event only has immediate consequence. For instance, when TA doing practice in virtual learning environment, it may prospect a praise from students. In this case, the event doing practice in virtual world is a prospect relevant event. Thus,

**Step 4**: Identify $prospect\_relevant(event_i) \in \{True, False\}$

Based on those four steps, we can generate twelve goal-oriented emotions. The Affective Goal Net can be drawn as Figure 3.8.

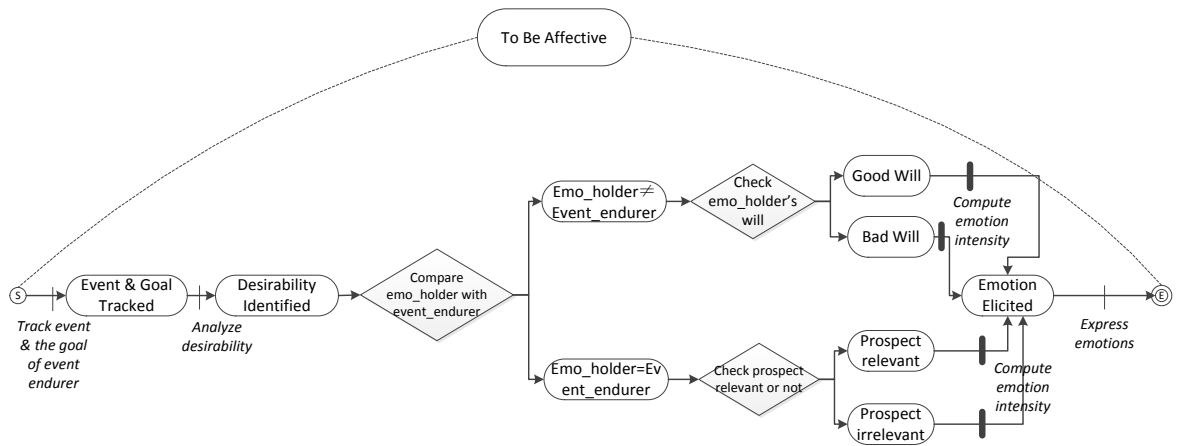

Figure 3.8: Model of ATA's Affectivability

The Goal Net of "to be affective" is described as follows,

Table 3.7: Goal States in ATA's Affectivability Model

| Goal State ID | Description | Type |
|---|---|---|
| $S_0$ | To Be Affective | Root |
| $S_1$ | Event & Goal Identified | Atom |
| $S_2$ | Desirability Identified | Atom |
| $S_3$ | Agent is not event endurer | Atom |
| $S_4$ | Agent is event endurer | Atom |
| $S_5$ | Good Will | Atom |
| $S_6$ | Bad Will | Atom |
| $S_7$ | Prospect Relevant | Atom |
| $S_8$ | Prospect Irrelevant | Atom |

| $S_9$ | Emotion Elicited | Atom |
| $S_s$ | Start State | Atom |
| $S_e$ | End State | Atom |

Table 3.8: Transitions in ATA's Affectivability Model

| Transition ID | Description | $(T_f, S_{pre}, S_{post})$ |
|---|---|---|
| $T_1$ | Track event and goal of event endurer | $(f_{perceive\_event\&goal}, S_s, S_1)$ |
| $T_2$ | Track desirability | $(f_{reason\_desirability}, S_1, S_2)$ |
| $T_3$ | Check relation with event endurer | $(f_{check\_identity}, S_2, S_3/S_4)$ |
| $T_4$ | Check agent's will | $(f_{check\_will}, S_3, S_5/S_6)$ |
| $T_5$ | Check relevance of the event | $(f_{check\_relevance}, S_4, S_7/S_8)$ |
| $T_6$ | Compute emotion intensity | $(f_{reason\_emotionIntensity}, S_5/S_6/S_7/S_8, S_9)$ |
| $T_7$ | Express emotion | $(f_{execute\_expression}, S_9, S_e)$ |

We have,

**Root** = $\{S_0\}$, **Atom** = $\{S_1, S_2, S_3, S_4, S_5, S_6, S_7, S_8, S_9, S_s, S_e\}$, **Comp** = $\phi$

Note that the transitions highlighted with four bold lines are the transitions for computing emotion intensity. To describe the expression of emotions, there are two crucial factors. One is what type of emotion it is; the other is the intensity of the emotion. Through the four steps of emotion elicitation, we can answer the first question, but we still do not know how strong the emotion is. To achieve this, we will do the quantitative calculation of emotions via a fuzzy tool FCM. The details will be introduced in next section.

Table 3.9: Correspondence between Emotions and the Eliciting Conditions

| Input: | event and emotion holder |
|---|---|
| Initialize: | goal and desirability |
| 1 | Check whether the event endurer is the emotion holder: yes, go to 4 no, go to 2 |
| 2 | Check the will of emotion holder towards event endurer: for good will, go to 3 for bad will, go to 9 |
| 3 | If desirable, ***happy-for*** If undesirable, ***pity*** End of algorithm. |
| 4 | Check the prospect is relevant or irrelevant for relevant, go to 6 |
|   | for irrelevant, go to 5 |
| 5 | If desirable, **joy**<br>If undesirable, **distress**<br>End of algorithm. |
| 6 | If desirable, **hope**, check whether hope is confirmed and go to 7<br>If undesirable, **fear**, check whether fear is confirmed and go to 8 |
| 7 | If confirm, **satisfaction**<br>If disconfirm, **disappointment**<br>End of algorithm. |
| 8 | If confirm, **fear-confirmed**<br>If disconfirm, **relief**<br>End of algorithm. |
| 9 | If desirable, **resentment**<br>If undesirable, **gloating**<br>End of algorithm. |

Twelve types of emotions can be generated based on the above Goal Net, and the correspondence between the conditions and the emotion type are shown in Table 3.9.

## 3.6.2 Quantitative Computation of Emotion Intensities

Using the OCC-based Affective Goal Net, we are able to find out what emotions the agent generates according to current situations in the virtual world. However, only identifying what emotions are likely is not enough. We also need to compute the specific intensity of emotions to achieve the precise control of agent behaviors. In this section, we will propose an approach to transform the Affectivity Goal Net to a computable causal graph, which can compute the emotion intensity through a fuzzy tool. We aim to depict the dynamic features in the emotion elicitation process and the causal relationships between all kinds of emotional elements.

To select the computational tool for computing emotion intensity, two most important requirements need to be fulfilled. First, the tool should be easy to transform from the Goal Net structure to the computational presentation. Second, the tool has the capability to compute the intensity value based on the causal relationships among different emotional elements. By considering both of the two requirements, we choose Fuzzy Cognitive Map for the emotion intensity computation. A Fuzzy Cognitive Map (FCM) is a fuzzy-graph structure, which can simulate the complex systems in the world through causes, effects, and the causal relationships in between. A FCM as an efficient fuzzy tool can be defined as a trio $(C, R, W)$, in which $C = \{C_1, C_2, ..., C_n\}$ is the Concept set. Each element, concept, is represented as a node in FCM graph, and the causes and effects are all defined as concepts in this set. $R = \{R_1, R_2, ..., R_m\}$ refers to the Causal Relation set. Each element $R_k = \overrightarrow{C_i C_j}$ refers to the causal relation between concepts $C_i$ and $C_j$. The relations are represented as arcs in the FCM graph. Each causal relation has a weight to depict the influential degree from the former concept to the latter one, and it is defined as $W = \{W_1, W_2, ..., W_m\}$. All weights of the causal relations can be compactly represented as a connection matrix $\mathbf{W} = \begin{bmatrix} w_{ij} \end{bmatrix}$. An example of a FCM graph and its corresponding weight matrix is shown below.

$$W = \begin{bmatrix} 0 & +2 & -1 & -1 \\ 0 & -1 & 0 & +1 \\ -1 & 0 & 0 & 0 \\ 0 & 0 & +1 & 0 \end{bmatrix}$$

A FCM as a fuzzy tool, on one hand, can describe the causal elicitation process of emotion generation; on the other hand, it can set values of the emotion intensity for the TA to express its emotions. The advantage of the FCM can be summarized as four aspects,

1. The FCM represents knowledge in a symbolic manner. The relationship between each concept can be directly signed by inter-linkages. The graphical representation of FCM is convenient for representing the Affective Goal Net.

   The OCC model is a cognitive model, which uses eliciting conditions to induce emotions, and the relationship between elements in this model is a causal relationship. As modeling causal relationship is just the most important feature of the FCM, the FCM has the ability to represent the computational model of the TA's emotion intensity.

2. Apart from the graphical representation, the FCM also provides the mathematical approach to analyze the problem. Each concept can be defined as a fuzzy set. In addition, the causal strength and the interactive relations can be depicted by weighted values for each connection. The symbolic and numeric transformation in FCM is straightforward.

   In order to clearly analyze the emotions elicited by the OCC, each element used to elicit emotion needs to have its own value, and the induced relationship between elements needs to be computable. These two requirements can be met by the state value of concepts and the weights of causal relationships respectively.

3. For building a dynamic system, the FCM is efficient to describe a complex dynamic process, because FCM makes the complex operation of the whole system transparent through defining the causal relationship in each concept pair. The whole system can be established by identifying each partial relationship. Therefore, no matter how complicated the elicitation process of emotions will be, FCM is a powerful tool for designers to model such process.

4. Finally, different FCMs can be easily combined into one, as long as the two FCMs share a common concept. Considering the complexity of modeling emotion, using the FCM can bring much greater flexibility for merging or splitting the inference structure.

Using the FCM to analyze the dynamic system of emotions and compute the emotion intensities, we need to do three phases of integration. Phase 1 is to map the inputs to the outputs, and to map the related concepts of the OCC-based Goal Net to the concept set $C$ of FCM. This phase is to model the emotional elements as a collection of concepts, and these concepts will then be used for simulating the dynamic process of emotion intensity elicitation.

The concepts belonging to set $C$ should contain all the factors that are necessary for computing the intensity of emotions. Based on the OCC theory, the factors that affect the intensity of emotions include

1. the degree of desirability of an event, and

2. the degree of unexpectedness that an event will occur (expectation).

Besides these factors, some important varying concepts also need to be considered in the FCM. These are

1. the impact of the causational event,

2. the impact of reactions of the emotion holder, and

3. the intensity of emotions

where the first factor can reflect the environmental changes of the system, and the second and the third factors reflect the behaviors of the affective agent.

**DEFINITION 1**: FCM Concepts in Affective Teachable Agent are defined as a set **Concept** $= \{C_1, C_2, C_3, ..., C_n\}$ , and $C_i$ Î **CausationSet** È **EvaluateReferenceSet** È **DesirabilitySet** È **ExpectationSet** È **EmoSet** È **ActionSet** , in which,

**DEFINITION 1.1**: **CausationSet** refers to the causation of emotions. According to OCC emotions arise from the evaluation of three aspects of the world: events, agents, and objects. We have,

$\qquad$ **CausationSet** $= \{cause_1, cause_2, ..., cause_n\}$

$\qquad$ where,

$\qquad\qquad cause_i$ Î **EventSet** È **AgentSet** È **ObjectSet**

$\qquad$ where,

$\qquad\qquad$ **EventSet** $= \{event_1, event_2, ..., event_n\}$
$\qquad\qquad$ **AgentSet** $= \{agent_1, agent_2, ..., agent_n\}$
$\qquad\qquad$ **ObjectSet** $= \{object, object_2, ..., object_n\}$

In light of this causation classification of emotions, emotions can be categorized as event-based emotion, agent-based emotion, and object-based emotion three types.

**DEFINITION 1.2**: The three causations of emotions correspondingly have three evaluating references for eliciting emotions which is defined as **EvaluateReferenceSet** .

$\qquad$ **EvaluateReferenceSet** $= \{e\_r_1, e\_r_2, ..., e\_r_n\}$

$\qquad$ where,

$\qquad\qquad e\_r_i$ Î $\{$**GoalSet**$|cause_i$ Î **EventSet** $\}$
$\qquad\qquad\qquad$ È $\{$**StandardSet**$|cause_i$ Î **AgentSet** $\}$
$\qquad\qquad\qquad$ È $\{$**AttitudeSet**$|cause_i$ Î **ObjectSet** $\}$

$\qquad$ where,

$$\textbf{GoalSet} = \{goal_1, goal_2, ..., goal_n\}$$

$$\textbf{StandardSet} = \{standard_1, standard_2, ..., standard_n\}$$

$$\textbf{AttitudeSet} = \{attitude_1, attitude_2, ..., attitude_n\}$$

According to OCC, the three causations of emotions correspondingly have three evaluating references for eliciting emotions. Specifically, elements in **EventSet** are compared with the relevant goal in **GoalSet** for judging the desirability of events, elements in **AgentSet** are compared with the relevant standard in **StandardSet** for judging the agreement of agents, and elements in **ObjectSet** are compared with the relevant attitude in **AttitudeSet** for judging the attraction of objects.

**DEFINITION 1.3**: **DesirabilitySet** refers to the result of the evaluation of the three aspects of world. For example, if the event is consistent with the goal, one's desirability is desirable, otherwise undesirable. Therefore, we have,

$$\textbf{DesirabilitySet} = \begin{cases} desirable & |event_i \text{ consistent with } goal_i \\ undersirable & |event_i \text{ inconsistent with } goal_i \end{cases}$$

**DEFINITION 1.4**: **ExpectationSet** refers to the level of people's expectedness or unexpectedness of events. In general, unexpectedness is positively correlated with the intensity of the emotions.

**DEFINITION 1.5**: **EmoSet** refers to the output emotions of the elicitation process.

$$\textbf{EmoSet} = \{22 \text{ types of emotions in OCC}\}$$

**DEFINITION 1.6**: **ActionSet** refers to the actions the emotion holder does. The elements in this is set can be defined according to the real emotional scenario.

$$\textbf{ActionSet} = \{action_1, action_2, ..., action_n\}$$

When designing the emotional model using FCM, the concepts can be generated by identifying whether these types of concepts are involved in the situation. If involved, then draw each concept as a node. After drawing the nodes, we need to find out the causal relations between them. The FCM can be generated as Figure 3.9.

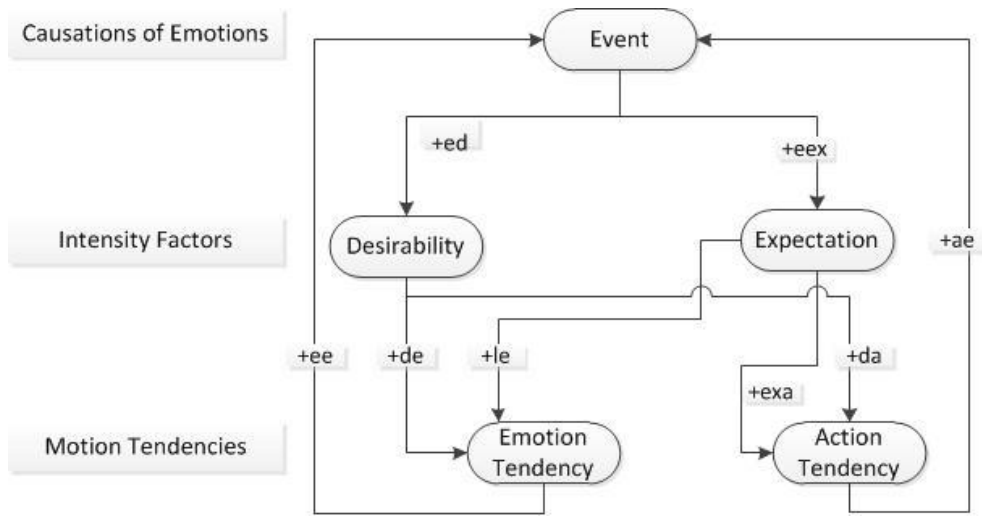

Figure 3.9: FCM for Affective Modeling

The second phase, after collecting all the concepts, is to find out the causal relations between these concepts and represent them in set $R$ .

**DEFINITION 2**: *Causal relations* in FCMs are defined as a set **RelationSet** $= \left\{ r_1, r_2, ..., r \right\}_n$ , in which each $r_i = \overrightarrow{C_i C_j}$ referring to the causal relation between concepts $C_i$ and $C_j$ , graphically represented as an edge.

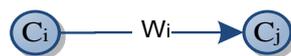

Figure 3.10: Causal relation $r_i$ between $C_i$ and $C_j$

**DEFINITION 3**: Each *Causal Relation* has a weight to represent how much one concept influences another which is defined as:

**WeightSet** $= \left\{ w_1, w_2, ..., w_n \right\}$

The whole causal relation set can be represented compactly as a connection matrix $W_{ij}$. When connecting all the relevant concepts with edges according to the causalities between concepts, an emotion elicitation process based on FCM is built up. It has the potential to have dynamic features and feasible interactions.

In the third phase, we draw the concepts as nodes and connect all the concepts based on their relations with arcs. Moreover, we represent the weights of all the relation arcs in a matrix format. If there is no causal relation between two concepts, we define the weight as zero. With the weight matrix we can use simple linear algebra to do the computation iteratively as $C_{i+1} = C_i \cdot W$. The details will be explained with an example in the case study.

We set the weight values based on the formula for emotion intensity computation generated by [92], which are shown in Table 3.10.

Table 3.10: Formula for Computing Emotion Intensities

| Emotion | Formula for Intensity |
|---|---|
| Joy | $Joy = (1.7 \times expectation^{0.5}) + (-0.7 \times desirability)$ |
| Sadness | $Sad = (2 \times expectation^2) - desirability$ |
| Disappointment | $Disappointment = Hope \times desirability$ |
| Relief | $Relief = Fear \times desirability$ |
| Hope | $Hope = (1.7 \times expectation^{0.5}) + (-0.7 \times desirability)$ |
| Fear | $Fear = (2 \times expectation^2) - desirability$ |

## 3.6.3 Case Study for Affective Elicitation

In this section, we will illustrate how to analyze a scenario with concrete environmental conditions and how to use it to do the intensity computation through the case study in the Dino project.

*Example:  Given that a learning companion dinosaur called Dilong (ID: avatarA) notices that a giant carnassials dinosaur is coming (event_content ID: event_content002), and Dilong realizes that maybe he is in danger of being eaten by the giant dinosaur.*

Analyzing by the OCC-based elicitation process we have,

Inputs:

1. $Emo\_Causation_t = event_t = \langle event\_content002, avatarA \rangle \in Event$
2. $Emo\_Holder = avatarA$  (the agent dinosaur)

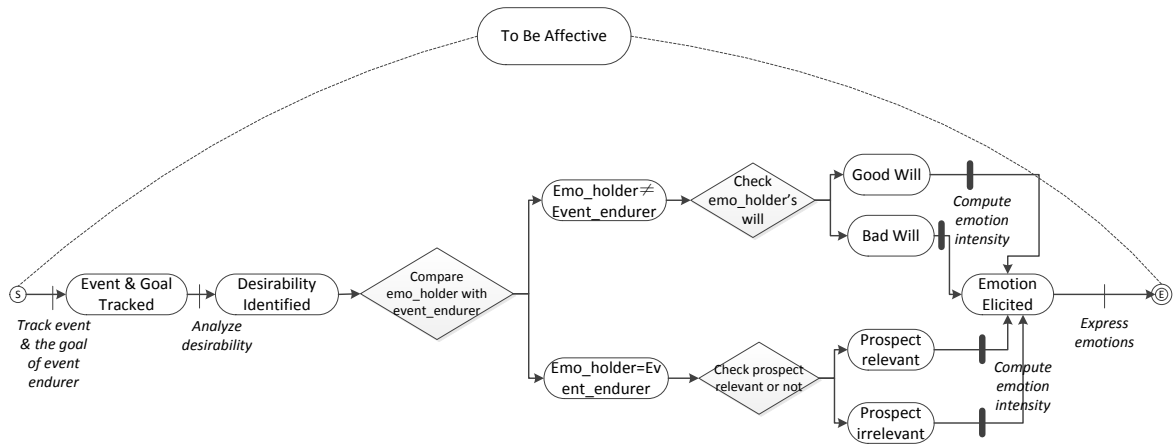

Figure 3.11: Model of ATA's Affectivability

Details:

1. Track $event_t = \langle event\_content002, avatarA \rangle$, which is $'giantDino\_come'$
   $Goal = goalof(avatarA)$, which is $'avoid\_giantDino'$.
2. Analyze $desirable(event_t, Goal)$ which returns $False$.
3. Check whether $event\_endurer_t = Emo\_Holder$, which return $True$.

4. Check $prospect\ relevant\ event\ content()$, which returns $True$, because being eaten by the carnassials dinosaur is the "prospect" of this learning companion dinosaur. Also, because of the undesirability of this event, we straightforwardly come to the conclusion.

5. Output: $emotion_t = Fear$ that Dilong feels fear when the giant dinosaur is near.

After the qualitative analysis via Goal Net, we generate the corresponding FCM to map the rules. First, collect relevant concepts of FCM. The concepts affecting the intensity of event-based emotions include *desirability* and *expectation*, the Dilong's desirability about the event, and the degree of Dilong's belief of being eaten by giant dinosaur. The impact of the causal event is defined as the distance between Dilong and the giant dinosaur. The output concept Emotion is fear, and the concept Action is a sequence of Dilong's frightened behaviors, which controls Dilong's reaction in the application level. To sum up, all the concepts are listed in Table 3.11.

Table 3.11. List of FCM Concepts for Dilong

| Concepts | Definition |
| --- | --- |
| C1 Event | Distance between Dilong and the giant dinosaur |
| C2 Desirability | Dilong's desirability |
| C3 likelihood | The degree of Dilong's belief of being eaten |
| C4 Emotion | Dilong's emotion "fear" |
| C5 Action | Dilong's frightened actions |

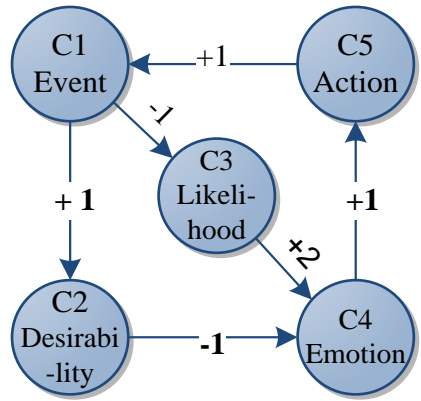

Figure 3.12: FCM for Dilong Being Eaten Situation [106]

According to the distance between the giant dinosaur and Dilong C1, Dilong's desirability C2 is changed correspondingly. At the same time, C1 also affects Dilong's belief of being eaten, the likelihood of being eaten C3. Furthermore, the desirability C2 and the likelihood C3 will together determine Dilong's emotion C4. Emotional state will influence the agent's reaction to the event, and thus Dilong's emotional state C4 will regulate its reaction C5. Finally, this reaction directly changes the distance between the giant dinosaur and Dilong in return. The weights define the degree of the causal effect. Based on the relations between each concept, the FCM is drawn as Figure 3.12.

After the graphical FCM is represented, we need to focus on how to do the intensity computation. In light of OCC, this model is related to the emotion "fear". By using the formulation of the emotion fear [92], we have $Fear = 2 \times Expectation^2 - Desirability$. As we have mentioned in Section 4.2, each concept or weight can use a predefined function, and we adopt three functions for concepts C1, C3 and C5 to simulate the causal relations of fear emotion. $C_k^{'}$ denotes the original concept values before passing through the concept function. For C1, the function is $f_{C_1} = C_1^{'} / d_{max}$, where $d_{max}$ is the maximum distance defined for cases in the virtual world. By dividing by $d_{max}$, each real distance value between Dilong and giant dinosaur in the system is normalized between 0 and 1.

For C3, the function is defined as $f_{C_3} = \begin{cases} (1-|C_3'|)^2, & C_3' \in [-1,0) \\ C_3', & C_3' \in [0,1] \end{cases}$, because the likelihood of being eaten is decreased with the increase of the distance; we use $(1-|C_3'|)^2$ to keep the value positive.

Concerning C5, it refers to the reaction of Dilong. In this model, Dilong's emotion C4 directly influences its reaction C5. We simply define the variation of reactions to be Dilong's running speed

$$v_{Dilong} = (emotion/emotion_{max})v_{Dilong\,max} = (C_4/C_{4max})v_{Dilong\,max}$$

Then, we use function $f_{C_5}$ to define the effect of Dilong's speed on the changes of distance. We assume, during the period of Dilong making running decision, the speed of the giant dinosaur remains the same. Thus we have,

$$f_{C_5} = C_1 - (v_{giantdino} - v_{Dilong})/d_{max}$$

With all these functions and the weight matrix, this approach can do the matrix computation with an initial concept vector. There is only one input in the vector, that is, the distance between Dilong and giant dinosaur C1, which can be read from the game engine directly. Given the distance $d$, the input vector of the five concepts is

C1  C2  C3   C4  C5

$$\vec{C_1} = [\,d \quad 0 \quad 0 \quad 0 \quad 0\,]$$

The matrix computation is an iterative process, and computations for iteration $i$ are summarized as follows,

$$
\begin{cases}
\overline{C_i} \times \mathbf{W} = \overrightarrow{C'_{i+1}} = \begin{bmatrix} C'_{1(i+1)} & C'_{2(i+1)} & C'_{3(i+1)} & C'_{4(i+1)} & C'_{5(i+1)} \end{bmatrix} \\
C_{1(i+1)} = f_{C_1}(C'_{1(i+1)}) \\
C_{3(i+1)} = f_{C_3}(C'_{3(i+1)}) \\
C_{5(i+1)} = f_{C_5}(C'_{5(i+1)}) \\
\overline{C_{i+1}} = \begin{bmatrix} C_{1(i+1)} & C_{2(i+1)} & C_{3(i+1)} & C_{4(i+1)} & C_{5(i+1)} \end{bmatrix}
\end{cases}
$$

We use Matlab to simulate this computation, and got the result shown below. Given $v_{giantdino} = 10$, $v_{Dilong\,max} = 8$, $d_{max} = 80$ and the initial vector $\vec{C_i} = [0.9 \ 0 \ 0 \ 0 \ 0]$, the computation results are shown as Figure 3.13. The figure shows how the distance between giant dinosaur and Dilong changes (blue line), and how the fearful feeling of Dilong changes (red line). When it runs to the 13th round, the distance is 0, and the whole computation terminated meaning Dilong has been caught. If we assume $v_{Dilong\,max} = 20$, the corresponding results are given in Figure 3.14. It shows that if we change the maximum speed of Dilong to 20, the distance between giant dinosaur and Dilong will reach a steady state at the 34th step. In this way, Dilong will never be caught, and its emotion of fear will remain at a constant level.

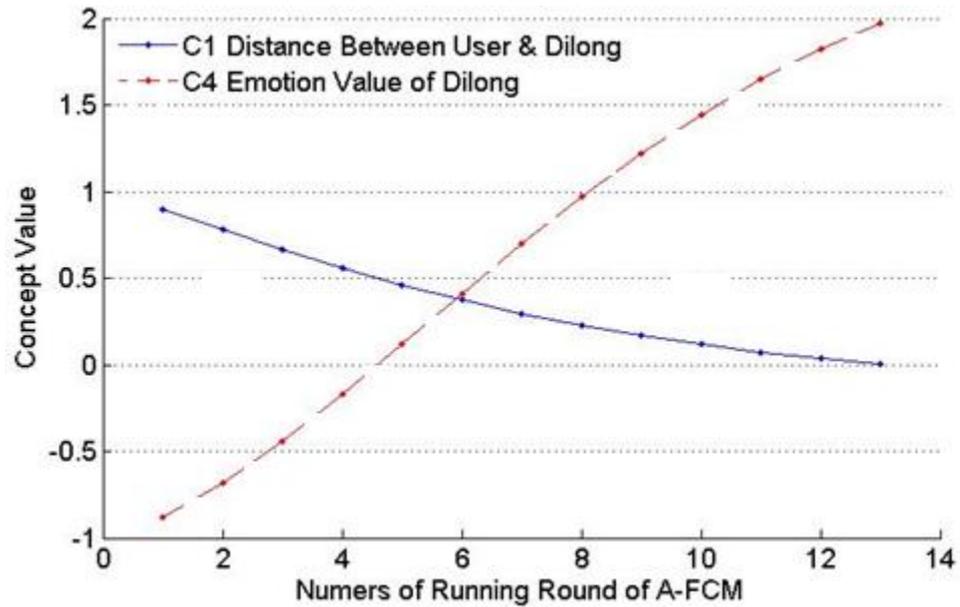

Figure 3.13: Simulation1—Giantdino faster than Dilong

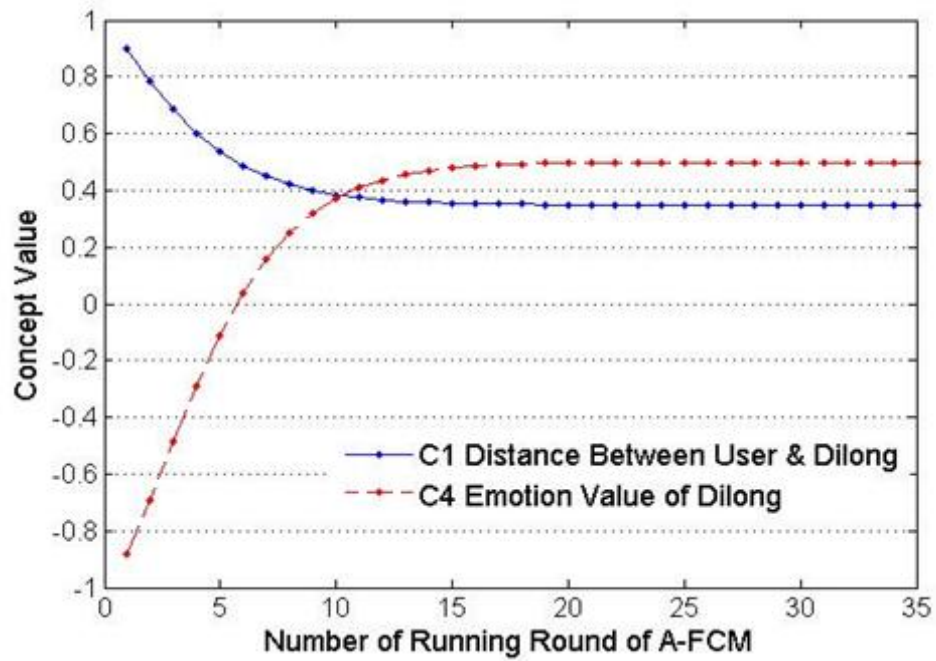

Figure 3.14: Simulation 2—Giantdino slower than Dilong

From the simulation results, we find the relationship between distance and Dilong's fearful feeling, and these emotion values can control the Dilong's emotional expression in virtual world. We can also alter the value of $v_{giantdino}$ in real time and do the computation similarly. In this way, we can adjust Dilong's fearful emotions in the virtual world with dynamic changing variables.

From this case study, it can be seen that this quantitative approach is easy to use, and owing to FCM's symbolic representation, it is very convenient to modify the FCM if there is any design mistake.

## 3.6.4 Discussion

The OCC model has many extensions and applications. Making a brief summary, there are five types of related work. These are 1) the rule-based emotion inference [9] which directly transforms the elicitation process of OCC into hard coded rules to control an agent, 2) the fuzzy logic approach [10] to map elements in OCC, 3) the dynamic decision networks [11] to depict the relationships in OCC, 4) the PAD (Pleasure-Arousal-Dominance) emotional state [12] to measure emotions, and 5) the logical formulization [13] of OCC. Except the first work, the rest of the research projects all can do the quantitative computation, and they used different ways to represent OCC model. However, these models do not have a straightforward representation so that educational designers cannot easily understand the reasoning process. If the design has some mistakes, it is hard to quickly modify it. As the research on affective agent needs to involve cross-disciplinary experts from education and computer science, an easy understanding and simple use of the model is very important. Hence, we propose to use a FCM for representing OCC-based Affective Goal Net and realize the computation of intensity. Its straightforward formulation and simple computation can meet the design requirements of a teachable agent.

Another obvious advantage of the FCM is that the output is not a single value. Rather, the value of all the nodes that a FCM has can be computed at the same time and available for further usage. For an emotional system, many results need to be recorded for the elicitation of following emotions. For example, the anticipation of an event needs to be recorded because it will affect the anticipation next time. Therefore, besides the node value of emotion intensity used to guide the agent's emotional behavior, all other node values are also useful. FCM can get all these results together without additional computations.

At the same time, our approach still has limitations. As we have mentioned in the introduction of OCC model, OCC discussed three types of emotions – event-based emotions, agent-based emotions and object-based emotions. Among the three groups, event-based emotions are the most complicated, yet they are the most frequent to occur in a virtual learning environment. Since virtual worlds are commonly modeled as event-driven (implemented by event listeners), and a virtual agent's behavior is usually considered as a certain reaction to virtual world events, in this book we only consider the event-based emotions. However, the eliciting processes of agent-based and object-based emotions are quite similar to the event-based emotions, but the difficulty lies in the combination of multiple types of emotions at the same time. For example, the agent may at the same time have emotion A because of a current event, and emotion B on an object, and emotion C on another teachable agent. To elicit the emotions in this situation, one solution is that we use a function to filter out the strongest emotion, or we merge current emotions as a compound emotion type. All these analyses are left for the future interest of space.

# 3.7 Summary

In this chapter, we proposed a new type of TA called an Affective Teachable Agent. From the educational requirements, we abstracted two major capabilities a TA should have, teachability and affectivability. In order to make the whole TA system proactive when interacting with students, we proposed an ATA with a goal-oriented modeling approach. We built up a routine Goal Net structure to control the TA system as a whole, which can call a sequence of Sub Goal Nets to realize the specific tasks such as to learn from students, to practice what has been learnt from students, and to affectively interact with students. The routine Goal Net may select the proper Sub Goal Net according to student's learning progress and their actions in a virtual world. The Sub Goal Nets will run the corresponding actions to control the TA. For Teachability, the system needs to clearly express the request of student's help/teaching, and select the proper teaching panel for students to begin the teaching task. Each teaching panel can be represented as a Sub Goal Net too, which is designed based on a certain topic and will be further described in next chapter. For Affectivability, two parts are discussed. First is the qualitative elicitation of emotions, which has been modeled through the Affective Goal Net structure. By representing the OCC theory, a cognitive structure of emotions, the Affective Goal Net can generate twelve types of emotions according to the situation that the TA encounters. The second procedure is to quantatively compute the intensity of the emotions, because to describe an emotion requires information from two facets, what type of emotion it is and how strong it is. To do this computation, we adopt a fuzzy tool, the FCM to graphically represent the related emotional factors and compute the value based on the causal relationships among the factors, which may influence the emotion intensity of the agent. The generation of emotions relates to many aspects. Hence, it requires the flexibility for extension and modification. With the development of deeper understandings of emotions and its components, the model should be capable of combining all the new elements and changes. The structure of a FCM brings with it the potential for further extension and incorporation of new elements. As long as the new elements have a common node with the existing FCM, it can directly combine with the existing one, which provides an ideal opportunity for our further study. Based on the proposed ATA model, the next chapter will introduce the proposed authoring tool for designing educational games with TAs. The benefit of the proposed graphical modeling of an ATA will be highlighted during the game authoring process.

# Chapter

# 4

# ATA Game Authoring Approach & System Design

In Chapter 3 we introduced the proposed ATA model. Apart from facilitating student learning, another important goal of an ATA is to facilitate teacher's involvement in the game design process, since teachers are experts who can contribute to the design of teaching content and domain knowledge in educational games. The component for domain knowledge input and update is an indispensable part of the whole system. To this end, we develop an authoring tool for the TA system to encapsulate the technical details and provide educational experts an easy way to do the TA game design. When teachers contribute to the game design, they can use explicit presentations to organize the domain knowledge.  If they need to update the learning contents regularly in the long term maintenance of the game system, they can perform as another type of teacher for the TA, and the "teaching" process will encapsulate the game programming details given by the educators, and ensure them to focus on the curriculum related features. The evaluation will be specified in Section 5.4.1.

Meanwhile, we propose a new feature for the TA to support student's self-selection learning path, which is to provide students with choices to control their learning progress and to work at their own pace. In this way, the game designers bring students learning materials within the game scenarios and the students can perform as co-constructors to decide the game flow and take the responsibility of learning.

In this chapter, we also specify how to design the system architecture to realize the proposed ATA model and the authoring approach. As we utilize Goal Net modeling approach to design an ATA, the system design framework of Goal Net, Multi-Agent Development Environment (MADE), is the best option to support the running of the ATA system. We will introduce the MADE framework and depict how to deploy the agent model based on the architecture.

# 4.1 ATA Game Authoring

Concerning the educational game design, two main requirements should be fulfilled. First, game players may come from different schools and be in different grades. Thus the game system should have the flexibility to customize the learning content for different students. Second, for game designers, the design of learning content is not a process in which we can puts things right once and for all. Instead, the learning content should be able to be easily updated regularly. The game developers need to coherently work with educational experts throughout the game design and the long-term maintenance. Therefore, an authoring tool is indispensable to smoothly connect the design and development process. In this section, we will propose an easy-to-use authoring tool for the ATA design.

## 4.1.1 Why Goal Oriented Authoring Approach

To design an educational game, the first task is to identify the learning objectives, and answer the question "what to learn?" then according to the identified learning objectives, we design the learning interventions and the concrete game scenarios. These two processes must be tightly integrated with each other so that the game scenarios can precisely serve the learning objective, and the learning objectives can be fully covered by game scenarios. We aim to make sure that the students can have a clear understanding of their learning purpose so as to avoid the unnecessary distraction. The playing experience can really help them to obtain a deep understanding and a long lasting retention of the related knowledge. How can we achieve this objective? In our opinion, it can best be realized through the setting of learning goals for game scenarios. Learning goals are used for specifying the knowledge or the understandings that students are able to obtain by the end of a learning process[107]. A clear setting of learning goals contributes to a well-defined game structure since it works as a map that tells developers what contents should be involved, and indicates to students what tasks they need to work on.

The setting of learning goals needs help from both computer developers and pedagogical researchers. So we need an easy-to-use tool to encapsulate the programming features and hide them from the educational experts to allow them to concentrate on the setting of learning goals or other pedagogical features. In the field of intelligent tutoring systems, authoring tools are widely used. According to [108], the authoring tools can be divided into seven categories, which include the tools used for curriculum sequencing and planning [109], tutoring strategies [110], device simulation and equipment training [111], domain expert systems [112], multiple knowledge types [113], special purpose [114] and intelligent hypermedia[115]. In order to author learning goals, our attention narrows down to the use of curriculum planning and work-flow management. To achieve this, the system should support two main tasks, which are supporting authors to identify domain knowledge and designing pedagogical strategies with practical scenarios [108].

Several studies [113, 116, 117] have focused on this perspective, but they have two common drawbacks. One is applying very constrained settings for data entry, and the other is asking a long list of questions to create the content such as sequentially asking what the learning focus are for all different sections. To overcome the limitations, we propose to use an interactive authoring tool based on Goal Net Designer for setting learning goals in the TA system.

## 4.1.2 Game Authoring and Learning Content Design

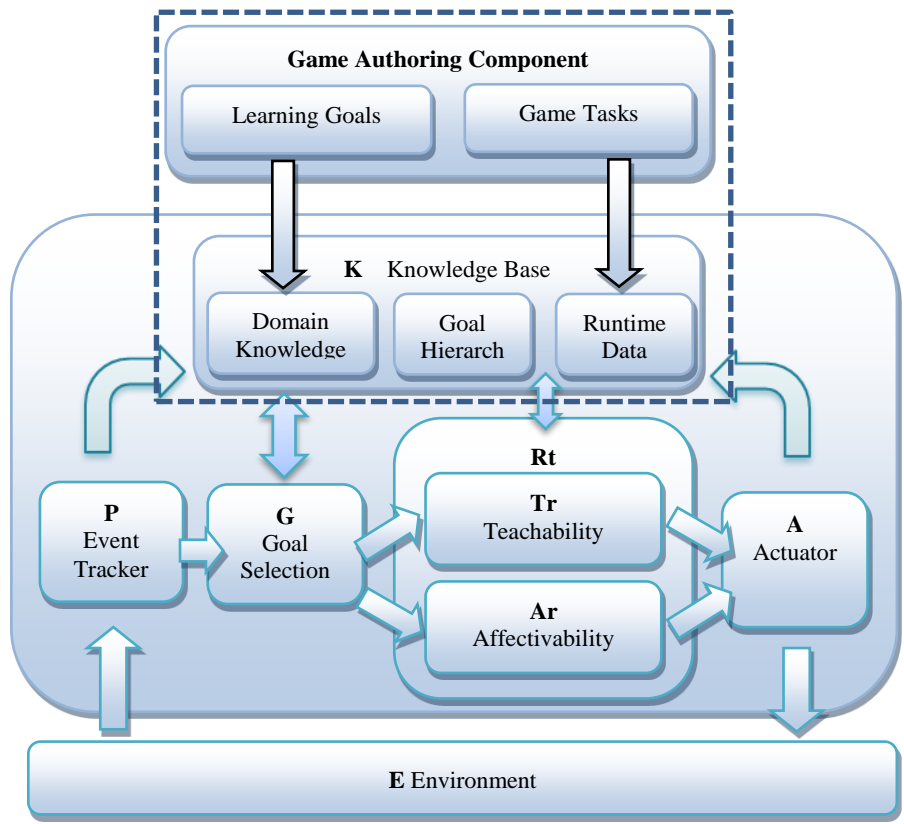

Figure 4.1: The Game Authoring Component for ATA Model

The component for domain knowledge input and update is an indispensable part of the whole system. The proposed game authoring tool is designed to support the learning content entering and updating. The relationship between the authoring component and the existing ATA model is shown as Figure 4.1. The game authoring process is organized by different learning goals. According to the requirement of each learning goal, the educator will design corresponding game tasks. After the design by educator, game programmers will based on the input to modify the system setting or build new game tasks. The authoring process involves three parties: the educational designers (teachers) who abstract the domain knowledge into learning goals and author the detailed learning contents, the game programmers who implement the game according to the educational designer's input; and the students who use the system to learn knowledge. The game authoring and development is an iterative process during the whole life cycle of the educational game. The details are summarized as Table 4.1.

Table 4.1: The Summary of the Proposed System with Different Parties [106]

| Game | Educational Designer (Teachers) | Program Developer | Student |
|------|------|------|------|
| System | Authoring learning content 1. Identify learning goals (first by topic, then by difficulty level) 2. Define task list in transitions for detail | Do the game programming based on the authoring results 1. Build up game flow based on learning goals 2. Realize game based on the task list in transitions | Teach ATA through: 1. Tutoring ATA via Concept Map, demonstrative experiment, etc. |
| Goal Net Designer | Generate learning goal structure to depict learning materials | Implement the transition functions into game system | Execute the game according to the Goal Net and trigger game scene |

Authoring learning goals with Goal Net Designer includes two aspects of work. The first is identifying learning goals. Two aspects are considered to capture the learning goal from domain knowledge, which include the learning level and the learning topic. Learning level refers to the difficulty of the learning content which divides learning materials into different teaching sequences, such as concepts, examples, transferred situations, etc. Learning topic refers to the knowledge correlation which divides learning materials into subjects and themes. Each learning level or learning topic is a learning goal which serves the overall goal, or the learning purpose of the whole game. All the learning goals are represented as nodes in Goal Net. If a learning goal has sub-components, it is a composite goal with sub-goals. Otherwise, it is modeled as an atomic goal. In a Goal Net structure, all the lower level goals serve their higher level goals.

Once learning goals are identified, we need to design the game scenarios to achieve these goals. The detailed scenario implementation highly depends on the game engine program. With the help of Goal Net Designer, educational experts can set the scenario plots and manage the game progress by designating the actions within the Goal Net transitions (rectangles shown in Figure 4.2). The transitions of Goal Net are used for depicting the relationships between goals. Each transition is associated with a task list which defines the possible tasks that the system needs to perform in order to transit from the input goal to the output goal. The game author can set the tasks or actions within transitions to design the game scenarios.

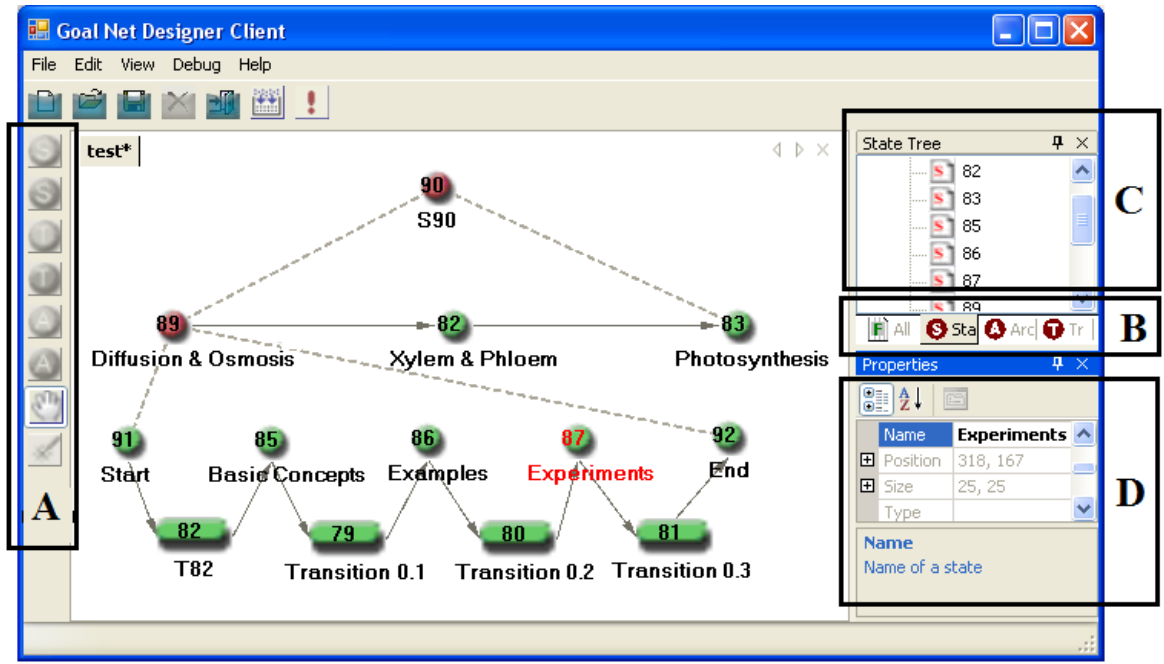

Figure 4.2: The Goal Net Designer Interface for Authoring Learning Content [106]

Figure 4.2 shows the interface of the authoring tool. In order to exemplify the functions, we use a topic in secondary school science course as an example, namely the transport in plants.

The authoring of learning content is structured based on the learning goals. Through specifying learning goals of the curriculum, teachers can depict their purpose and further ensure the educational ideas to be deliverable. All the learning goals can be drawn as round dots in Goal Net Designer which are defined as states in a Goal Net structure. The relationships between learning goals can be drawn as lines with arrow, which are defined as transitions of Goal Net. If the fulfillment of a learning goal needs several tasks, we can add a task list to the subsequent transition of the current goal state. The transitions with task list are presented as rounded corner rectangles.

The buttons for drawing Goal Net are listed on the left panel (Figure 4.2A). Teachers can edit the learning contents through dragging nodes to the canvas. The dot with "S" is used to add states; the dot with "A" is for adding transition without task list; and the dot with "T" is for transition with task list. On the upper right panel all the used elements such as goal states, transitions, and tasks in transitions are listed in Figure 4.2C with a catalogue as Figure 4.2B. The corresponding properties of the activated element will be displayed in Figure 4.2D. Users can edit the details through fill in the property blanks.

When teachers do the authoring, they can follow and repeat the two steps below.

1. Identify the skeleton of learning content through setting learning goals: Before the game design, the first thing we need to do is to identify the underpinning educational purpose – what students are going to learn. Thus, the teacher needs to capture all the knowledge key points from the corresponding curriculum and represent the key points through drawing the hierarchy of learning goals. To be practical, teachers can follow two sub-steps.

   a. Specify learning topic: Learning topic refers to the knowledge differences which divides learning materials into subjects and themes. Each learning topic is a learning goal which serves the overall goal, to teach the knowledge on "transport in plants". The teacher can add the learning goals through dragging the round dot with "S" to the canvas, and edit the description through the property window in Figure 4.2D. Then the teacher can add the transitions between learning goals through dragging the round dot with "T" to the canvas. At this time the system will pop up an editing window for entering the start and end dots of the transition. The whole structure of learning goals can be completed by repeating this process. For the dots with number 89, 82, and 83 in Figure 4.2, the example authoring has three learning topics:

- Xylem and Phloem of Root, Stem and Leaf: the cross section and functionalities of xylem and phloem inside the plant.

- Osmosis and Diffusion: different movement patterns of the water and mineral molecules.

- Photosynthesis: the way in which the energy and oxygen are generated inside the leaf with water, light and carbon-dioxide.

b. Depict each learning goal at different difficulty levels: Difficulty level refers to the difficulty of a learning content which divides learning materials into different teaching sequences, such as concepts, examples, transferred situations, etc. For each learning topic, designers can edit its sub-goals according to the learning levels. Taking the sub-goal "Diffusion and Osmosis" ( the node 89 in Figure 4.2) as an example, the author set three different learning levels:

- Learning basic concepts

- Obtaining concrete understandings

- Doing experiments

These difficulty levels are shown as the nodes 85, 86, and 87.Once the setting from both learning topic and difficulty level has been finished, the system will provide the option for adding "details", which comes to the next step for game scenario design as the following.

2. Fill up the detailed domain knowledge for each identified learning goals: This process is achieved within the Goal Net transitions. Teachers can set the scenario plots and manage the game progress by designating the actions in transitions through double clicking the transition rectangle, and there will be a small editing window for user inputs. Currently, we support two means of detailed input.

a. Upload Concept Maps related to the key knowledge: Concept map is a popular approach to represent knowledge. The system supports teachers to upload the Concept Maps, and the agent system will convert the Concept Map to a blank-filling problem for students to use.

b. Design the detailed experimental procedure through editing a task list: in our system the authoring of detailed design is organized as the editing of task list within the transition rectangle. A task-list editing window can pop up if the user double clicks any round corners rectangle. Taking the sub-goal "87 Experiments" in Figure 4.2 as an example, the setting details of the "experiments about diffusion and osmosis" are shown in Figure 4.3 (a).

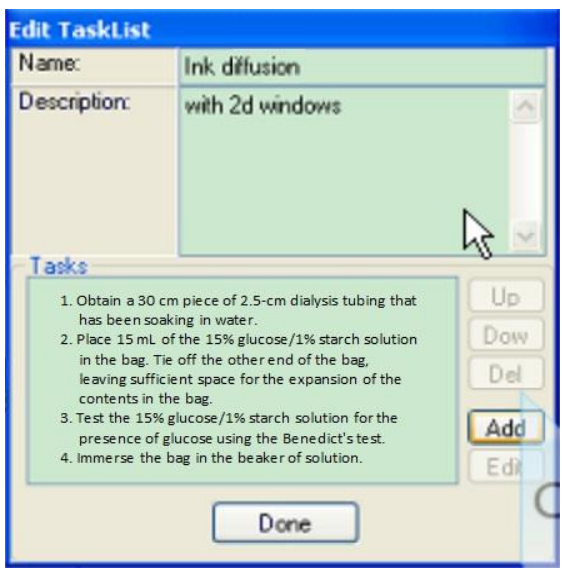
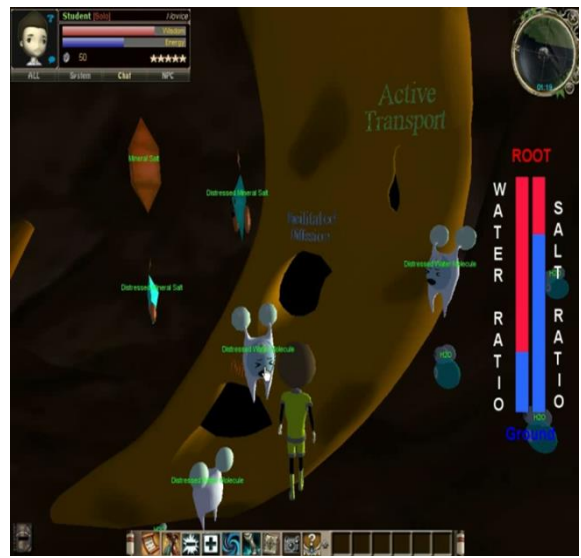

(a)                                        (b)

Figure 4.3: (a) Task list editing window; (b) Meeting ATA in game scene

All these tasks in transition will be organized in a task library on the game engine side. The game developers will create the program based on the task lists to realize the game scenario in 3D virtual world, provided that the task implementations do not exist before.

The build-in tasks will be recorded as functions in the task library, and prepared for reuse in the future. As illustrated in Figure 4.4, the standby functions are stored in the Dynamic-Link Librarys (DLLs). Once a TA needs to do the related actions, the system will call the functions from DLLs.

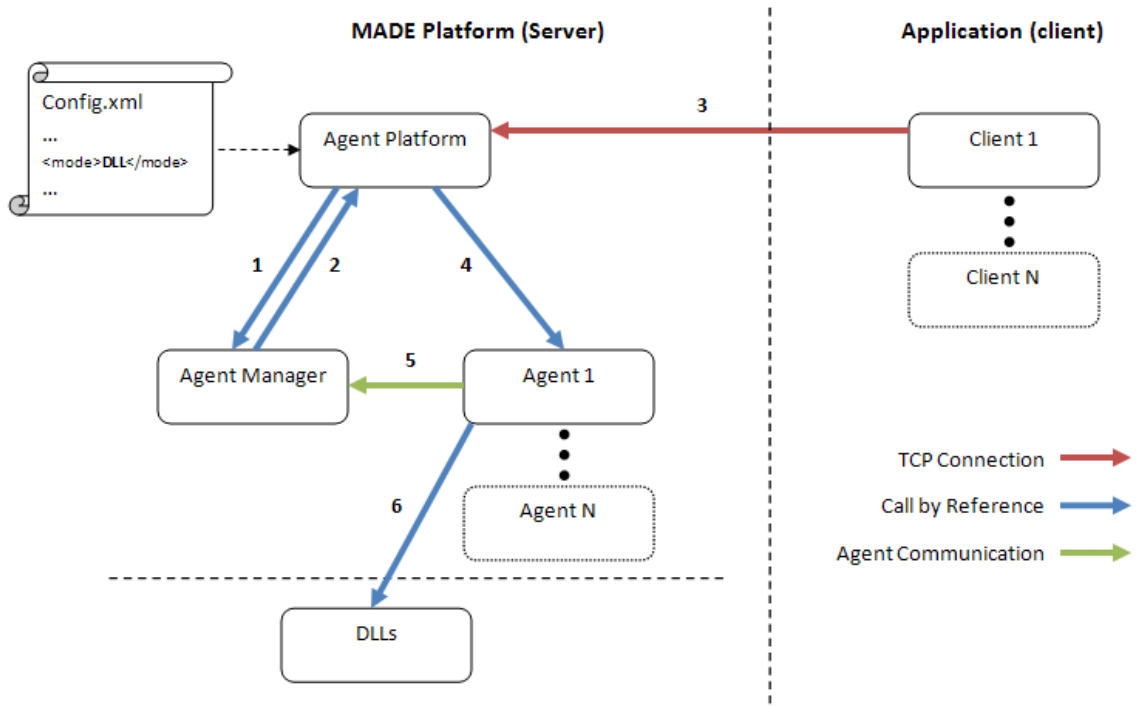

Figure 4.4: MADE Storing Agent Functions in DLL

The detailed scenario implementation still highly depends on the game engine program, but with the help of the authoring tool, educational experts can deliver their ideas through repeating the above procedures and write the learning content to the system designer and game developers.

With the input from teachers, the ATA model can select different learning contents for different students at different learning stages. The selection is based on three aspects. First, the system omits the content which has been learnt before. Second, the system sets high priority for the content that is related to the mistaken tasks. Third, the system gives hints for connecting learnt knowledge point with current learning content. The Goal Net with root goal "To Learn from User" will use the build-in functions to provide students with an interactive platform to teach the agent.

## 4.1.3 Learning Content Updating

Owing to the characteristic of Teachability, a teachable agent can potentially be used in different perspectives when it faces different "teachers": when a **student** teaches the agent, the agent performs as a "naive **learning companion**, but when a **teacher** teaches the agent during the learning content updating process, the agent works as an **authoring tool** which can bring teachers a natural way to convey domain knowledge to the agent's knowledge base. Therefore, we propose to extend existing ATA authoring capability to be used by teachers in their daily maintenance of the educational game.

We have introduced three life cycles of the ATA in Chapter 3, which are Learning Cycle, Acting Cycle, and Affective Cycle. The extended features in this chapter only affect the learning cycle. The learning cycle represents agent's learning process, which includes two components, Perceiving and Storing Knowledge. Current ATA not only perceives the knowledge taught by students, but also perceives the authored content from the Goal Net Designer interface from teachers. Teachable agents set all the perceived knowledge from Goal Net Designer as default correct knowledge. In this way, we simply provide teachers an easy way to update the learning content regularly.

# 4.2 Self-Selection of Learning Path

We have mentioned that this system tries to provide students with the capability of controlling their learning progress. We aim to encourage students to get rid of the habit of relying on the decisions made by external authorities such as teachers and parents. The first step is to allow students to control their learning pace and take the responsibility of playing this educational game. We believe that asking students to select learning goals and game flows can relieve them from the feeling of anxiety and uncertainty, and further enhance their motivation for learning. Therefore, our authoring system provides students the capability to select learning goals from the "learning goal pool" which has been set up by game authors. When students begin to play the game, the system will allow students to select one learning goal or arrange all the learning goals by dragging the Goal Net nodes to the goal setting interface. Through assembling these learning goals, students can decide and monitor their learning progress, as the Goal Net will run based on the rescheduled flow.

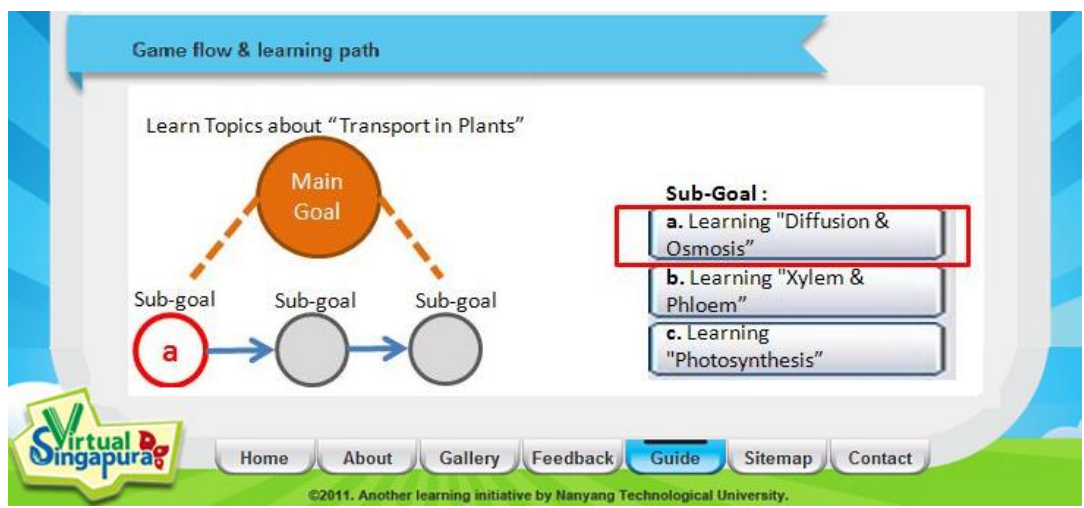

Figure 4.5: Sample of a Student's Learning Goal Setting Interface [118]

In Figure 4.5, the available learning goals sorted by topics are listed on the right side, and students can drag their preferred goals (the rectangles) to the blanks (the circles). Similar to the process of designing learning goals by game designers, the assembling of learning goals by users also includes two steps, 1) choosing learning topic and 2) choosing learning level. Once the two step selections are complete, the system will teleport the user directly to the corresponding areas in the virtual learning environment to do the game tasks.

Take the subject of transport in plants as an example. There are three learning topics and if a student begins with setting the learning topic, then the system will generate a learning topic selection panel as Figure 4.5. The student can select their preferred learning sequence through dragging the related topic to the sub-goal blank. Next, for each learning topic, students can edit its sub-goals according to the difficulty levels. Taking the sub-goal "learning diffusion and osmosis" in Figure 4.5 as an example, it contains four different difficulty levels, which are learning basic concepts, obtaining concrete understandings, doing experiments, and teaching abstracted knowledge respectively (Figure 4.6). The difficulty levels are increased gradually. Once the setting from both learning topic and difficulty level has been done, the system will begin the game from the place where the student selected.

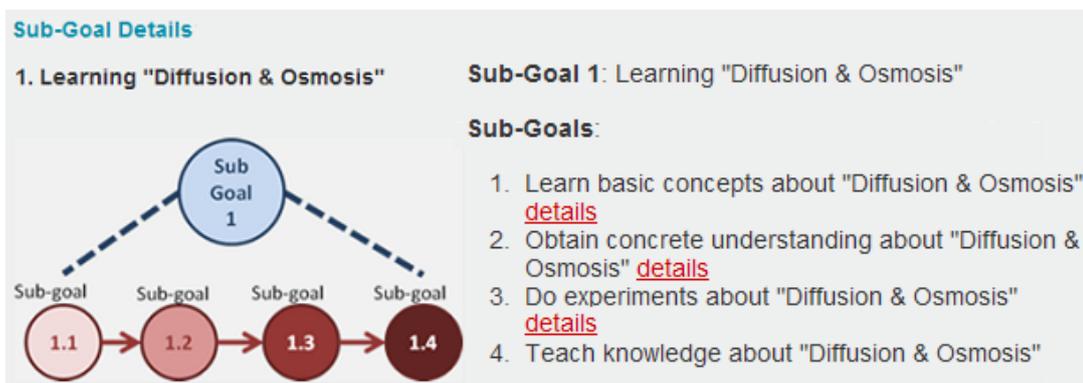

Figure 4.6: Goal Net for Learning Diffusion and Osmosis

Through the game authoring by educational experts and the detailed scenario programming by developers, the system has obtained all the learning goals and corresponding transition functions. Students can work on assembling these components into a goal hierarchy to customize their unique game flows. With this feature, the authoring process of the whole game system can be summarized as Table 4.2.

Table 4.2: The Summary of Authoring Processes for Both Designers and Users.

| Authoring | Educational Designer | Program Developer | Student |
|---|---|---|---|
| Processes | Identify learning goals (first by topic, then by difficulty level) & Define task list in transitions | Program task functions for game scenarios based on the task list in transitions | Customize game flow by assembling learning goals |
| System | Generate goal net structure to depict learning materials | Merge the transition functions into game system | Connect the user interface with goal net and trigger game |

The goal oriented structure will bring students a pre-mindset on what they need to learn. This pre-mindset will assist students to concentrate on learning materials and meanwhile reduce their uncertain feeling about the game flow.

## 4.3 ATA development Platform

The implementation of the ATA model is based on a platform named Multi-Agent Development Environment (MADE) [25], which is specifically designed for the Goal Net modeling approach to build up the system framework. The main components of the MADE framework include two parts. The *Goal Net Designer* helps an agent designer to design the TA's Goal Nets in a drag-and-drop manner through a graphical user interface (GUI). The *MADE Runtime* interprets the Goal Net design into the real agent operations by running through Goal Net from the starting state to the end state. These two parts shares a central database, *Goal Net Database*, which stores the Goal Net structures in a format that can be easily loaded and interpreted at runtime. Hence, this platform supports the goal oriented design for TA and helps shift the design to implementation.

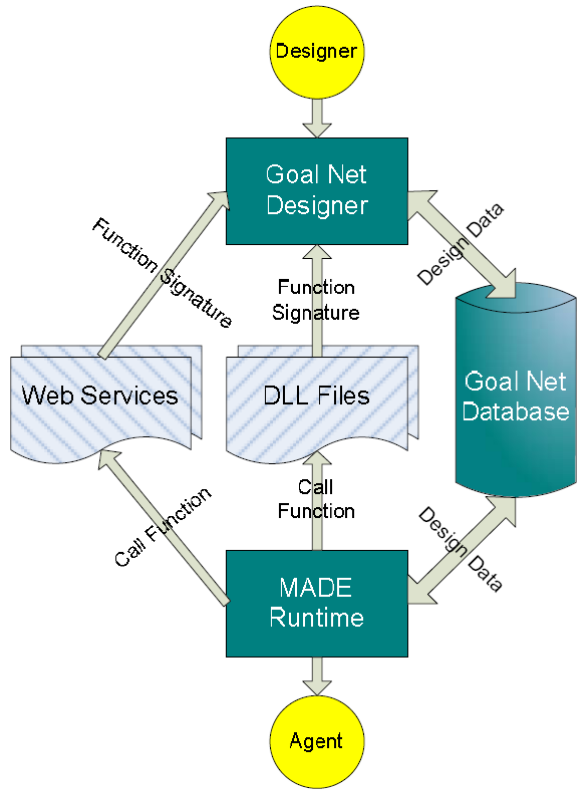

Figure 4.7: System Architecture of MADE [119]

For the Goal Net Designer, the details about the TA model designing and authoring will be specified in section 4.2. For the MADE Runtime, the data flow is shown in Figure 4.8.

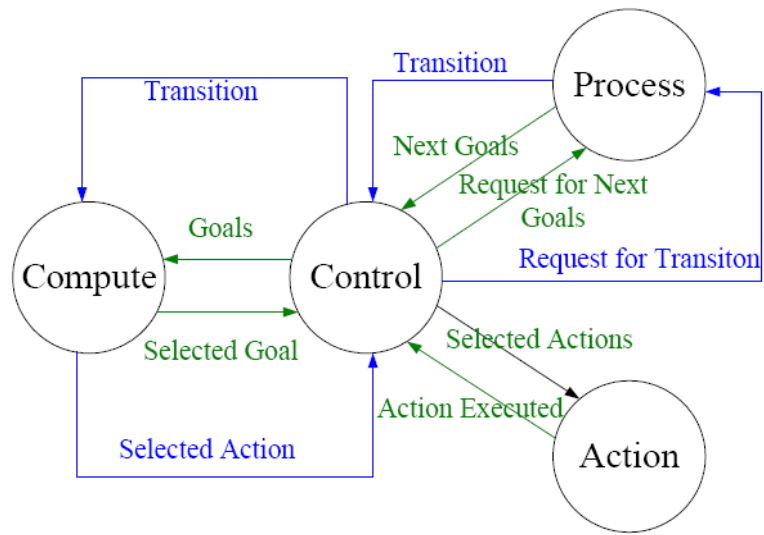

Figure 4.8: Data Flow in MADE [25]

- Control Unit requests the next goals from Process Unit (there are more than one next goal available when there is a choice transition).

- Process Unit informs Control Unit about all the possible next goals.

- Control Unit sends these goals to Compute Unit

- Compute Unit does a computation on these goals according to the goal selection algorithms, and sends the selected goal back to Control Unit.

- Based on the selected next goal, Control Unit ask Process Unit about the transition that can reach the selected goal.

- Process Unit sends Control Unit the corresponding transition.

- Control Unit sends the transition to Compute Unit for action selection.

- Compute Unit finds the best action to execute, and sends this action back to Control Unit.

- Control Unit sends the selected task to Action Unit for execution.

- Action Unit executes the given task by dynamically invoking the task class name, and sends an acknowledgement to the Control Unit once the task is done.

- Control Unit transits from its current goal, and asks Process Unit about the next goals.

With the cooperation of these components in MADE Runtime, the Goal Nets can be executed according to the design based on Goal Net Designer.

When operating the agent system, MADE runtime (setup on game server) will retrieve the TA model (Routine Goal Net of TA) from the database. In a real TA-enhanced educational game, it is quite possible to have more than one TA in the game server. Thus, for a certain client (a game program running on a personal computer used by a student), the game request will send to the agent platform in MADE Runtime (as Figure 4.9).

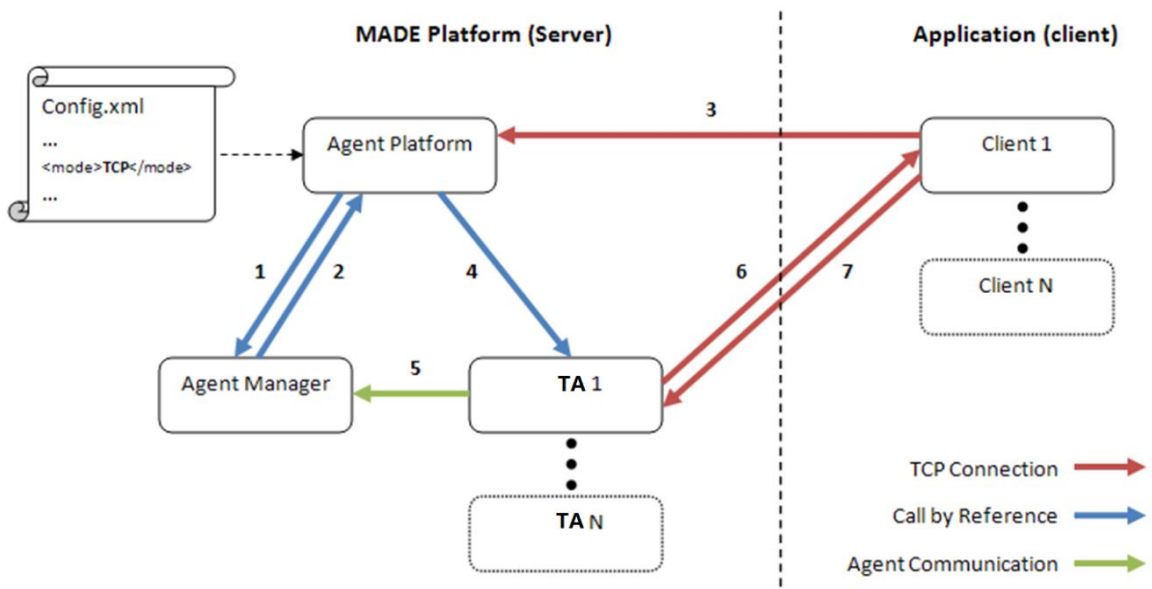

Figure 4.9: The Communication between MADE and User's PC

The *Agent platform* will trigger a TA, such as TA 1, to interact with students. Once a TA is triggered, the TA sends a message to the *Agent Manager*, and the Agent Manager will load the stored Goal Net data to the agent's knowledge base, which allows the agent to have a plan to pursue its goal. If a Sub Goal Net in the routine Goal Net is triggered, the program will jump to execute the Sub Goal Net and jump back to the routine when the Sub Goal is reached.

# 4.4 Summary

In this chapter, we introduced the game authoring, learning path selection in TA design, and the system architecture. The key feature of our system is using goal oriented modeling and authoring approaches to tightly integrate the domain knowledge with the game scenarios. As an indispensable component of the TA system, it is used for entering the domain knowledge which needs to be embedded into the system.

Learning goal as a continuous thread runs through the game authoring process, bringing both game designers and students a clear mindset for maximizing the educational profit. The idea of involving students to select their learning sequence enabled students to organize their own learning process. It will motivate them to take on the learning responsibilities and build up their self-esteem. The essence of learning is from the critical acceptance of knowledge. So we try to encourage students to critically analyze the learning materials and further establish their own knowledge structures and sense of understandings. This internalization process is what developmental theorists call self-authorship [120], or "the capacity to define one's beliefs, identity, and social relations" [121]. The attempt of offering students capability to arrange game flow to a certain degree is a first step to practice the self-authorship.

Chapter

# 5

# Case Study: Affective Teachable Agent in Virtual Singapura

In this chapter, we will introduce a virtual learning environment project - Virtual Singapura (VS), which is equipped with the agent-augmented features based on the proposed Affective Teachable Agent (ATA) modeling approach. We aim to build up a flexible virtual environment with authentic contexts and case-based problems for students to explore. After an overview of VS project in Section 5.1, we specify how to apply the proposed ATA model in VS. At the beginning of this section, how to build the ATA routine Goal Net in VS is introduced. It is an example of the overall ATA model mentioned in Section 3.4; Section 5.2.1 instantiates the model in Section 3.5 on how to realize Teachability in VS, through two types of interactions – concept maps and virtual experiments. Section 5.2.2 instantiates the Affectivability modeling approach mentioned in Section 3.6. Section 5.2.3 further explains the content of Section 4.1 on how to author the domain knowledge and game scenarios in VS system, and Section 5.2.4 elaborates Section 4.2 on learning path customization. Formative and summative assessment results of the ATA system are also provided in Section 5.3 and 5.4, showing that this type of educational agent modeling approach is not only promising but also practical.

## 5.1 Virtual Singapura Project

The Virtual Singapura (VS) project aims to build a 3D virtual learning environment which provides secondary school students in Singapore with a culturally familiar environment to learn science lessons (especially the knowledge about the transport in living things). Using the ATA modeling approach, two types of teachable agents "Little Water Molecules" and "Little Mineral Salt Molecules" were developed in the VS project. Rather than using illustrated pictures to depict how water molecules are transported from the roots to the leaves of a plant, VS project brings students on an exciting journey together with water molecules to explore the inside of a banana tree.

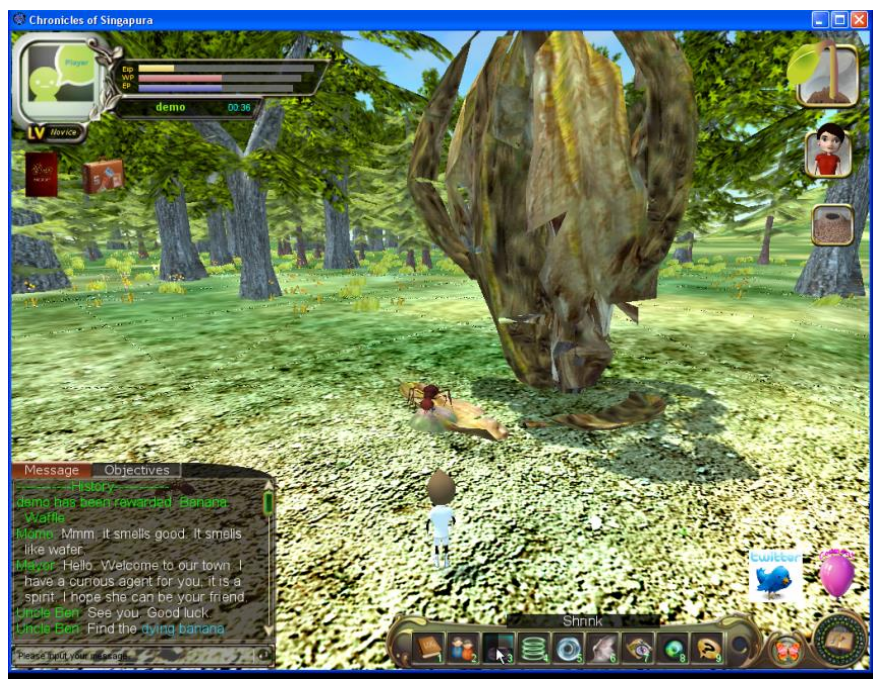

Figure 5.1: The Game Interface of VS Project

The game scene is set on Blakang Mati Island. For some mysterious reasons, the banana trees of Uncle Ben are getting sick (as Figure 5.1). Students need to get into the banana trees and find out what makes the banana tree unhealthy. While playing the game, every student is assigned a computer, and they can control the game and explore the virtual environment using the mouse and keyboard, as shown in Figure 5.2.

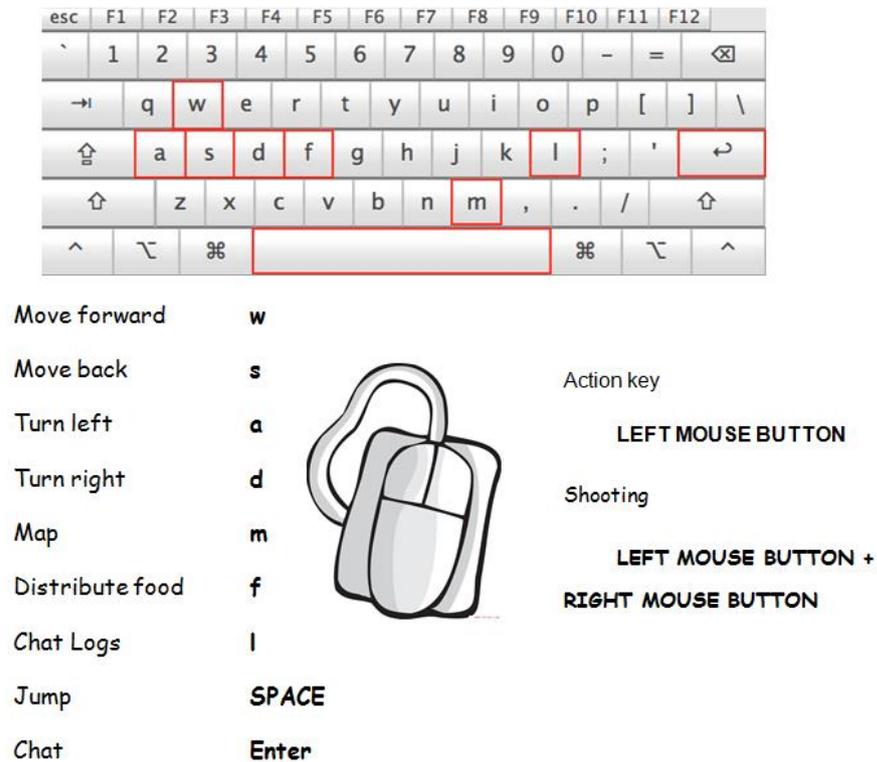

Figure 5.2: Game control in VS Project

At the beginning of this adventure, students need to "shrink" themselves to the size of an ant, and then enter into the underground through an ant hole (as Figure 5.3 & 5.4). Under the ground, the student will meet several "Little Water Molecules" and "Little Mineral Salt Molecules" (as Fig 5.5). These molecules are our affective teachable agents.

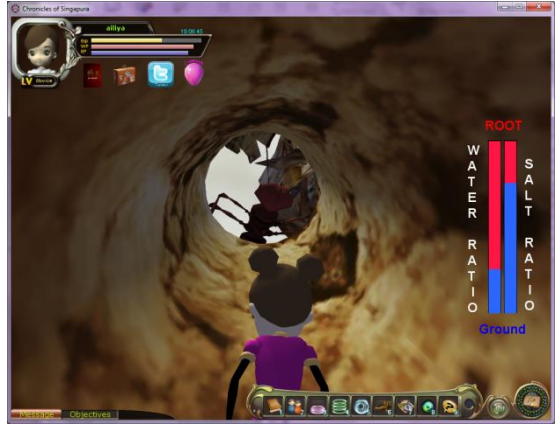

Figure 5.3: Ant Hole View 1

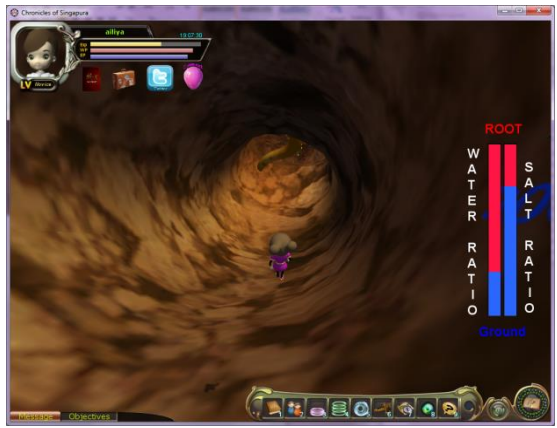

Figure 5.4: Ant Hole View 2

As shown in Figure 5.5, water molecules are begging student's help since they cannot enter the root of the banana tree. If the student wants to help the water molecule, he/she needs to teach the molecules through a teaching panel.

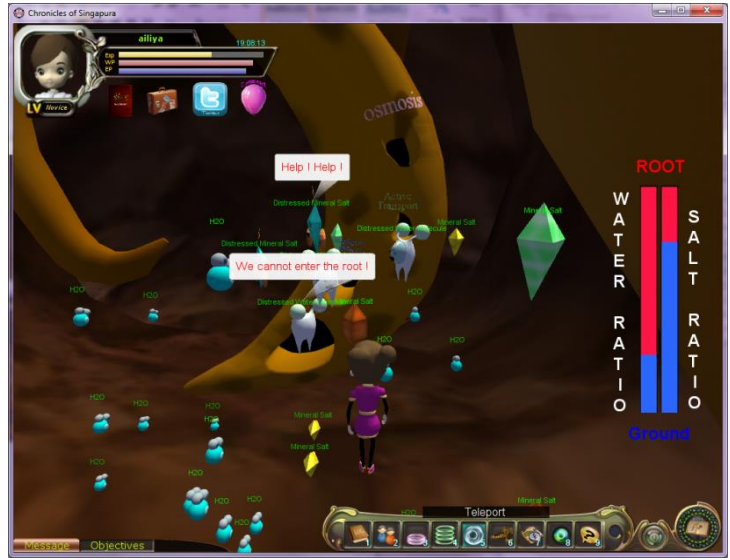

Figure 5.5: Meeting Water/Mineral Salt Molecules

Student can teach the water molecule or mineral salt molecule with a concept map or an experiment (as Figure 5.6).

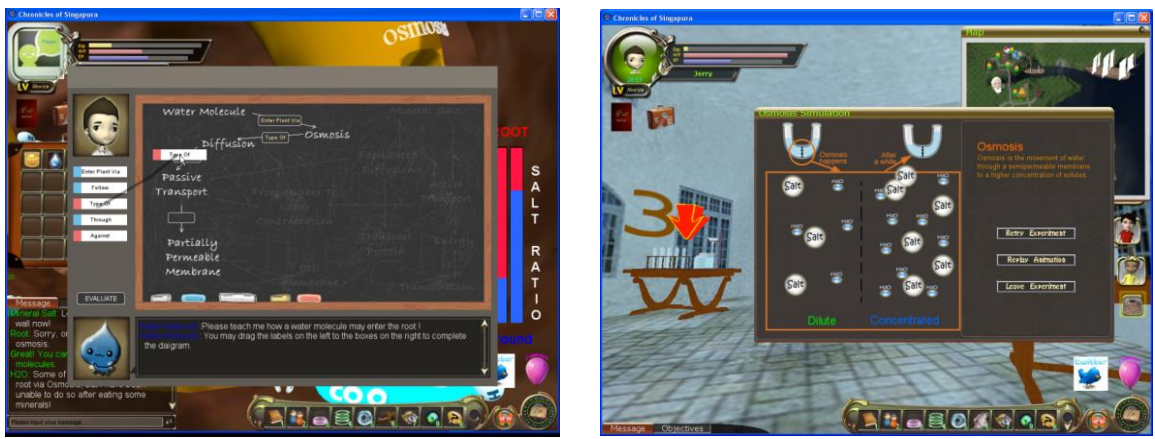

Figure 5.6:  Two Teaching Panels – Concept Map & Experiments

The related knowledge points are shown below:

**Concept: Osmosis to transport Water**

**Osmosis** is the net movement of water molecules through a partially permeable membrane, form a region where there is a higher concentration of water molecules to a region where there is a lower concentration of water molecules.

**Concept: 3 ways to transport Mineral Salts**

**Transport 1: Diffusion**
Diffusion is the movement of particles (atoms or molecules) from a region of higher concentration to a region of lower concentration.
No need to consume energy.
When there is a higher concentration of mineral salts in the soil than in the root hair cells, the mineral salts enter the root hares by diffusion.
No transport proteins

**Transport 2: Facilitated Diffusion**
Facilitated Diffusion is a special kind of diffusion which involves transport proteins.
No need to consume energy.
There are transport proteins in cell membrane as carriers and facilitator to take mineral salts into the root.

**Transport 3: Active transport**
Active transport happens when a cell uses energy to transport something against a concentration gradient.
Need to consume energy from the respiration of the plant.

According to the student's teaching, water molecules will react correspondingly and practice in the virtual world according to the knowledge taught by students. After that, water molecules will show students the outcome, such as entering the root smoothly or keeping outside of the root pitifully. The prototype of the VS project has been deployed based on the proposed ATA modeling approach. In the next section, we will illustrate how to apply an ATA in the virtual learning environment.

# 5.2 Applying Affective Teachable Agents in Virtual Singapora Project

This section explains how the proposed Teachable Agents are designed and implemented in a computer-based learning project, Virtual Singapora (VS). As we have mentioned, the water / mineral salt molecule wants to enter the root of a banana tree, and students are expected to help it by teaching it the knowledge of plant transport systems. By interacting with a student, the TA is motivated to achieve Teachability and Affectivability. The course of teaching forms a story line in the project setting, and it includes following scenes (with events):

- Game Scene1: Greeting

   o E1: The student meets the water molecule for the first time.

- Game Scene 2: Learning

   o E2: The student agrees or disagrees to teach the water molecule.

   o E3: The student starts teaching or becomes idle.

   o E4: The student inputs knowledge without or with syntax error.

- Game Scene 3: Practicing

   o E5: The water molecule has learnt from students, and attempts to be transported through the tree.

   o E6: The water molecule has been successfully transported through the root or has been rejected.

The whole functionality of ATA is controlled by the Main Routine Goal Net (introduced in Section 3.4), in which the Sub Goal selection or the Sub Goal Net selection is the most important task. We can design the goal selection mechanism of module G in Figure 3.5 (in Chapter 3) to decide when and which Goal to pursue.

We consider the goal selection mechanism GS as a function, mapping current events E to the set of goals G, i.e. $f_{GS}: \mathbf{E} \to G$. In VS, the mapping is described as follows,

- E1 → "To learn from user"

- E3 → "To practice"

- E1-E6 → "To be affective"

The Main Routine Goal Net systematically integrates various agent behaviors and proactively takes the initiatives of the next step.

## 5.2.1 Teachability Design

The first time the water molecule detects a new student is approaching, it starts begging for help as it cannot enter the root of a banana tree. If the student accepts the request, he needs to teach via a concept map or an illustrated experiment. There are two types of teaching panels in VS. The basic teaching panel is an interface for drawing a concept map, while the advanced teaching panel is for doing experiments in a laboratory.

Concept Map is an easy-to-use tool for students to represent the knowledge points and the relationships among the concept points. Many research projects have focused on how to generate concept maps and how to use them effectively [122]. We implemented a tool such that the student can drag and drop components and relations to finalize the concept map. The agent converts the concept map into pieces of logic rules, which consist of premises and conclusions. After that, the generated rules are stored as agent's knowledge for the agent to do practice later on.

The first version of our concept map-based teaching panel brings students a big concept map which includes many knowledge points. Students need to fill in the relationship between concepts by distinguishing provided tags between relevant and irrelevant information. The interface is as Figure 5.7. c such as Figure 5.8. We let students teach the simplified concept maps before chaining all the concepts together. This design may allow students to achieve the final teaching goal through a series of steps.

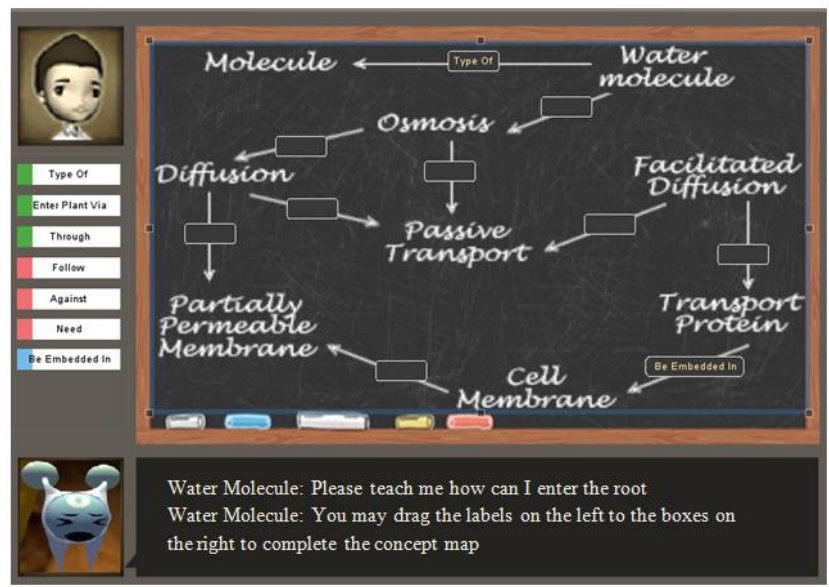

Figure 5.7: Interface of Concept Map

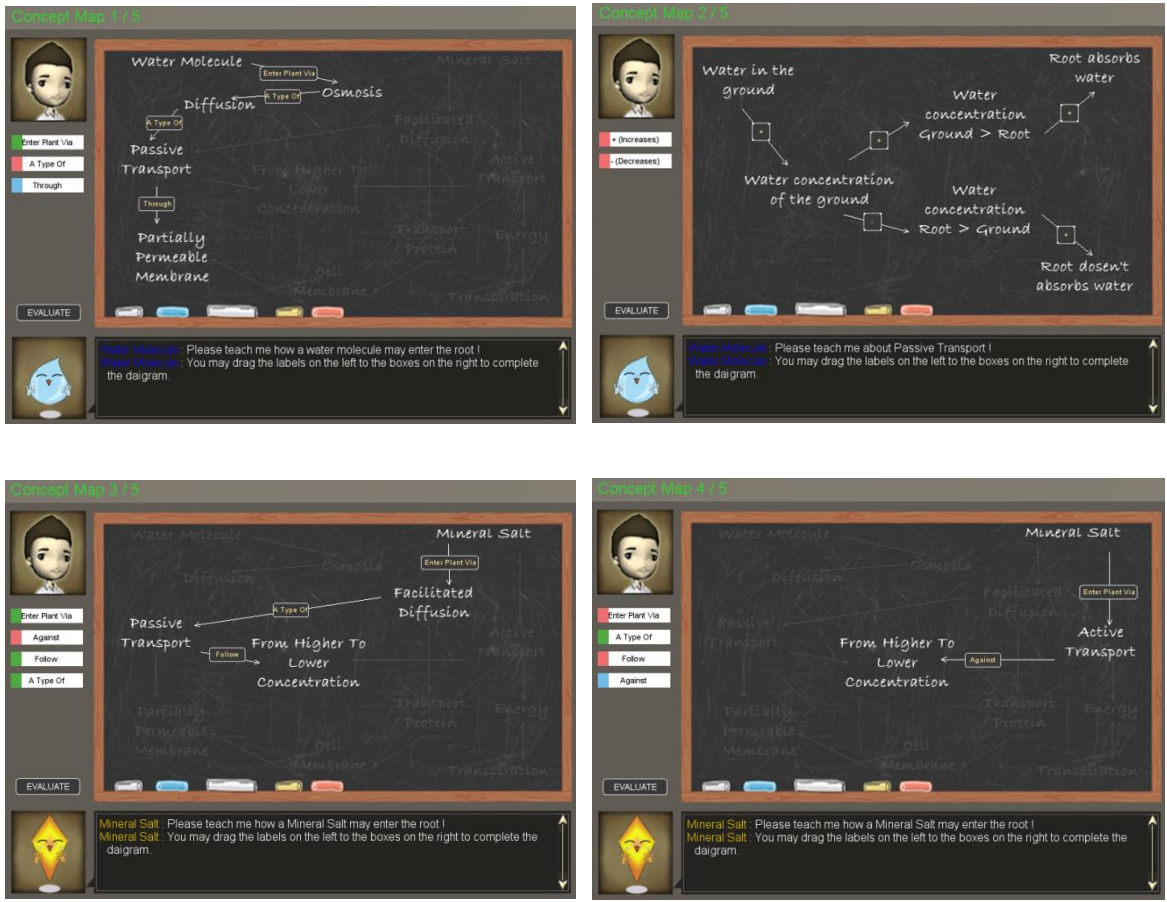

Figure 5.8: Four Simplified Concept Maps

Another way of teaching is by using an experiment to illustrate the related phenomenon. For the topic about diffusion and osmosis, five experiments were designed. In the north of the mainland of the virtual environment, we build up a house as a laboratory. The system teleports the student to laboratory (as Figure 5.9) to do experiments. The five experiments and the related knowledge points are listed as Table 5.1.

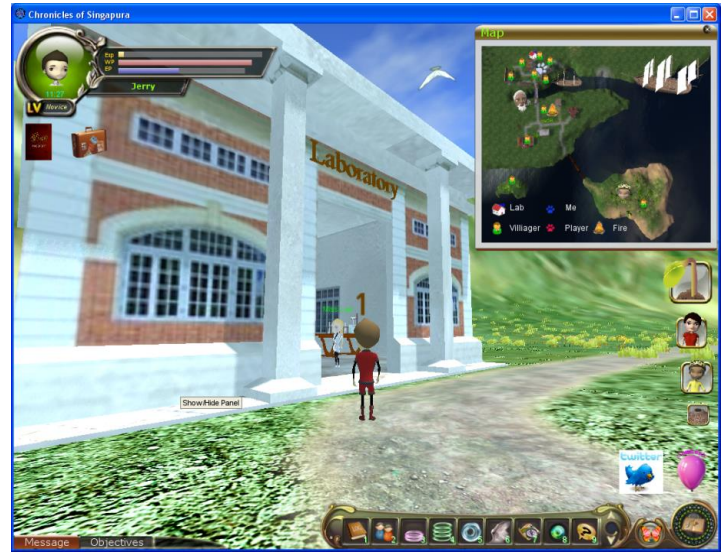

Figure 5.9: Laboratory in VS Project

With the above two types of domain module, concept map teaching panel and experiment teaching panel, the ATA can learn from students with the explicit knowledge representation platforms. The system runs the Goal Net with root goal "to learn from user" (Fig. 3.6) step by step.

Table 5.1: Five Experiments and Related Knowledge Points

| Knowledge Points | Environment Settings | Teachable Agent's Note Book |
|---|---|---|
| Viewing the diffusion effect through an ink solution.<br>• Hitting can speed up the movement of molecules and the diffusion phenomenon | | Invention Table showed: **Diffusion** is a **net** transport of molecules from a region of **higher concentration** to one of **lower concentration**. |
| Simulation of Osmosis effect<br>• What type of molecule can pass the semi permeable membrane<br>• Water moves to a higher concentration of solutes | | Invention Table showed: **Osmosis** is the **diffusion** of **water molecules** through a **partially permeable membrane**. |
| The solution concentration in Osmosis effect<br>• What is Cell Membrane<br>• Osmosis is the movement of water through a semi-permeable membrane to a higher concentration of solution. | 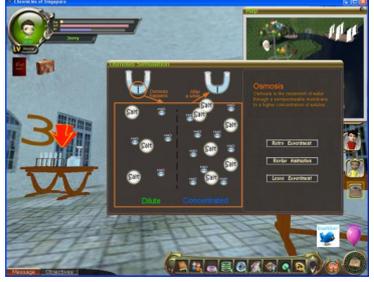 | 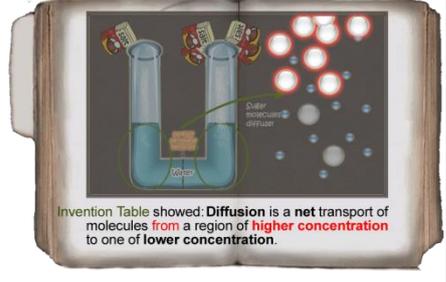 |
| Comparison between diffusion and osmosis effects<br>• The difference between diffusion phenomenon and osmosis phenomenon | 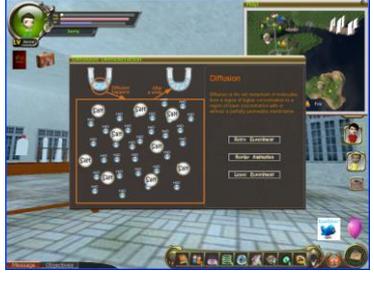 | 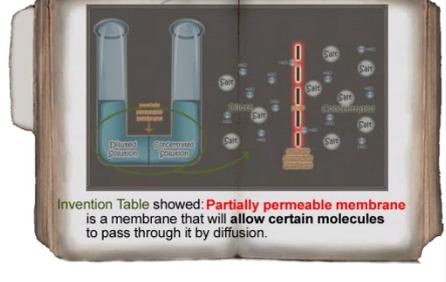 |
| Another form of experiment 3<br>• Egg inner membrane is a type of Cell Membrane | 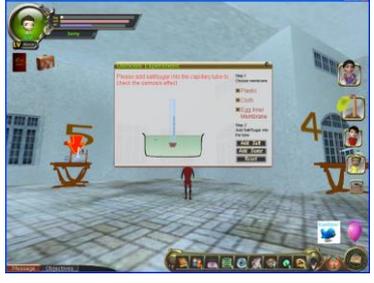 | 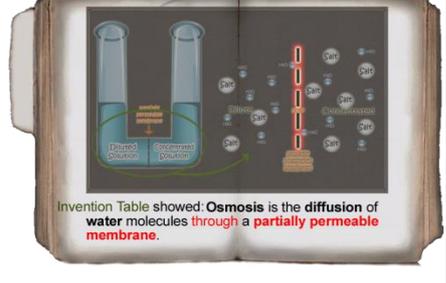 |

The most important task in this Goal Net is at the transition "Choose a teaching approach". This transition works on selecting different learning contents for different students at different stages of learning. We set the teaching panel with concept maps at difficulty level 1, and the teaching panel of experiments at difficulty level 2. The selection is based on three rules.

• Display teaching panels from easy to difficult

• Omit contents which have been learnt before.

• Set high priority for contents which are related to previous mistakes.

Apart from the teaching panel, the system also pops up hints for students when they are stuck in a step for more than three minutes. The hints are given by connecting current learning content with a previous learning scene which had involved the same knowledge point before. The interface is presented as a TA's learning note book. The system coded eight knowledge points, linking with all the game scenes. Once a student is stuck in a scene, the TA's note book will turn to the page, containing the related knowledge point, with a brief description and the screen capture of previous learning scenario. This design aims to encourage students to reflect on what they have learnt before with visualized clues.

Table 5.2: Summary of Knowledge Points in VS

| ID | Knowledge points |
|----|------------------|
| 1 | Diffusion (Passive transport) |
| 2 | Osmosis |
| 3 | Partially permeable membrane |
| 4 | Facilitated diffusion (Passive transport) |
| 5 | Active transport |
| 6 | Absorption of water |
| 7 | Absorption of mineral salts |

After student's teaching, the main routine will control the TA to pursue the goal of "to practice" (Fig. 3.7).

As we hope the TA can perform actively during the interaction with students, the request on practicing the learned knowledge will be sent automatically by the system when a TA has learned all the knowledge points in a teaching panel. In this game scenario, the performance goal of the water molecule is to enter the root. In order to achieve this, the agent is required to have a correct action plan, generated in the transition "Generate action plan". For entering the root, the water molecule needs to enter from the "Door of Osmosis", and the ground water ratio must be greater than that of the root. Failing to do so will result in prohibition from entering.

As mentioned above, the representation of knowledge learnt takes the form of a logic rule, i.e. $A \rightarrow B$ where A and B are predicates. For example, if the student conveys a correct concept map:

- The agent will learn

  $$\left( through\_osmosis \right) \wedge \left( water\_ratio(ground) > water\_ratio(root) \right) \rightarrow entering\_root$$

- By default, we build in some rules as agent's prior knowledge,

  $enter\_hole(osmosis) \rightarrow through\_osmosis$ , and

  $rain \rightarrow \left( water\_ratio(ground) > water\_ratio(root) \right)$.

- By rules of inference, the agent forms its action plan by entering the osmosis hole and waiting for rain.

- By carrying out the plan, the agent successfully enters the root (shown in Figure 5.10 (a)). Otherwise, the agent will be blocked (shown in Figure 5.10 (b)).

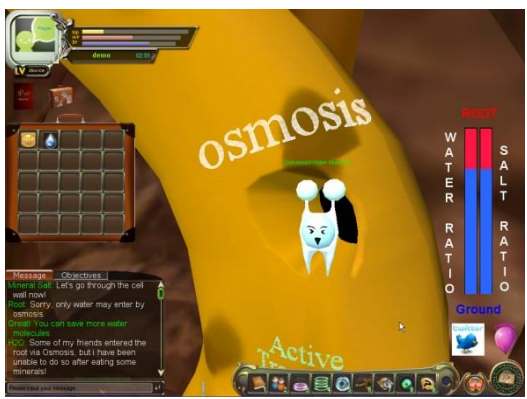
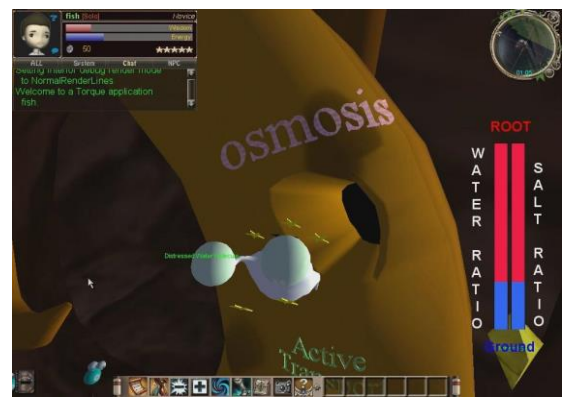

(a)                                            (b)

Figure 5.10: A Water Molecule (a) Successfully Entered the Root, and (b) Was Blocked

## 5.2.2 Affectivability Design

The scope of Affectivability of a TA is in line with the Goal Net with root goal "to be affective" (Fig. 3.8), which can elicit the TA's emotions according to current event. It is natural to allow emotion elicitation to be together with agent's learning and practicing. According to the Main Routine Goal Net of a TA, the thread for Affectivability is in parallel with the thread for Teachability.

The qualitative elicitation of emotions is achieved through the Affectivability Goal Net. As we have mentioned, with the input of current event ID and the emotion holder ID, the Goal Net may generate a type of emotion for TA. For the qualitative computation of emotions, the system uses the FCM to go through three phases.

**Phase 1:** Map the concepts into the concepts of FCM and draw them as nodes.

   1) Inputs: the impact of the causal events (system settings or students' actions)

   2) Outputs: agents' emotion tendency and action tendency

   3) Emotion intensity factors:

      a. the degree of **desirability** of an event,

      b. the degree of belief that an expected event will occur (**likelihood**),

      c. the degree to which resources were exhausted in hopes of perceiving or avoiding an expected event (**effort**), and

      d. the degree to which an anticipated event actually occurs (**realization**).

**Phase 2:** Find out the causal relations between these concepts and connect them with different weights according to the influential degrees in OCC (as Figure 5.11).

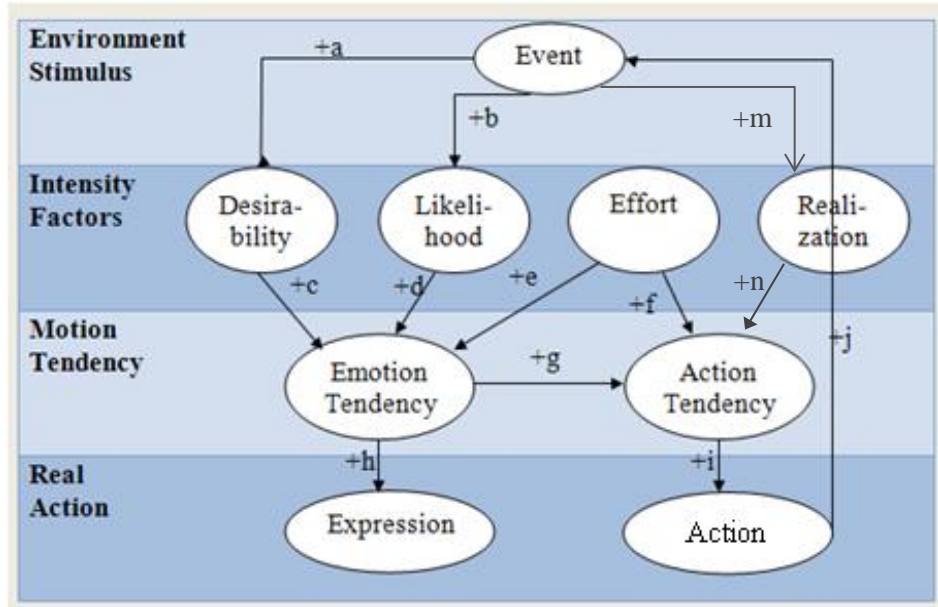

Figure 5.11: Causal Relationships among Emotional Factors

***Phase 3:*** Based on the concrete situation, set the activation functions of concepts and do the matrix computation $C_{t+1} = f(C_t \cdot W)$

Note that since an event has its counterpart, e.g. agree/disagree, the negation of an event denotes the one with negative effect. Taking E2 as an example, the event depicts that the student submitted his concept map which has no syntax error. The goal of the agent is to have good performance, and thus the event is desirable; the event endurer is the student, instead of the agent; and the event has only an immediate consequence, and thus it is prospect irrelevant. According to OCC, the emotion is happy-for, implying the agent is happy for the student's progress. The illustrated expressions of "Little Water Molecule" and "Mineral Salt Molecule" in VS project are shown in Figure 5.12.

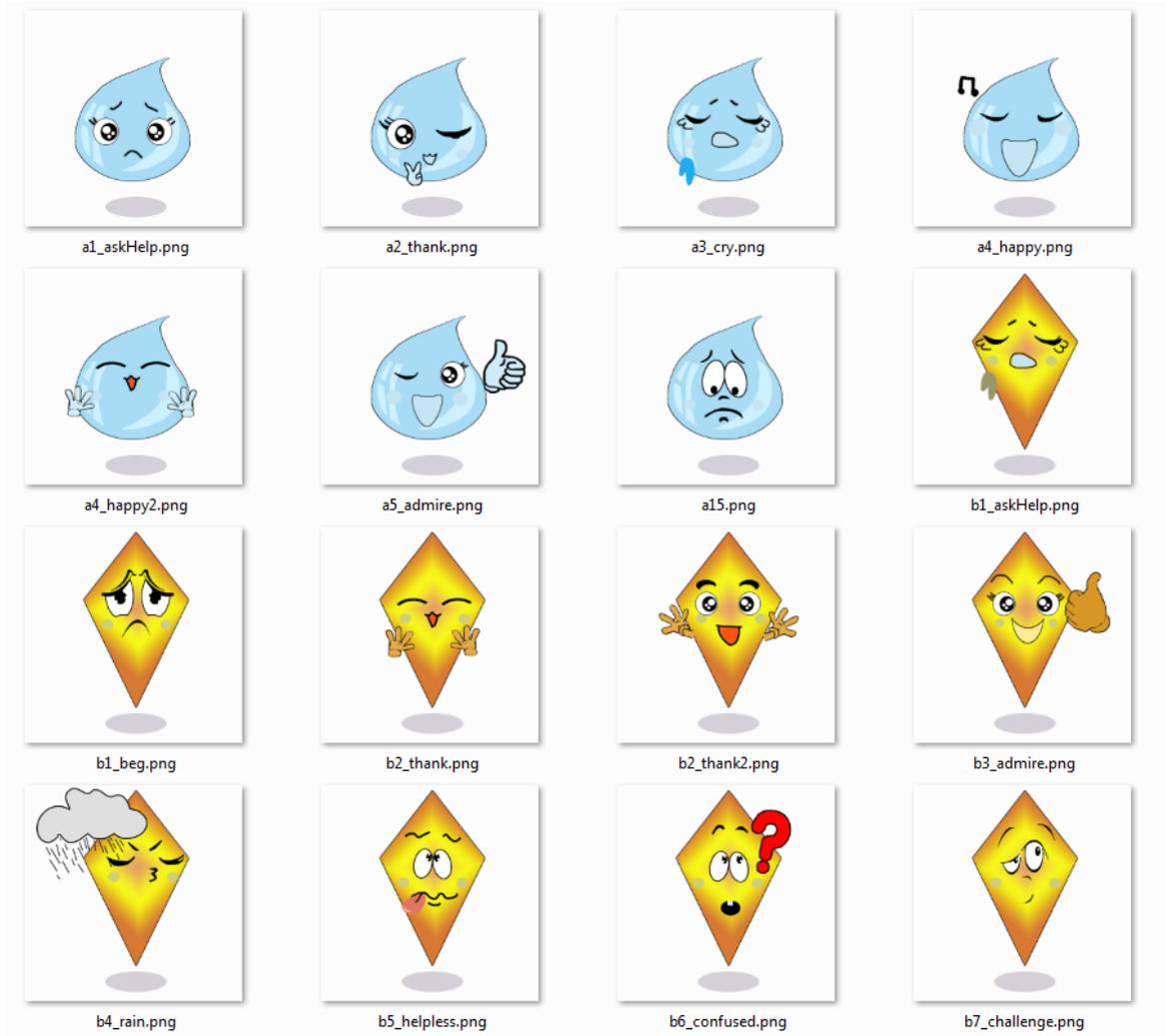

Figure 5.12: Emotion Expressions of Water Molecule and Mineral Salt Molecule

### 5.2.3  Authoring Domain Knowledge & Game Scenario

We developed an authoring tool for educational experts to be involved into the game design and maintenance. The relationship among teachers, game developers, and students are shown in Figure 5.13. The teachers can use the authoring tool to author the learning content. Based on the input from teachers, game developers realize the design in a 3D game engine. The teachers and game developers cooperate with each other to build up the educational game, providing students a TA-enhanced learning experience.

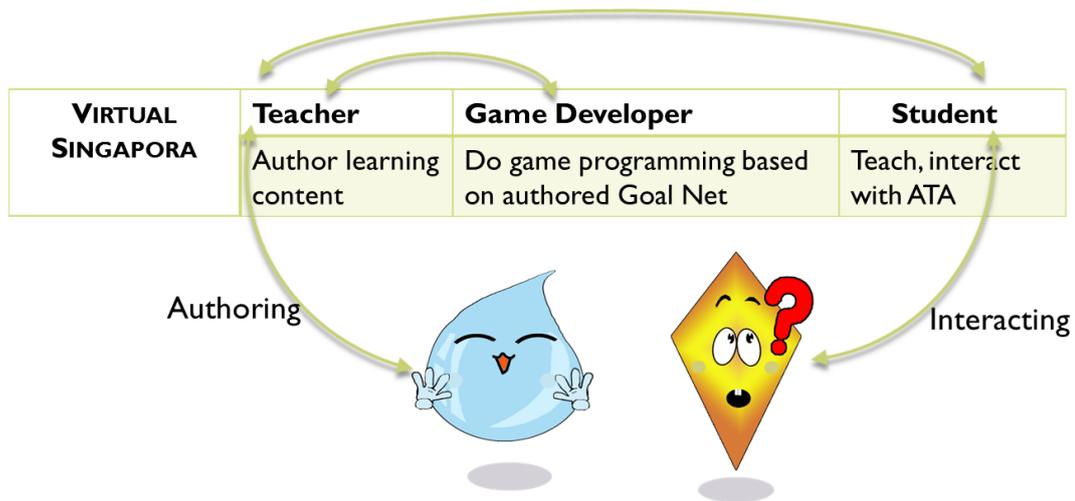

Figure 5.13: Cooperation among Teachers, Game Developers & Students

For the domain knowledge authoring, the teachers can input learning content by identifying learning goals through 1) listing various learning topics, and 2) setting different difficulty levels. This may help teachers to build up a broad structure of the game, such as Fig. 5.14 that shows the learning content authored in VS project. The teachers identified three big topics in transport in plant:

- Osmosis and Diffusion: different movement patterns of the water and mineral molecules

- Xylem and Phloem of Root, Stem and Leaf: the cross section and functionalities of xylem and phloem inside the plant

- Photosynthesis: the way the energy and oxygen are generated inside the leaf with water, light and carbon-dioxide

For each topic, three difficulty levels are also set as the sub-level. The difficulty level 1 is using concept map to teaching basic concepts; difficulty level 2 is giving examples to show the concrete structure; and difficulty level 3 is conducting experiments to examine the related phenomenon. For each difficulty level, teachers can define multiple tasks by editing the transition which is towards the related node. For example, we add a node 155 in Figure 5.14, and edit this node as "Experiment 1" with the description "Ink diffusion". In this way, the task "Experiment 1" is established. The interface fields with gray words are purposely faded in order to remove the inactivated functions so as to facilitate users to ignore the unrelated functions.

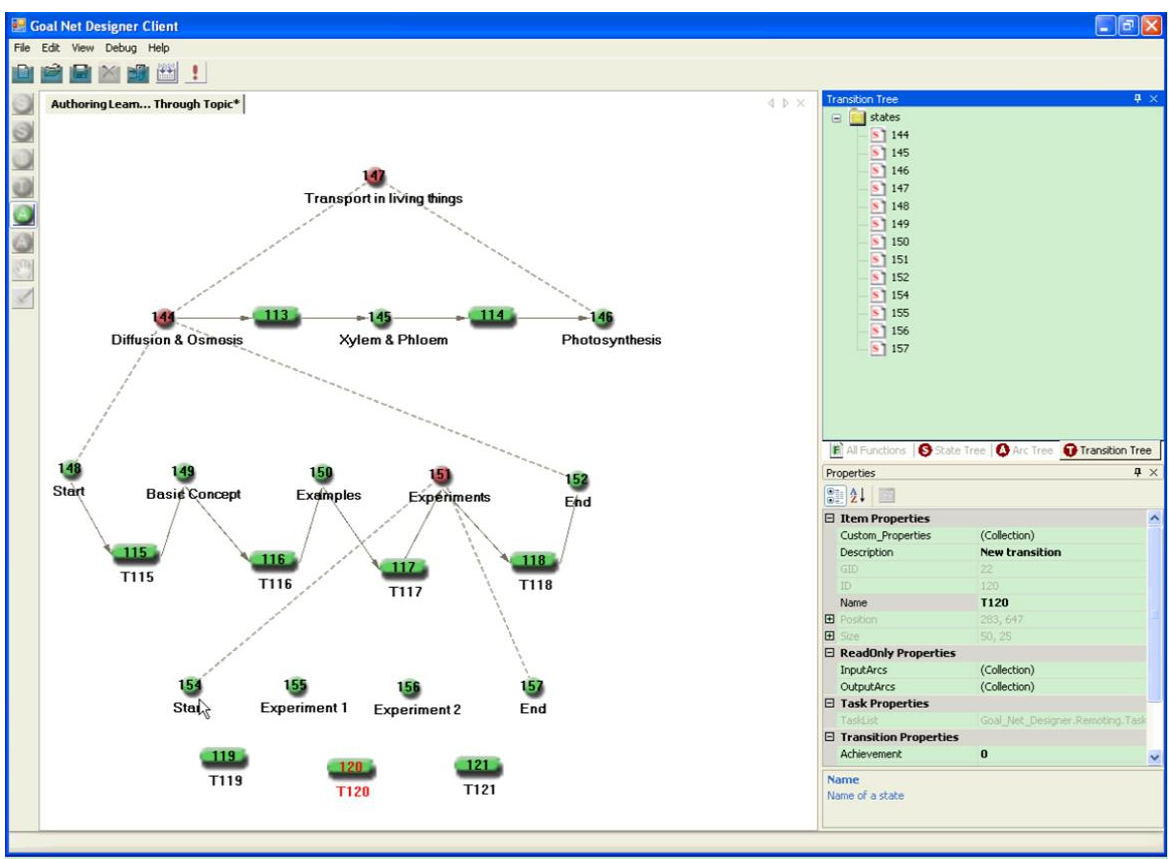

Figure 5.14: ATA Authoring for VS Project

For designing the detailed game script, the teachers can add function descriptions into the task by right clicking the transition. An editing window will pop up for writing the task functions. The teacher can write a layman word description for game developers to do coding. If some reusable functions are already existing, teachers can drag the function from the function list to the transition rectangle, and the system will automatically add the function into the corresponding task. This scenario has been designed as an osmosis experiment in a laboratory in a virtual world (Figure 5.15). The task list in Goal Net transition includes four tasks:

- Come to the place (the third experiment desk)

- Trigger the corresponding task (the third experiment)

- Check whether the task is successfully completed

- Display the learning goal setting page

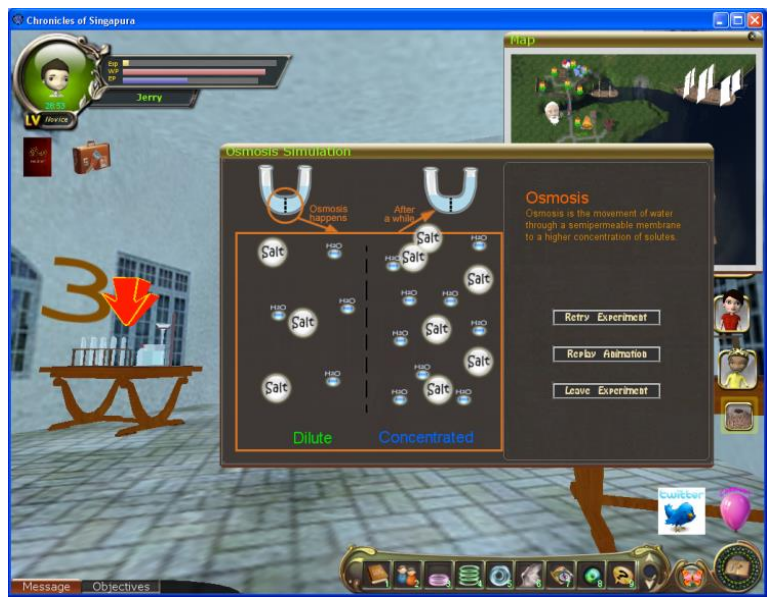

Figure 5.15: Interface of Experiments

## 5.2.4 Learning Path Customization

After the educator's design, the game developers create the tasks in transitions via editing transition functions. Through establishing the connection between Goal Net Designer and the game engine, the students can customize their learning path through selecting existing learning goals in the system at the beginning of each game scenario.

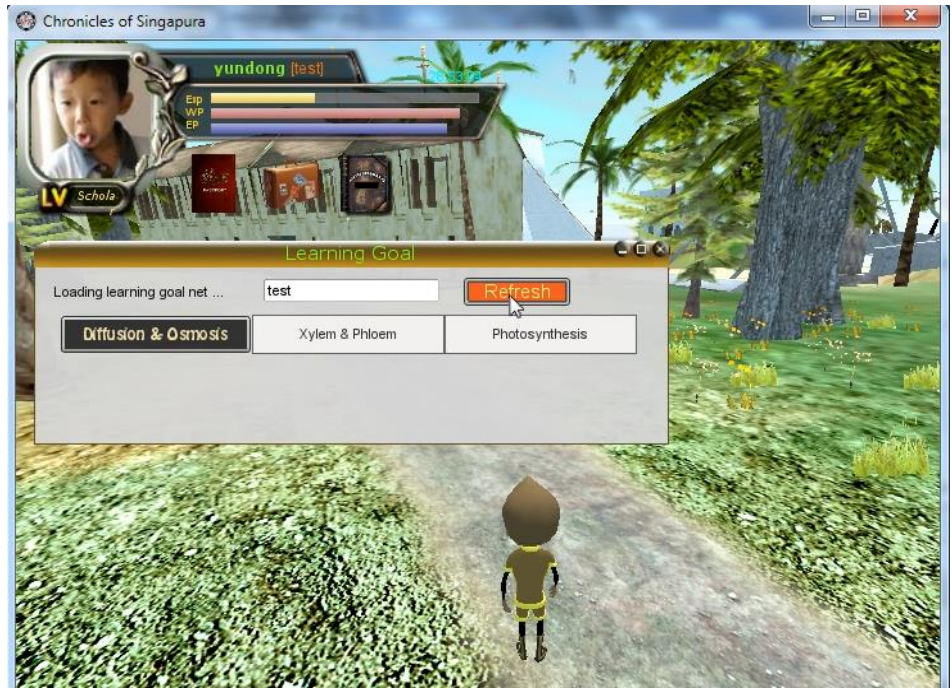

Figure 5.16: Learning Goal selection 1

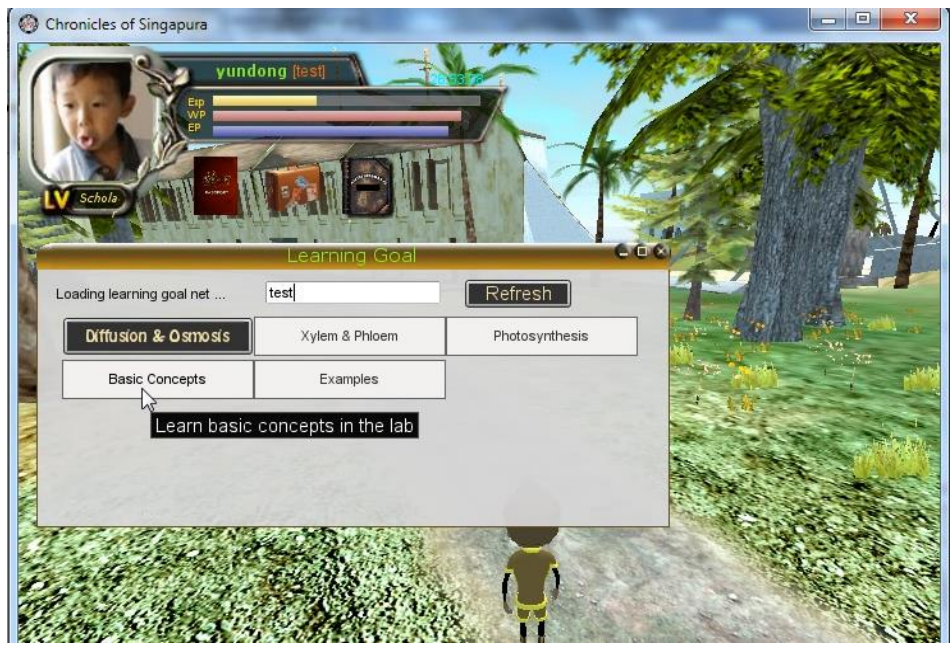

Figure 5.17: learning Goal Selection 2

From the virtual world interface, we provide an operation window to let students select their own learning path. After selecting a certain learning goal, the related learning tasks will be triggered, and the system will teleport the student to the task location in the virtual world. Figure 5.18 shows samples of student generated learning goal structures for VS project.

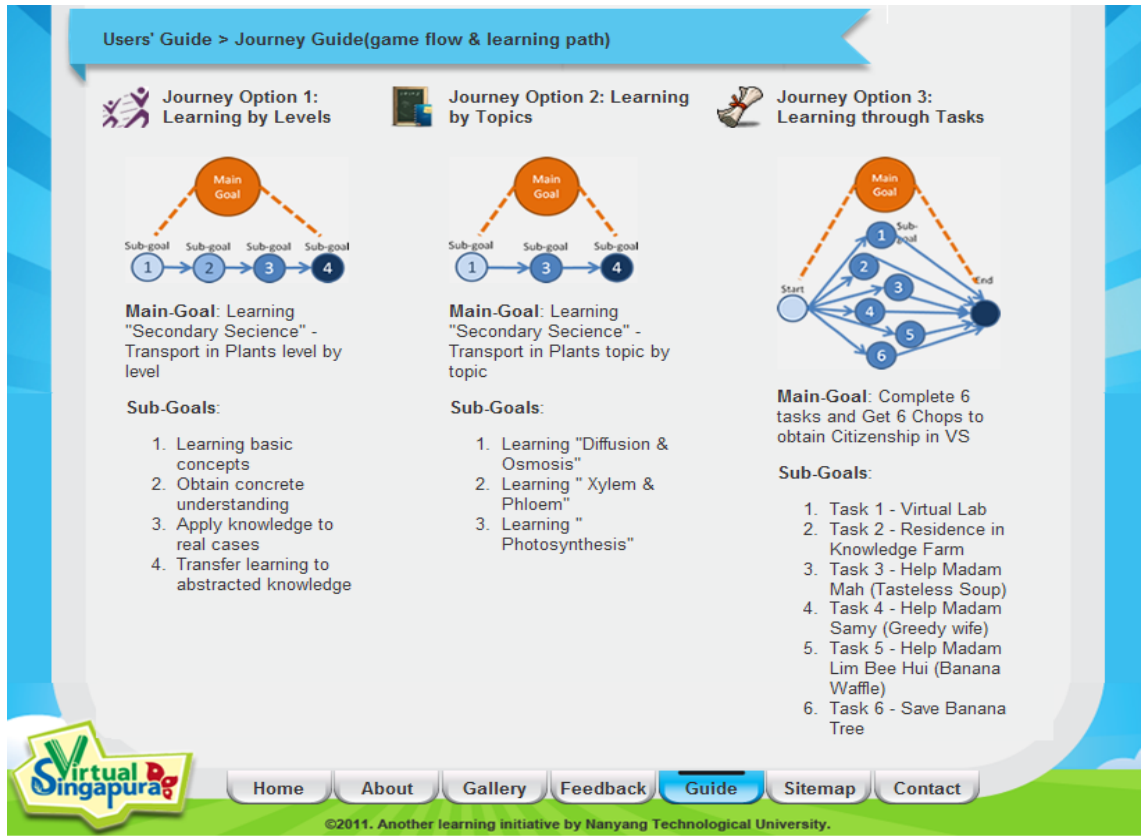

Figure 5.18: Samples of student Generated Learning Goal Structures

# 5.3 Assessment Design

Before designing assessments for the ATA system, the first step is to clarify the goals of evaluation. In this research, the goals of evaluation include three perspectives. Specifically, 1) whether a student's learning outcome improves; 2) whether a student's learning motivation and self-efficacy improve; and 3) whether the authoring tool helps in the design of the TA. To achieve these assessment goals, we involve both formative and summative approaches.

### 5.3.1 Formative Assessment

Formative assessments are used for "taking the pulse" of student's learning, which can be graded as ungraded, low-stakes assignments. We aim to evaluate the effectiveness of the TA design and reveal the insights for further modification. The assessment focuses on subjective and descriptive evaluations through questionnaires or observations. This type of assessment should carry out throughout the life cycle of the TA system design and attempts to find improper design and limitations. We use questionnaires with multiple-choice questions (MCQ) and open-ended questions to collect student's feedbacks from four aspects:

- Cognitive walkthrough of the TA

- Information about the interface

- Information about students' learning outcome

- Cognitive walkthrough of the authoring tool

### 5.3.2 Summative Assessment

The summative assessment is used for evaluating the ultimate effectiveness of the ATA system with quantitative approaches. We use concept map, pre- & post-test, and questionnaires to do the summative assessment.

A Concept map, according to [123], can be used as an assessment tool to reflect student's knowledge. The authors also recommended using a master map to assist scoring. Therefore, in our test analysis, we use a concept map to examine the learning outcome through pre- and post-test. We developed the detailed procedure based on the coding scheme from [42] to evaluate the quality of student maps.

First, we group concepts and relationships in student maps to four categories. These are:

1) Master: the concepts and relationships in the master concept map which is provided by school teachers.

2) Relevant: the correct or relevant concepts and relationships according to the evaluation scheme found in the learning materials

3) Irrelevant: the concepts and relationships which are irrelevant to the learning materials

4) Uncodable: the concepts and relationships which are incomprehensible or ambiguous

Then we count:

nMConcept: the number of the master concepts

nRConcept: the number of the relevant concepts

nMProposition: the number of the master relationships

nRProposition: the number of the relevant relationships

With these counting data, we calculate the mean values and the standard deviations of students in pretest and posttest, and do the comparison with a t-test.

Apart from the concept map drawing, the pre- and post-tests also include MCQ questions and open-ended questions to examine student's domain knowledge, which is part of the prescribed curriculum.

We also use a questionnaire to survey student's attitude. Our game evaluation survey is generated based on two assessment instruments. First, the students' learning motivation and self-efficacy is derived based on the scale of "Self-efficacy" and "Intrinsic-value" in [124] and the scale "interested-disinterested" in [125]. Second, the effectiveness of intelligent agents is measured by a questionnaire derived from the Agent Persona Instrument (API) [126], which includes 4 sub-measures as shown in Figure 5.19.

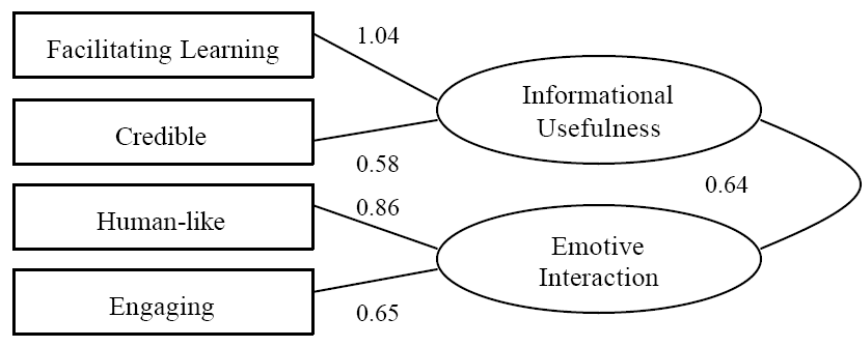

Figure 5.19: Factors and Correlation Coefficients between Factors in API Model [126]

The questionnaires used in High School are 7 point scales, from 1 (strongly disagree) to 7 (strongly agree). The questionnaires for Xingnan Primary School are changed to 5 point scales because of the lower age of the students. According to [127], "Smileyometer" design is used as Figure 5.20.

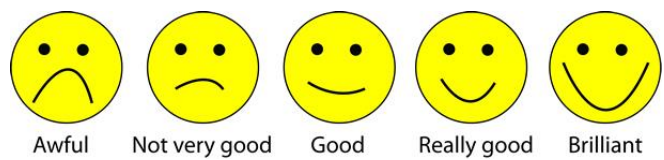

Figure 5.20: The Design of Smileyometer

## 5.4 ATA Assessment in VS Project

## 5.4.1 Formative Assessments during Design Time

The prototype of the VS project has been deployed in the Catholic High School in Singapore from 2009. The TA prototype in 2009 has no emotional expressions (i.e. the "Affectivability" features) but only the features related to "Teachability". The field study was used for lower secondary level science classes. A total of 71 Secondary Two students participated in our study. The students were divided into two groups according to their respective classes. A class with 34 students was set as a control group, who learnt the same topic in standard school classes with the same amount of learning time. The other class with 37 students formed sub-groups. Each group consists of 3 or 4 students, who collaboratively explored the VS learning environment with group mates.

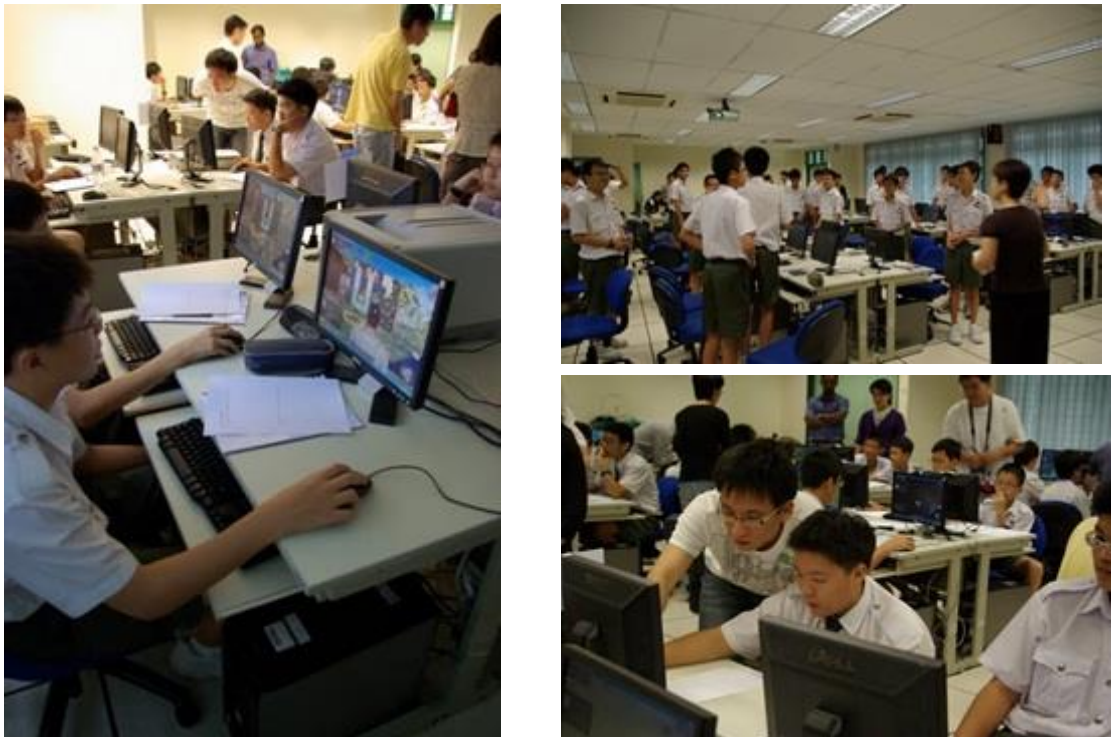

Figure 5.21: The Field Study at Catholic High School Singapore

The study included 4 sessions – two separate sessions of 45 minutes each were conducted to allow the students to go through all the learning activities in the prototype, one session of real world experiment and one session for post-test assessment.

The post-test consisted of 15 multiple choice questions (MCQ) and 3 open-ended questions that were part of the prescribed curriculum. This test design served as a benchmark, which measured the learning outcome of students by comparing the TA approaches with conventional instructional approaches. The treatment group worked with the TA, and the control group proceeded in a traditional classroom. From the results, there was no significant difference between the learning outcomes of the students with the TA and that of students without the TA. As it was a formative study to collect insights on improving the TA design, we focused on the lesson learnt from the unsuccessful design. This type of assessment had a limitation that it did not entail the prevailing features of the TA. That is, this kind of evaluation approach provided a limited analysis on each of agent modules. The effectiveness of each agent feature deserved a full analysis in detail.

Table 5.3. The Domain Knowledge Test Results in Post-Test

| Group | Statistics | Multiple Choice Questions (MCQ) | | Open-ended Questions |
|---|---|---|---|---|
| | | Basic Questions | Deep Understanding Questions | |
| Treatment Group | N | 37 | | |
| | Mean | 4.47 | 8.37 | 3.8 |
| | SD | 1.64 | 2.95 | 2.50 |
| Control Group | N | 34 | | |
| | Mean | 4.30 | 8.97 | 5.68 |
| | SD | 1.45 | 3.16 | 2.87 |

The students also completed a questionnaire rating each statement on the scale of 1 (strongly disagree) to 7 (strongly agree) and 3 open-ended questions to understand their learning experience in the system prototype.

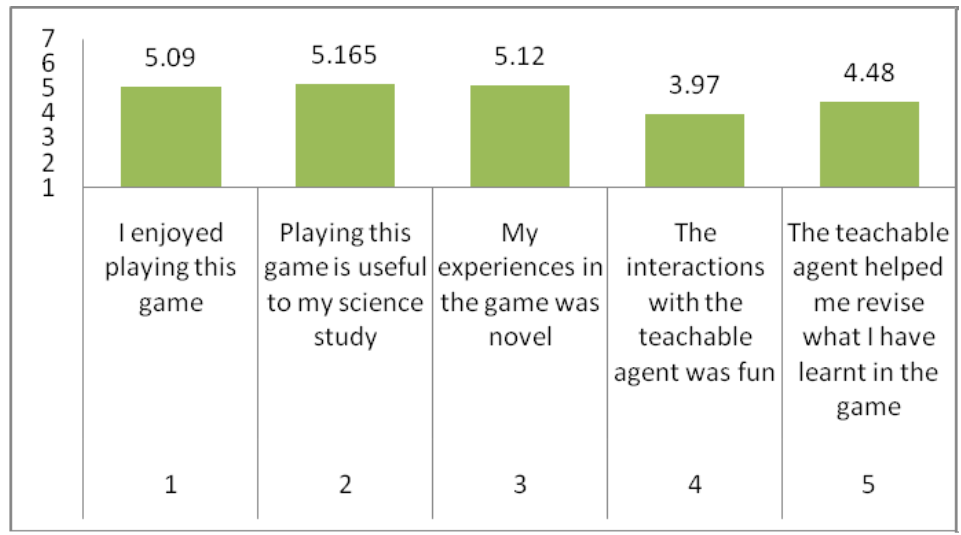

Figure 5.22: The Average Rating Score of the Questionnaire Organized by Categories

The overall rating for the game prototype showed that the agent-augmented learning environment offered students an engaging learning experience, but the interaction with the TA was not quite satisfactory. From an interview session, we found that students wanted to interact with a more believable TA, which would be more inviting. Therefore, we added emotional expressions to the existing TA and built up the ATA model in 2010. On the other hand, the concept maps used for teaching the TA were complicated for the students. The feedback indicated that splitting the big concept map into several smaller parts would be more suitable, since students could focus on a small number of concepts more easily rather than many of them. Thus, we re-designed the corresponding modules with several smaller concept maps in the new version.

On another perspective, we found some students were excited about the free exploration in 3D world, such as the comment: "The teleport function is very cool, dazzling and fantastic. When I teleport, I am excited and very curious where I will go." However, some of students argued that the open exploration in the virtual world made them feel uncertain and they had difficulty to link the game exploration with learning materials. In order to solve these practical problems, we developed a new version of VS with learning goal setting. During the subsequent design and development process, several informal tests among game researchers, school teachers, and young students have been conducted, and have received positive feedbacks.

For the assessment of the authoring tool, we used a small-scaled testing to get indications of the system design. We interviewed three teachers, two game designers and two program developers. Each interview lasted around two hours. Due to the small numbers of interviewees, the evaluation does not allow for statistical significance analysis. But as a formative evaluation, it is valuable to summarize the useful features and identify areas for improvement. The feedback is summarized as follows.

Table 5.4: The Interview Results from Teachers and Game Developers [128]

| | |
|---|---|
| From teach -er side | Advantages: |
| | • The graphical presentation of learning goals is clear and direct to depict the domain knowledge. |
| | • The drawing with Goal Net is easy to learn and convenient to use. |
| | Areas for improvement: |
| | • The authoring tool should provide more constraints to guide users how to complete the authoring. |
| | • Provide several basic structures for teachers to choose and fill in the blanks rather than totally drawing by themselves. |
| From game deve - loper side | Advantages: |
| | • The structured presentation is easier for programmers to resemble the object-oriented design which encapsulates different parts and functions into small groups. |
| | • The strategy for function reuse can reduce workload of coding. |
| | Areas for improvement: |
| | • If the authoring advances to the game development, it is ok to bring teachers all the flexibility to do the designing; but if the authoring is for regular updating of learning content, teachers should design based on existing 3D models and game logic. |

From the results we may find that the problems are focused on the tradeoff between the flexibility of user control and the constraints from built-in system settings. On one side, more flexibility of user control may allow teachers to do authoring with less rules and constraints. They may express their ideas more freely. But this is at cost of more workload for programmers and more expenses for game maintenance. Meanwhile, some of the teachers who are not very familiar with the authoring interface feel that there is insufficient guidance to follow. On the other side, more built-in constraints may hedge the author's creativity, but on the contrary the programmers can implement the game with a well-understood structure.

One way to solve the problem is to design a general framework as a knowledge carrier. For example, in VS we have a mission system which is used to assign tasks for students to complete. The general framework for loading learning content can be set in the mission format. For example, to complete a mission the student has to collect several virtual objects; students need to fully understand certain concepts by successfully filling in all the blanks of a designed concept map. The mission of collecting virtual objects is a general framework, but the concept maps embedded in with domain knowledge are authored by teachers. With the framework, the learning content updating can be easily achieved without a big change from the game development side.

In 2010, with the new features of TA design (affective modeling and separated concept maps), we did another round of tests in Catholic High School.

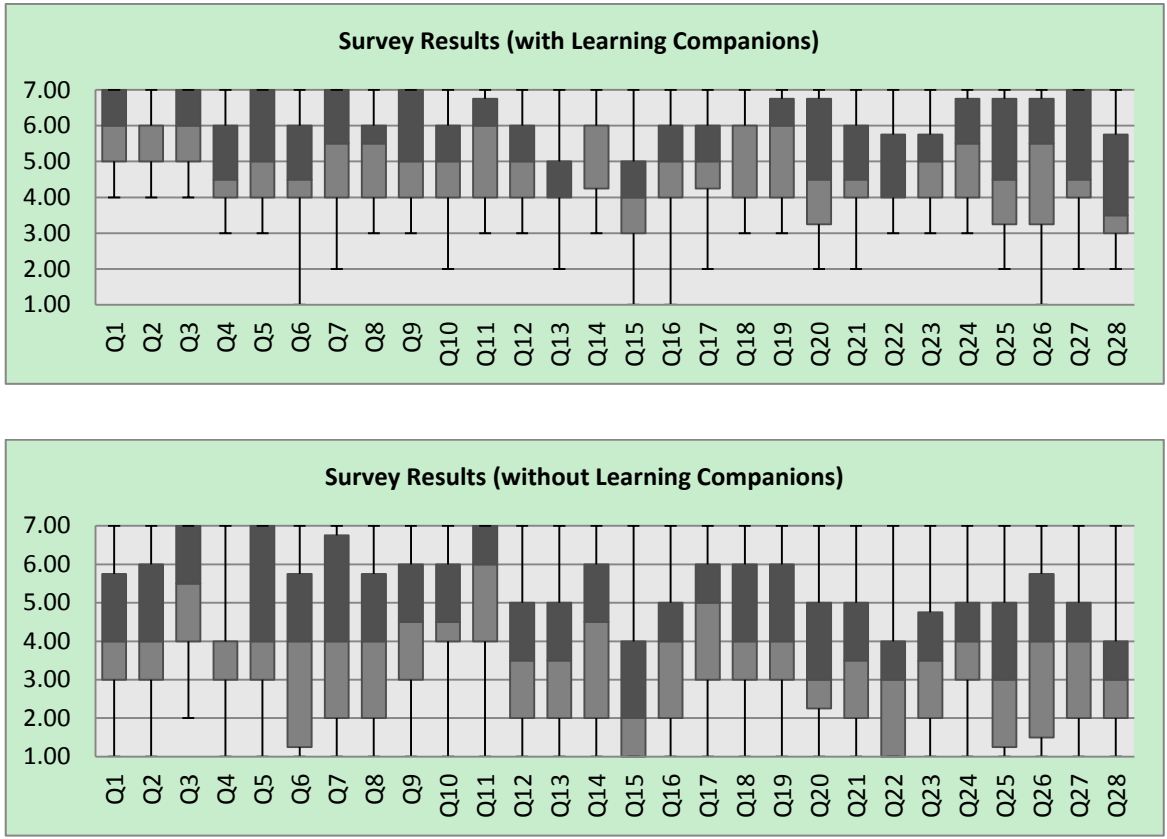

Figure 5.23: Box-plots (max, 75 percentile, median, 25 percentile, and min values)

The survey results consisted of 28 rating questions regarding the perceived ease of use of the prototype system, the interested retention effectiveness of the learning companions and the helpfulness of the learning companions in completing learning tasks. Figure 5.23 compares the distributions of the results from the students using a version of the prototype without the learning companions against the students using a version of the prototype with the learning companions. The x-axis represents the 28 questions and the y-axis represents the 7-scale scores. The box-diagram depicts the median and spread of the score obtained for each question. It can be observed that the survey results from the group with learning companions were better than the survey results from the control group without learning companions with statistically significant differences. A more clear way for showing the changes is as Figure 5.24.

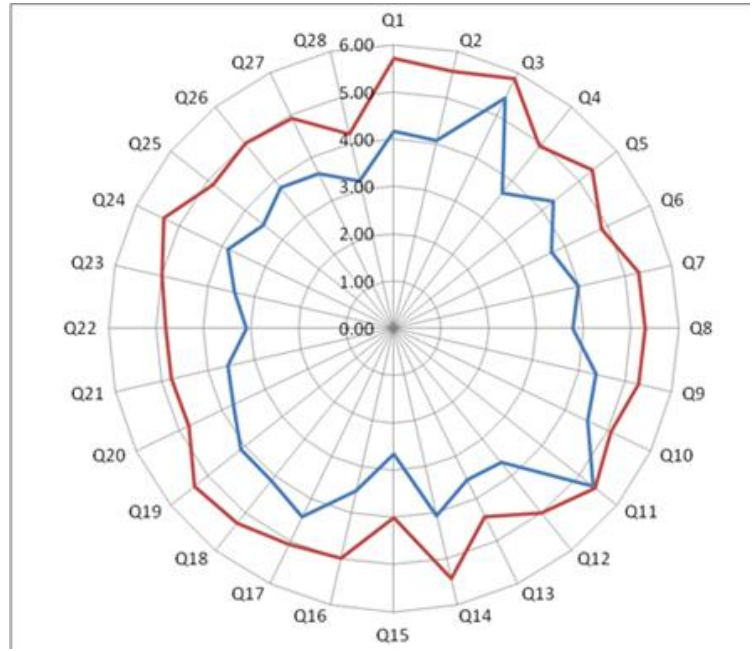

Figure 5.24: Survey Results Comparison between groups using prototypes with ATA and without ATA

## 5.4.2 Summative Assessments for ATA System

In 2011, we conducted the assessment in Xingnan Primary School in Singapore. The field study included two groups of students from 10 to 11 years old. The treatment group had 24 students who used our prototypes to learn diffusion and osmosis through two separated sessions of forty-five minutes each. The 24 control group students learnt the same topic on standard school classes with the same learning time. Both groups conducted a pre-test and a post-test with 20 multi-choice domain questions and 2 discussion questions related to diffusion and osmosis, which were approved and later examined by the teacher. For the multi-choice domain questions, one point was given to each correct answer, zero otherwise. In the treatment group, the scores of the students obtained were (Mean = 9.32, SD = 2.90) in pre-test and (Mean = 11.84, SD = 2.13) in post-test. We used paired samples t-test to compare the means of pre-test and post-test within the treatment group, and the result showed a significant difference with (t = 2.2, p < 0.01). For accessing the difference between the two groups, we analyzed the results with Analysis of Covariance using the pre-test scores as the covariate. However, the analysis showed the treatment was not significant (F (1, 21) = 0.97, p > 0.3). Nevertheless, with respect to the discussion questions, the treatment group had many more creative ideas with diverse explanations, e.g. they illustrated various approaches to extract salt from sea water.

In our opinion, the questions we used for the pre- and post-test were more for assessing academic results in traditional pedagogy. For the modern classroom, the virtual learning environment could be beneficial in improving a student's competencies, such as reflection of one's previous thinking, information filtering, and finding solutions from failures. However, the improvement over learning competencies cannot be directly inferred from the current design. In the future work, we aim to propose a new type of assessment, particularly for the virtual learning environment.

We also used concept maps to assess whether there were improvements of students' learning outcome in terms of number of concepts, relationships, elaborations and pictures. The results showed a significant increase, form pre-test maps to post-test maps, through the evaluations of three categories: concepts, elaborations and pictures (Table 5.5).

Table 5.5: Mean for Student Concept Map Components

| Map Components | Pre-test Maps1 | Post-test Maps1 | t-test | Effect Size (r) |
|----------------|----------------|-----------------|--------|-----------------|
| Concepts | 8.40 (5.23) | 13.50 (4.52) | -2.84 | 0.46 |
| Relationships | 7.70 (6.46 ) | 11.30 (8.55) | -0.94 | 0.23 |
| Elaborations | 4.10 (2.02) | 10.20 (6.34) | -4.12 | 0.54 |
| Pictures | 1.50 (2.72) | 6.40 (3.81) | -2.45 | 0.59 |

1 Values in parentheses denote standard deviation

2 $p < 0.05$

Significant results [105] were obtained for students' concept quantity (t = -2.84, $p < 0.05$), elaboration quantity (t = -4.12, $p < 0.05$) and picture quantity (t = -2.45, $p < 0.05$) from pre-test maps to post-test maps. The effect size (r = 0.59) for pictures represents an encouraging substantive finding for student's ability to convert information from short- to long-term memory and recall the information.

The treatment group also took a questionnaire of five rating questions on the scale from 1 (strongly disagree) to 5 (strongly agree) with the "Smileyometer" design [127] to evaluate the TA avatars from the 3 designing perspectives. The results are shown in Table 5.6.

Table 5.6: Students' Rating on TA through Questionnaire [105]

| Tested Agent Abilities | Examined Aspects in Questionnaire | Average Rating | SD |
|---|---|---|---|
| "To Learn from students" | Interest in teaching TA | 4.25 | 0.60 |
| "To Practice in 3D World" | Interest in entering the underground world | 4.75 | 0.43 |
| | Preference to view TA's practice and exploration after the teaching process | 4.67 | 0.22 |
| "To be Affective" | Favorable attitude towards TA | 4.67 | 0.22 |
| | Experience on interacting with TA | 4.33 | 0.52 |

The treatment group students chose multiple ways to seek the solution after they accepted to help the water molecule. 5 students searched from textbook after the trial-and-error of the concept map drawing; 1 searched from Google; 4 asked their classmates; and 2 continuously did the trial-and-error playing. We believe that all these behaviors are beneficial for students to learn the domain knowledge and find the effective way to acquire knowledge.

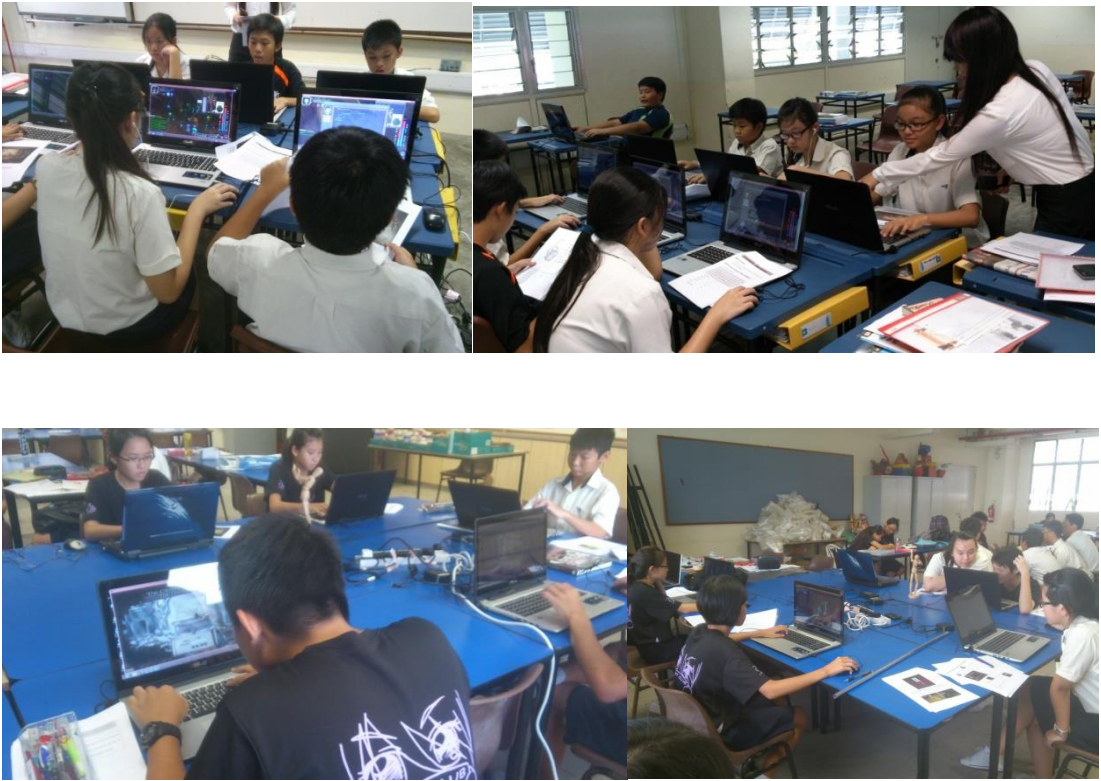

Figure 5.25: VS-II assessment system field study in Xinmin Secondary School

We did another study in Xinmin Secondary School in August 2013. The field study aims to examine whether the TAs with Affectivability and without Affectivability may have different impact on students' learning. We conducted 3 sessions of school test. 26 pairs of students finished the test. All the tests were deployed during student's Co-Curricular Activities (CCAs), and we randomly asked for volunteers in Class 208 each time. Before the study commences, students have not learnt the chosen topic at the secondary school level (but they did encounter this topic during their primary school years). During the study, each student was provided a user menu which included the instructions on game control and the introduction of the tasks they need to complete. Each session lasted for 40 minutes. When the test started, students sat at individual computers and log into the student portal. They signed up the system with their own account, and opened the assessment system which was immediately allowed to choose what their avatar look like.

They selected an avatar and entered the world. The students were supposed to finish three tasks in three game scenes (root, stem, and xylem). If they have more time when completing the three tasks, they can continue to another two game scenes (leaf and phloem). At the end of the study, students filled in a self-report questionnaire on their self-efficacy and the opinion on the TA. Four screenshots of the field study in 2013 are shown in Figure 5.25.

Table 5.7: Students' Rating on TA and ATA

| Tested Perspective | ATA | | TA | | t-test | $P_{two-tail}$ | $P_{one-tail}$ |
| | Average Rating | SD | Average Rating | SD | | | |
|---|---|---|---|---|---|---|---|
| Informational usefulness | 3.6000 | 0.6291 | 3.0974 | 0.3960 | 2.6185 | 0.0154 | 0.0077 |
| Emotive Interaction | 3.7259 | 0.4931 | 2.4861 | 0.7717 | 5.2434 | 2.55541E-05 | 1.27771E-05 |
| Students' self-efficacy | 3.3667 | 0.7841 | 2.8833 | 0.5334 | 1.9739 | 0.0600 | 0.0300 |

The treatment group had 15 students who used the version which the TAs (water molecules or the mineral salt molecules) have emotional reactions. The control group also had 15 st6udents who used the version which the TAs have no emotional reactions. The difference between the two versions of game is only on having emotions or not. The 26 post-test survey questions include three categories. As we have mentioned in Section 5.3.2 –summative assessment design, 13 questions focus on the informational usefulness, 9 questions focus on the emotive interaction, and 4 questions on students' self-efficacy.

Significant results were obtained for students' attitude towards TA's informational usefulness (t = 2.6185, p < 0.05) and attitude towards TA's emotive interaction (t = 5.2434, p < 0.05) from the ATA (the TA with Affectability) to TA (the TA without Affectability). The questions on emotive interaction focus on whether the interaction with TA is engaging and human-like, and the questions on informational usefulness focus on whether the TA is credible and facilitates students' learning. From the results, we can find out that the feature of Affectivability on one side can make students more engaged in the game and on the other side can facilitate students to more effectively interact with the interface design. When the TA does not have the affective capability, students' attitudes on the whole system decreased correspondingly. Students' self-efficacy did not obtain significant difference between the two groups (t = 1.9739, p > 0.05). As the self-efficacy is affected by many factors throughout a long term period, we plan to track student's self-efficacy data for a longer tested period in the future work.

This experiment design is the so called "intervention versus ablated intervention" design, in which one group works with the complete TA and another group works with the same TA in which the Affectivability is deleted. We removed the design feature, compared the user experience differences, and identified that whether the Affectivability contributes to TA system. The advantage of this design is that it is a more rigorous way to evaluate the ATA; it measures the added benefit for Affectivability module and measures student aptitude by treatment interaction. But the disadvantage is that this kind of comparison is resource intensive, and the data size is limited.

From the assessment results, we also found that the current system lacked an assessment tool to analyze students' behavior data, which can take the advantage of the virtual learning environment to evaluate student's learning skills by collecting behavior data. Currently, the system assessment results were derived by academic tests and questionnaires from students' learning outcome and their preference for TAs. However, since our project is a computer-based system, it is convenient to collect student learning achievement data and the learning behavior data during their play. We believe the new types of data will reveal more information which cannot be discerned from traditional assessments. For instance, during the same period of time, students who attempt more tasks may possess a greater desire in learning than those who attempt fewer. Similarly, students who deal with tasks of higher difficulty levels may have greater curiosity and may be more willing to face challenges. In future work, we plan to incorporate new functions into the existing system to record all types of user behavior data in the virtual learning environment. The data will then be analyzed to assess student characteristics related to self-directed learning, such as self-regulation, learning motivation, reflective thinking skills. We will also conduct more experimental and classroom field studies to assess students using both questionnaires and the collected behavior data.

## 5.5 Summary

Chapter 5 exemplified the realization of the proposed ATA model from Teachability design and Affectivability design, to game authoring and learning path customization. The VS project is used as a show case to illustrate how an ATA-enhanced educational software can be implemented based on the agent design and the Learning-by-Teaching theory. From the formative assessments, we summarize our findings in four perspectives,

- The explicit representation of knowledge is crucial for the teaching panel design, and big, complex tasks should be broken down into smaller, simpler components with series of steps to integrate with the broad knowledge network.

- The affective interaction between TA and students has significant influence on the motivations of students. They prefer human-like and believable interactions from the computer-based avatars in order to achieve close emotional binding with agents.

- Students want to receive clear guidance on what tasks need to be completed and have the feeling of direct control about the learning pace in an open environment.

- For domain knowledge related game authoring, the key problem is at the tradeoff between the flexibility of user control and the constraints from built-in system settings. We need to develop a generalized platform for educators to do authoring on case-dependent game scenarios.

From the summative assessments, we can find that the proposed ATA model can stimulate student's learning motivation and foster self-efficacy, and improve their learning outcome to certain extent. However, the traditional assessment approaches such as pre- and post-tests with MCQ, open-ended questions, and questionnaires cannot truly reflect all the benefits of computer-based educational software. We need to develop a new type of assessment approach to evaluate student's performance, and elicit useful information from students' behavioral data. The ATA system as an extendable educational tool is an ideal platform to potentially conduct the behavioral assessment. The details will be recommended in future work.

Chapter

# 6

# Conclusions and Future Work

## 6.1 Conclusions

In this book, we proposed a new type of teachable agent—the Affective Teachable Agent, and used a goal-oriented approach to design and implement the agent system, allowing agents to proactively interact with students with affective expressions. This design endowed TAs with abilities to learn and practice in a virtual world in order to assist students to reflect on their learning process, and consequently improved students' learning experience and encouraged them to take learning responsibility.

In Chapter 1, we introduced the research background of teachable agents and its significance for pedagogical systems in practical education. Through analyzing the challenges and educational requirements, we specified research problems and objectives. In Chapter 2, we reviewed existing TA systems and highlighted the problems associated with current TA design. From these reviews, we found that existing TAs lacked enough proactive interactions with students and also lacked believability to arouse a student's empathy on TA's poor performance.

As a consequence, in Chapter 3, we began with analyzing the teaching goals of designing TA system, and proposed a new type of TA, ATA, with two prospective capabilities. Teachability focused on learning new knowledge from students and doing practice in a 3D virtual world. Affectivability focused on generating proper emotions based on current events. A goal-oriented approach, Goal Net, was used to allow TAs to reason and act according its goal structures. The routine Goal Net selected proper Sub Goal Nets according to students' learning progress and their actions in a virtual world. The Sub Goal Nets then run the corresponding actions to control the TA. For Teachability, the system needs to clearly express the request of student's teaching, and selects the proper teaching panel for students to complete the teaching tasks. For Affectivability, one task is to qualitatively elicit what emotions are to be presented, and the other task is to quantitatively compute the emotion intensity. As emotions are difficult to quantize, we involved a fuzzy tool, Fuzzy Cognitive Map, to interpret the emotional model OCC to achieve the emotion generation process.

To practically realize the theoretical model of ATA, Chapter 4 introduced the system design framework of Goal Net, Multi-Agent Development Environment (MADE), to deploy the ATA model. We also developed an authoring tool for educational experts to use for game design. This brought an easy way for educators to add updated domain content during the game maintenance process. We also provided a feature for students to select their personalized learning path when playing the game.

In Chapter 5, the Virtual Singapura project was used as a show case to exemplify the development of ATA in real educational software. Formative and summative assessment approaches and results were elaborated from two perspectives. On one side, the results showed the usefulness of the ATA system. It can improve students' learning motivation and learning outcome. On the other side, we introduced the design heuristics, which can be used as guidelines for future implementation of ATA system.

## 6.2 Future Work

This research focused on the TA modeling and described the evolution of a new type of TA from theoretical modeling to practical implementation. Regarding future work, we emphasize and discuss new trends in assessing agent-augmented virtual learning environment.

As we have mentioned at the end of Chapter 5, traditional assessment approaches cannot reflect all the benefits of computer-based educational software. We need to develop a new type of assessment approach to evaluate students' performance, and elicit useful information from students' behavioral data. The ATA system as an extendable educational tool provides an ideal platform to potentially realize this new assessment approach.

### 6.2.1 The Changing Scope of Assessment

Sound assessment can be a useful learning tool which is both a barometer of how well things are progressing as well as a compass indicating future direction [129]. The traditional academic assessment approaches cannot thoroughly reflect the student competencies which are crucial for students to thrive in a fast-changing world. According to [130], the core values and the related capabilities are more focused on the non-academic dispositions of a child. A sound assessment not only requires a clear purpose for assessment, but also proper methods, an appropriate sample of the target, and elimination of bias and distortion in measurement. Therefore, the new criteria of educational assessment urge new assessment methods.

Educational data mining is an emerging discipline concerning the development of methods for analyzing types of data that are uniquely from educational settings, and utilizing those methods to foster a better understanding towards the relationship between students and the learning environment [131]. Methods for mining educational data have been found from a variety of literature work, such as machine learning and data mining, psychometrics and statistics, and information presentation with visualization, as well as computational modeling. [132] divided educational data mining into two schools, namely web mining and statistics with visualization. Another viewpoint on educational data mining from [133] specified this field into five perspectives, which were prediction, clustering, relationship mining, model discovery, and distillation of information for human justification. Academics and educational researchers have found that education systems based on games that are highly interactive can facilitate the realization of recommendations stated by researchers from education [134-136]. This kind of system can well support the learning of students. Indeed, Squire [135] wrote that the question was not "whether educators can use games to support learning, but how we use games most effectively as educational tools."

To the best of our knowledge, the largest ongoing research project to date on content learning in a 3D virtual environment is being conducted by Harvard University's Chris Dede group [137]. A recent paper on this work describes the use of the RIVER CITY MUVE system by approximately 700 students in grades five to eight in two different U.S. cities. It was found that the experimental group students who used the RIVER CITY MUVE over a two-week period had a significantly higher gain in science content knowledge and science inquiry skills compared to the comparison condition students who used a paper-based version of the science inquiry curriculum. Another important finding in this study was that compared to controls, students in the experimental group were highly engaged in their learning activities with the system, had improved attendance and less disruptive behavior, and made significant learning gains. Our VS project, which was developed based on the experience of the RIVER CITY project, also has the potential to integrate the educational data mining based assessment approach into the agent-augmented virtual environment.

Conducting assessments in a virtual learning environment has many potential benefits:

- It provides opportunities to collect multiple types of student learning behavior data such as location-based data, time-based data, mouse clicking data, and keyboard input data.
- It has the flexibility to provide students with a highly immersive learning experience.
- It can be scaled up to support a massive number of users relatively easily.

For future work, we plan to re-construct the VS project to record a wide range of user-generated behavior data during their learning process in the virtual environment, and analyze the data with advanced educational data mining techniques to assess skills which cannot be evaluated in conventional assessment, such as the skills in the framework of 21st Century Competencies [130].

## 6.2.2 Evidence-Centered Design for Behavioral Assessment

An assessment serves the purpose of finding related inference evidence, regarding a particular expectation from students [138]. Educational researchers have won great achievement in the design of assessment methods. Like Evidence-Centered Design (ECD) [139] and Assessment Triangle [140] methods offer rigorous frameworks to combine the theories and interpretations. ECD framework is a great choice to provide guaranteed validity of an assessment. There are four phases of design in ECD framework, including problem analysis, case modeling, conceptual assessment and compilation. Domain analysis and modeling emphasize the assessment goal, the nature of knowledge, and the structures of experiments as well as knowledge organization. Conceptual assessment exploits the Assessment Triangle method, for which the experiment designers pay attention to the skill model that are to be evaluated, the evidence of which behaviors can elicit the skills addressed, and the task for which the desired behaviors can be generated. Similar to the above method, all the models of the assessment design are related to each other. During the compilation of conceptual assessment, the framework creates multiple tasks. The aim of compilation is to identify and define models for task authoring with schemas and protocol developing for examining and analyzing psychometric models. The delivery model considers the presentation of tasks and the evaluation of tasks.

As an extendable system, ATA also can be used for collecting student data. Conceptual system architecture can be illustrated as Figure 6.1. Three types of student data can be collected from the existing system as
Table 6.1.

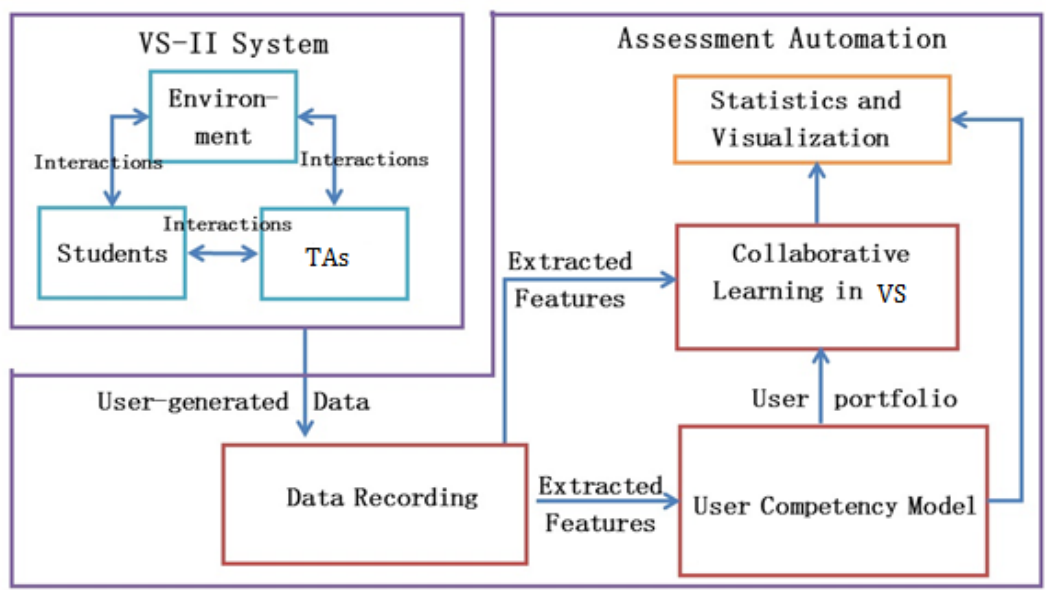

Figure 6.1 System architecture – Using ATA System for Collecting Student Data

Table 6.1: User Data Types to be Potentially Tracked in VS

| | |
|---|---|
| **Student Learning Behavior Data** | Location data ( position of user's virtual avatar in the 3D world) |
| | Timestamp (every 0.5 second) |
| | User's avatar posture data (walk, fly, facing direction) |
| | Mouse click on environmental events or objects (openning a window, collecting items, checking map, openning notes, chatting online, tweeting…) |
| **Student Learning Achievement Data** | Collected items which mark their experience level ("badges", "passport chops") |
| | Fulfilled missions and completed learning tasks (complete "experiment 1", "helped water molecule to be absorbed by tree") |
| | Students' performance in answering quiz questions |
| | Students' performance in drag-and-drop test |
| | Students' drawing of concept map |
| **Student Knowledge Data** | Hints requested, repetitions of wrong answers, correctness of responses, errors made by students, and time spent before making an attempt; |
| | Feedback of system (i.e., practiced skills, including skill types and previous opportunities) |

Table 6.2: Competencies to be Assessed and the Corresponding Learner-Generated Evidence

| Competencies for the 21st Century | Specific Skills to be Measured | Related User Learning Behavior Data in VS-II |
|---|---|---|
| **Self-directed Learning Skills** | Having desire to learn and explore | Track & measure the variety and scope of student's movement (exploration route in virtual world, number of clicked objects, fulfilled tasks…) |
| | Not be afraid to make mistakes and face challenges | Track the selection of difficulty level and the subsequent movement after making mistake; |
| | Persevere on learning tasks | Try multiple ways or attempt many times on one difficult task |

As a prototype, we will focus on the analysis of three skills for assessing the competency of self-directed learning. In the ATA system, the knowledge is embedded as learning objects and learning tasks in predefined locations. Student's historical behaviors (such as route of exploration, number of learning tasks undertaken) can reflect a student's desire to learn and explore. For example, over the same period of time, a student who has a complex route of exploration with broader breadth and depth may be considered to demonstrate a higher desire to explore than a student who has a simpler route with a superficial exploration. Also, in the same period of time, students who attempted more teaching tasks may show a higher desire to learn than students who only attempted a very limited amount of teaching load. Similarly, the teaching tasks which have different level of difficulty may induce different level of challenges to the students. The students' teaching behavior has the potential to reflect their desire to face challenges. For example, for a difficult task, a student who makes several attempts to finish may show higher desire to face challenge than a student who just leaves the task. These intuitions are listed in **Error! Reference source not found.**. They are only the basis from which we will start ur investigation. More sophisticated educational data mining methods need to be designed to obtain accurate assessments on the related learning skills.

We know that the use of agent-augmented virtual learning environment for summative assessment in a standardized fashion is still in its infancy. It is a good time for us to work on this direction. With the help of the behavioral/performance assessment, we can finally find out the proper way to assess the ATA system and reversely use the findings of assessment to guide and enhance the system design.

# Author's Publications

## Conference

## Journal